\documentclass[]{aa} 
\usepackage[varg]{txfonts}
\usepackage{longtable,lscape}
\usepackage{rotating}
\usepackage{graphicx}
\usepackage{natbib}
\usepackage{txfonts}
\usepackage{color}

\begin{document}
\newcommand{\fe}{[\ion{Fe}{ii}]}
\newcommand{\fii}{[\ion{Fe}{i}]}
\newcommand{\Ti}{[\ion{Ti}{ii}]}
\newcommand{\tii}{[\ion{Ti}{i}]}
\newcommand{\si}{[\ion{S}{i}]}
\newcommand{\sii}{[\ion{S}{ii}]}
\newcommand{\siii}{[\ion{S}{iii}]}
\newcommand{\SiIII}{[\ion{Si}{iii}]}
\newcommand{\SiII}{[\ion{Si}{ii}]}
\newcommand{\oi}{[\ion{O}{i}]}
\newcommand{\oip}{\ion{O}{i}}
\newcommand{\pii}{[\ion{P}{ii}]}
\newcommand{\Ni}{[\ion{N}{i}]}
\newcommand{\Nii}{[\ion{N}{i}]}
\newcommand{\Nip}{\ion{N}{i}}
\newcommand{\caii}{\ion{Ca}{ii}}
\newcommand{\cai}{\ion{Ca}{i}}
\newcommand{\phoi}{[\ion{P}{i}]}
\newcommand{\cip}{\ion{C}{i}}
\newcommand{\he}{\ion{He}{i}}
\newcommand{\mgip}{\ion{Mg}{i}}
\newcommand{\mgiip}{\ion{Mg}{ii}}
\newcommand{\nai}{\ion{Na}{i}}
\newcommand{\brg}{Br\,$\gamma$}
\newcommand{\pab}{Pa\,$\beta$}
\newcommand{\fei}{\ion{Fe}{i}}
\newcommand{\feii}{\ion{Fe}{ii}}
\newcommand{\hei}{\ion{He}{i}}
\newcommand{\sip}{\ion{Si}{i}}
\newcommand{\mdot}{$\dot{M}_{jet}$}
\newcommand{\mjet}{$\dot{M}_{jet}$}
\newcommand{\mh}{$\dot{M}_{H_2}$}
\newcommand{\Ne}{n$_e$}
\newcommand{\h}{H$_2$}
\newcommand{\kms}{km\,s$^{-1}$}
\newcommand{\um}{$\mu$m}
\newcommand{\lam}{$\lambda$}
\newcommand{\msyr}{M$_{\odot}$\,yr$^{-1}$}
\hyphenation{mo-le-cu-lar pre-vious e-vi-den-ce di-ffe-rent pa-ra-me-ters ex-ten-ding a-vai-la-ble}
%

\title{POISSON project -- II -- A multi-wavelength spectroscopic and photometric survey of young protostars in L\,1641
\thanks{Partially based on observations collected at the European Southern Observatory
 La Silla, Chile, 082.C-0264(A), 082.C-0264(B)}}

\author{A. Caratti o Garatti \inst{1},
        R. Garcia Lopez \inst{2},
        S. Antoniucci \inst{3},        
        B. Nisini \inst{3},
        T. Giannini \inst{3},
        J. Eisl\"{o}ffel \inst{4},
        T.P. Ray \inst{1},
        D. Lorenzetti \inst{3},
        \and
       S. Cabrit \inst{5} 
}
\institute{
Dublin Institute for Advanced Studies, 31 Fitzwilliam
Place, Dublin 2, Ireland \\ 
\email{alessio;tr@cp.dias.ie}\\
\and
Max-Planck-Institut f\"{u}r Radioastronomie, Auf dem H\"{u}gel 69, D-53121 Bonn, Germany\\
\email{rgarcia@mpifr-bonn.mpg.de}\\
\and
INAF - Osservatorio Astronomico di Roma, via Frascati 33, I-00040 Monte Porzio, Italy\\
\email{antoniucci;nisini;giannini;lorenzetti@oa-roma.inaf.it}\\
\and
Th\"uringer Landessternwarte Tautenburg,
Sternwarte 5, D-07778 Tautenburg, Germany\\
\email{jochen@tls-tautenburg.de}\\
\and
LERMA, Observatoire de Paris, 
Avenue de l' Observatoire 61, 75014 Paris, France\\
\email{sylvie.cabrit@obspm.fr}\\
}

   \date{Received ; accepted }
\titlerunning{A multi-wavelength spectroscopic and photometric survey of young protostars in L\,1641}
\authorrunning{Caratti o Garatti et al.}

 
  \abstract
   {Characterising stellar and circumstellar properties of embedded young stellar objects (YSOs) is mandatory for understanding the early stages 
   of the stellar evolution. This task requires the combination of both spectroscopy and photometry, covering the widest possible wavelength range, 
   to disentangle the various protostellar components and activities.}
   {As part of the POISSON project (Protostellar Optical-Infrared Spectral Survey On NTT), we present a multi-wavelength spectroscopic and photometric 
   investigation of embedded YSOs in L\,1641, aimed to derive the stellar parameters and evolutionary stages and to infer 
   their accretion properties.}
   {Our multi-wavelength database includes low-resolution optical-IR spectra from the NTT and Spitzer (0.6-40\,$\mu$m) and
   photometric data covering a spectral range from 0.4 to 1100\,$\mu$m, which allow us to construct the YSOs spectral energy distributions (SEDs) 
   and to infer the main stellar parameters (visual extinction, spectral type, accretion, stellar, bolometric luminosity, mass accretion and ejection rates).}
  {The NTT optical-NIR spectra are rich in emission lines, which are mostly associated with YSO accretion, ejection, and chromospheric activities. 
  A few emission lines, prominent ice (H$_2$O and CO$_2$), and amorphous silicate absorption features have also been detected in the Spitzer spectra. The SED analysis allows us 
  to group our 27 YSOs into nine Class\,I, eleven Flat, and seven Class\,II objects.
  However, on the basis of the derived stellar properties, only six Class\,I YSOs have an age of $\sim$10$^5$\,yr, while the others are older (5$\times$10$^5$--10$^6$\,yr), 
  and, among the Flat sources, three out of eleven are more evolved objects (5$\times$10$^6$--10$^7$\,yr), 
  indicating that geometrical effects can significantly modify the SED shapes.
  Inferred mass accretion rates ($\dot{M}_{acc}$) show a wide range of values (3.6$\times$10$^{-9}$ to 1.2$\times$10$^{-5}$\,M$_{\sun}$\,yr$^{-1}$),
  which reflects the age spread observed in our sample well. Average values of mass accretion rates, extinction, and spectral indices decrease with the YSO class. 
  The youngest YSOs have the highest $\dot{M}_{acc}$, whereas the oldest YSOs do not show any detectable jet activity in either images and spectra.
  Apart from the outbursting source \#\,25 and, marginally, \#\,20, none of the remaining YSOs is accretion-dominated ($L_{acc} > L_*$).
  We also observe a clear correlation among the YSO $\dot{M}_{acc}$, $M_{*}$, and age. For YSOs with $t > 10^5$\,yr and $0.4\,M_{\sun}\leq M_*\leq1.2$\,$M_{\sun}$, a relationship 
  between $\dot{M}_{acc}$ and $t$ ($\dot{M}_{acc} \propto t^{-1.2}$) has been inferred, consistent with mass accretion evolution in viscous disc models 
  and indicating that the mass accretion decay is slower than previously assumed.
  Finally, our results suggest that episodic outbursts are required for Class\,I YSOs to reach typical classical T Tauri stars stellar masses.}
   {}

   \maketitle
%

\section{Introduction}
\label{intro:sec}

Our working model of low-mass star formation arises from a combination of the empirical
classification of the YSO SEDs~\citep[][]{lada-w,lada} with a theoretical picture of YSO formation, which
involves the collapse of an isolated rotating dense core, which then forms an accreting protostellar core and a disc~\citep[][]{adams86,adams87}. 
The empirical evolutionary sequence goes from Class\,0 to III objects. Class\,0 YSOs 
are the youngest sources ($\sim$10$^4$\,yr), while Class III (the so-called weak T Tauri stars) are the oldest ones
(10$^7$\,yr). The classification of Class\,I, II, and III objects is based on the slope of the SED between 2
and 20\,$\mu$m ($\alpha_{2-20\,\mu m}$), defined as $\alpha=dLog(\lambda F_\lambda)/dLog(\lambda)$~\citep{lada}.
On the other hand, Class\,0 YSOs, which are usually not visible at these wavelengths, 
are defined as having $L_{smm}/L_{bol}>$0.5\%, where $L_{smm}$ is measured longward of 350\,$\mu$m~\citep[][]{andre93,andre00}, and they 
have more than 50\% of their mass in the surrounding envelope. The youngest protostars (0, I) 
are thus extremely embedded, and are characterised by steeply rising SEDs from near to far-IR, mostly or
entirely coming from the emission of their surrounding envelope. 
According to this picture, most (in the case of Class\,I YSOs) or all (in the case of Class\,0 YSOs) 
the YSO luminosity is believed to come from accretion ($L_{bol}\sim L_{acc}\sim G M{_*} \dot{M}_{acc}/R_*$)
through a circumstellar disc. Part of the accreted material ($\sim$10\%) is ejected by means of powerful collimated jets. 
The more evolved Class II and III sources instead have smaller IR excesses, and 
their SEDs can be modelled well by pre-main sequence photospheres surrounded by circumstellar discs (i.\,e., classical T Tauri stars - CTTs). 
The luminosity of these ‘older’ objects mostly originates in their stellar photosphere, 
rather than from accretion processes. Thus, accretion and ejection activities are strongly reduced or even absent in the latest stages.

Our understanding of the YSO early evolutionary stages (Class\,0 and I), mostly relies on SED analysis or, indirectly, on 
studies of their jets and outflows. On the other hand, we have a clearer picture of the latest stages, 
for which the physical properties of the accreting protostars can be well studied and characterised.
Recent theoretical studies on SEDs~\citep[][]{whitney03-2,whitney03-1,robitailleFT} have shown that the SED analysis alone
may not be sufficient to disentangle the YSO evolutionary stage, and NIR spectroscopy is needed to characterise the embedded accreting protostellar core.
For example, geometrical effects may produce SED misclassifications, i.\,e. old objects observed edge-on may show Class\,I shapes, or
young YSOs face-on may appear older, showing Class\,II SED shapes.
Thus, a correct classification can be only obtained from the combined analysis of the YSO SED and the characterisation
of the embedded stellar object and its activity. Quantitative information on the various
phenomena characterising the environment of young stars can be derived from NIR spectroscopic
studies, which allow us to observe embedded objects, and investigate processes occurring in regions spatially unresolved,
like the accretion funnel flows, and the ejection of jets from the adjacent inner disc, by
using features related to the different emitting regions.

These considerations have recently triggered a series of observational studies, aimed at deriving the stellar physical properties and evolutionary 
status of the embedded YSOs, in particular of the Class\,I YSOs, which are visible at IR wavelengths.
As a result, these studies have indicated that only a fraction of Class\,I sources is
composed of highly accreting objects~\citep[see, e.\,g.][]{wh04,nisini05,doppmann,antoniucci}, which show evidence of jet activity. 
Several objects, classified as Class\,I, have indeed mass accretion/ejection rates similar to those of Class\,II.
Some of them are mis-classified, whereas others appear to be bona-fide young embedded objects~\citep{wh07}.
Indeed, the so far studied sample is still limited, biased, and mostly confined to a few selected star-forming 
regions~\citep[i.\,e. mostly Taurus, see, e.\,g.,][]{wh07,beck,prato}.

In this framework, we have undertaken a combined optical/IR unbiased spectroscopic survey on a flux-limited sample
of selected Class I/II sources, located in six different nearby clouds~\citep[namely Cha I \& II, L\,1641, Serpens, Lupus, Vela, and 
Corona Australis; see,][\textit{hereafter paper-1}]{antoniucci11}, using the EFOSC2/SOFI instruments on the \textit{ESO-NTT} 
(POISSON: Protostellar Objects IR-optical Spectral Survey On NTT).
In \textit{paper-1} we presented the results of the Cha I \& II regions, comparing $L_{acc}$ determinations from
the different tracers, and discussing the reliability and consistency of the different empirical relationships considered.
In this paper we report the survey results on L\,1641, largely complemented by archive and literature data, which allow us to characterise the 
studied YSOs.

At a distance of 450\,pc, the Lynds 1641 molecular cloud (L\,1641) is part of the Orion GMC complex~\citep[for a complete review see,][]{allendavis}. 
Southward of the ONC, L\,1641 extends from NW to SE for $\sim$2.5$\degr$, and contains hundreds of young stellar objects, ranging from high- to low-mass YSOs.
Moreover, the L\,1641 cloud has been producing stars for nearly 30\,Myr~\citep{allen95}, thus it harbours a very heterogeneous sample of YSOs,
which span from extremely active and young to old and quiet objects, making it the perfect candidate to study the different YSO evolutionary stages in 
an unbiased fashion.
Our observations were performed on 27 embedded YSOs selected in L\,1641 on the basis of the brightness and SED spectral index ($\alpha$). 
Our multi-wavelength database includes low-resolution optical-IR spectra from the \textit{NTT} and \textit{Spitzer} (0.6-40\,$\mu$m), as well as 
photometric data covering the spectral range from 0.4 to 1100\,$\mu$m, which allow us to construct the YSO SEDs 
and to characterise the object parameters (visual extinctions, spectral types, accretion and bolometric luminosities, mass accretion and ejection rates).

This paper is organised as follows.
Section~\ref{sample:sec} describes the selection criteria for our YSO sample. 
In Section~\ref{observations:sec} we define our observations, data reduction, and the collected literature data.
In Section~\ref{results:sec} we report on the results obtained from our photometry and spectroscopy, we describe
the detected spectral features and characterise the YSOs.
In Section~\ref{discussion:sec} we discuss the accretion properties of the sample, as well as the origin of the observed mass accretion evolution.
Finally, our conclusions are drawn in Section \ref{conclusion:sec}.

\section{Sample definition}
\label{sample:sec}

Our sample of YSOs in L\,1641 was selected from \cite{chen94} on the basis of the SED spectral index ($\alpha$) and NIR brightness of each source, 
because of the instrumental sensitivity.
In particular, we chose those objects showing $\alpha>$-0.4 and $K'\leq$12\,mag, implying that most of our targets are, in principle, Class I and flat YSO candidates.
According to te latest \textit{Spitzer} surveys, this brightness constraint limits our study to about 15\% of the 
entire embedded population in L\,1641~\citep[see,][and references therein]{allendavis}. 

It is worth noting that the $\alpha$ classification in \cite{chen94} is based on $K'$ and $IRAS$ 25\,\um\ photometry, and, as it will be shown in Sect.~\ref{photdata:sec} 
and Sect.\,\ref{phot:sec}, often differs from the 2MASS/Spitzer classification obtained in this paper (Sect.~\ref{phot:sec}).
This is mostly due to the low spatial resolution of the IRAS beam (1$\arcmin\times$5$\arcmin$ at 25\,$\mu$m) that is not enough to properly resolve different 
sources in crowded regions such as L\,1641. 
As a result, the 27 YSOs are grouped into nine Class\,I, eleven Flat, and seven Class\,II objects.
Additionally, some targets identified as single sources in \cite{chen94} turned out to be double YSOs after analysing 
the Spitzer images (namely \object{[CTF93]146}, \object{[CTF93]216}, \object{[CTF93]237}, and \object{[CTF93]245B}).

The targets, along with their ID, $SIMBAD$ name, and coordinates (J2000.0), are listed in Table~\ref{obs_spec:tab} (Columns 1--5). 
The sources have been named following the nomenclature of \cite{chen93} with the exception of those identified here as double YSOs, 
which have been labelled as \textquotedblleft\ target-1\textquotedblright\,, when coincident or closest to the literature coordinates,
and \textquotedblleft\ target-2\textquotedblright\,, the farthest.
The large separations ($\ge$40$\arcsec$) exclude that these sources are physically bound, with the exception of sources \object{[CTF93]237}, separated less than 5$\arcsec$ ($\sim$2250 AU
at 450\,pc). Another exception might be \object{[CTF93]216-1} and \object{[CTF93]216-2}. They might be part of a wide binary system (35$\arcsec \sim$15\,700\,AU, at d=450\,pc), and
both sources show two precessing jets in the \textit{Spitzer} images (see also Fig.~\ref{CTF216_img:fig} in Appendix~\ref{appendixB:sec}). However such a wide system would not produce precessing jets, thus
it is more likely that both sources are binary systems.

\begin{table*}
\caption{Targets and Observations.}
\label{obs_spec:tab}
\centering
\renewcommand{\footnoterule}{}  
\begin{tabular}{cccccccccc}

\hline \hline
ID  & Target name  & Other Name &    \multicolumn{3}{c}{$\alpha$\,(J2000.0)} & \multicolumn{3}{c}{$\delta$\,(J2000.0)} & Spectroscopy\tablefootmark{a}\\
    &              &             & ($^{h}$ & $^{m}$ & $^{s}$) & ($\degr$ & $\arcmin$ & $\arcsec$) &       \\
\hline\\[-5pt]
1   &  \object{[CHS2001]13811}  &  \object{2MASS-J05360665-0632171}  & 05&36&06.65  &  -06&32&17.1  & O,B,R,S \\
2   &  \object{[CTF93]50}       &  \object{NAME HH 147MMS}           & 05&36&25.13  &  -06&44&41.8  & O,B,R,S \\
3   &  \object{[CTF93]47}       &  \object{V* V380 Ori}               & 05&36&25.43  &  -06&42&57.7  & O,B,R  \\
4   &  \object{[CTF93]32}       &  \object{V* V846 Ori}             &   05&36&41.34  &  -06&34&00.3  & O,B,R,S \\
5   &  \object{[CTF93]72}       &  \object{IRAS 05350-0700}         & 05&37&24.47  &  -06&58&32.9  & O,B,R,S \\
6   & \object{ [CTF93]83}       &  \object{HH 43 IRS1}               &   05&38&07.44  &  -07&08&30.0  &  R,S \\
7   &  \object{[CTF93]62}       &  \object{V* V1787 Ori}            &   05&38&09.23  &  -06&49&15.9  & O,B,R,S \\
8   &  \object{[CTF93]79}       &  \object{V* V883 Ori}              & 05&38&18.10  &  -07&02&25.9  & O,B,R  \\
9   & \object{[CTF93]99}        &  \object{Haro 4-254}               & 05&38&52.36  &  -07&21&09.5  & O,B,R,S \\
10  &  \object{[CTF93]87}       &  \object{2MASS-J05390536-0711052}  & 05&39&05.36  &  -07&11&05.2  & O,B,R,S \\
11  &  \object{[CTF93]104}      &  \object{HBC 176}                   & 05&39&22.32  &  -07&26&44.5  & O,B,R  \\
12  &  \object{Meag31}          &  \object{2MASS-J05401379-0732159}   & 05&40&13.79  &  -07&32&15.9  &  B,R  \\
13  &  \object{[CTF93]146-2}    &  \object{2MASS-J05401494-0748485}   & 05&40&14.94  &  -07&48&48.5  &	R,S \\
14  & \object{ [CTF93]146-1}    &  \object{IRAS 05378-0750}          & 05&40&17.80  &  -07&48&25.8  &	R,S \\
15  &  \object{[CTF93]211}      &  \object{IRAS 05379-0815}           & 05&40&19.40  &  -08&14&16.3  & O,B,R,S \\
16  &  \object{[CTF93]187}      &  \object{V* V1791 Ori}              &   05&40&37.36  &  -08&04&03.0  & O,B,R,S \\
17  &  \object{[CTF93]168}      &  \object{IRAS 05389-0759}          & 05&41&22.13  &  -07&58&03.0  &	R,S \\
18  &  \object{[CTF93]191}      &  \object{V* DL Ori}                & 05&41&25.34  &  -08&05&54.7  & O,B,R,S \\
19  &  \object{[CTF93]246A}     &  \object{2MASS-J05413005-0840092}  & 05&41&30.05  &  -08&40&09.2  &	R,S \\
20  & \object{ [CTF93]186}      &  \object{IRAS 05391-0805}          & 05&41&30.22  &  -08&03&41.4  &	R,S \\
21  &  \object{[CTF93]246B}     &  \object{2MASS-J05413033-0840177}  & 05&41&30.33  &  -08&40&17.7  &	R,S \\
22  &  \object{[CTF93]237-2}    &  \object{2MASS-J05413418-0835273}  & 05&41&34.18  &  -08&35&27.3  &  R,S \\
23  &  \object{[CTF93]237-1}    &  \object{IRAS 05391-0836}          & 05&41&34.78  &  -08&35&23.0  &  R,S \\
24  &  \object{[CTF93]216-1}    &  \object{IRAS 05403-0818}          & 05&42&47.06  &  -08&17&06.9  &	R,S \\
25  &  \object{[CTF93]216-2}    &  \object{2MASS-J05424706-0817069}  &  05&42&48.48  &  -08&16&34.7  &	R,S \\
26  &  \object{[CTF93]245B-2}   &  \object{[BDB2003]G212.98-19.15}   & 05&42&50.48  &  -08&38&29.2  & O,B,R  \\
27  & \object{ [CTF93]245B-1}   &  \object{IRAS 05404-0841}           & 05&42&50.50  &  -08&39&57.6  &  B,R  \\
\hline
\end{tabular}
\tablefoot{
\tablefoottext{a} {Labels O, B, R, and S indicate EFOSC2 optical spectra, SofI blue-, and red-grisms, and Spitzer spectra, respectively. See text for details.}
}
\end{table*}


\section{Observations and data reduction}
\label{observations:sec}

Our multi-wavelength database on L\,1641 targets is made of:
\textit{i)} low-resolution optical and NIR spectroscopic observations from the \emph{ESO/NTT}, to infer most of the YSO 
main stellar parameters (spectral type, accretion, mass accretion and ejection rates);
\textit{ii)} archival \emph{Spitzer-IRS} spectra, mostly used to derive the visual extinction;
\textit{iii)}  \emph{Spitzer-IRAC} and \emph{Spitzer-MIPS} images from the \textit{Spitzer Heritage Archive}\footnote{http://sha.ipac.caltech.edu}, 
and literature photometric data, covering a spectral range from 0.4 to 1100\,$\mu$m, which allow us to construct the SEDs.

\subsection{Optical and NIR spectroscopy}
\label{sofoscspec:sec}
Our optical and near-infrared spectroscopic observations were obtained at the ESO New Technology Telescope 
with EFOSC2~\citep{buzzoni} and SofI~\citep{moor1}, respectively. The observations were carried out over a short period of time 
(10-13 and 14-16 February 2009, for the NIR and optical spectroscopy, respectively), thus possible YSO variability should not 
significantly affect the optical and NIR segments of our spectra.
The observational settings (spectral resolution and slit width) were chosen to obtain, as far as possible, homogeneous spectra, aiming to apply a combined optical/NIR analysis. 
The total integration time on source (\textit{I$_t$}) in both optical and NIR spectra differs from target to target, depending on the source brightness, 
which was chosen to obtain a signal to noise ratio $\ge$200 on the continuum and detect emission lines, which are 1/20 of the continuum, with at least S/N = 10. 
No preferential position angles for the slits were chosen.

The EFOSC2 spectra were taken with the grism N.16 and a 0\farcs7 slit width, covering a wavelength range from $\sim$0.6 to 1\,$\mu$m, and 
with a spectral resolution of R$\sim$700. \textit{I$_t$} ranges from 30\,s (targets with $V$=11\,mag) up to 1800\,s (targets with $V$=17\,mag). 
Only those targets having a $V$ band magnitude $\le$17  were observed (namely 14 out of 27 objects), and they are indicated in our 
observation log of Table~\ref{obs_spec:tab} (column\,6) with the label `O'. 

The SofI NIR spectra were taken with an AB offset scheme. 
We used a 0\farcs6 slit (R$\sim$900) for both the blue grism (ranging from 0.95 to 1.64\,$\mu$m, and labelled as `B' in column\,6 of Tab.~\ref{obs_spec:tab}) 
and the red grism (ranging from 1.51-2.5\,$\mu$m, and labelled as `R' in column\,6 of Tab.~\ref{obs_spec:tab}). Only those targets with J$\le$15\,mag were 
observed with the blue grism (namely 16 out of 27 targets), whereas all the targets were observed with the red grism. \textit{I$_t$} ranges 
from 20\,s (objects with J=8\,mag) up to 1200\,s (objects J$\ge$13\,mag) in the blue grism, and from 20\,s (targets with K=5\,mag) 
up to 1800\,s (K$\ge$12\,mag) in the red grism.

Additionally, telluric and spectro-photometric standards were observed to correct for the atmospheric spectral response and 
to flux-calibrate the spectra, respectively. 
In particular \object{HD\,289002} (B3 spectral type) was chosen for the optical spectra, whereas an F8V type star (\object{Hip 23930}) was selected for the NIR.

The data reduction was done using standard \emph{IRAF}\footnote{IRAF (Image Reduction and Analysis Facility) is distributed by the National
Optical Astronomy Observatories, which are operated by AURA, Inc., cooperative agreement with the National Science Foundation.} tasks.
Each spectrum segment was flat fielded, sky subtracted and corrected for the curvature derived from long-slit spectroscopy, while atmospheric features 
were removed by dividing the spectra by the telluric standard star. The wavelength calibrations were performed using a helium-argon lamp in the optical 
and a xenon lamp in the infrared. The resulting spectra were originally flux calibrated by means of the observed standard stars. 
We thus obtained for each target up to three individually calibrated spectra (optical, blue and red grism).
However, due to variable seeing conditions during our observations (from 0\farcs6 to 1\farcs3) and SofI instrumental problems (the
source did not remain correctly centred on the slit during the nodding cycle), often the stellar calibrated continua 
do not properly match in the three spectral segments. Thus, we performed aperture photometry on the $K_s$ band SofI acquisition images
(see Sect.~\ref{photdata:sec}), to get an absolute flux-calibration for the red grism spectra, and match to these the stellar continua of the 
two remaining segments, finally obtaining a single flux-calibrated spectrum from 0.6 to 2.5\,$\mu$m. Uncertainties on the $K_s$ photometry range from 
$\sim$5 to 10\% of the total flux.

\subsection{Archival Spitzer spectroscopy}
\label{irsspec:sec}
From the \textit{Spitzer Heritage Archive} we retrieved 20 spectra~\citep[the spectrum of \object{[CTF93]50} - \#2 - has already been presented by][]{fischer}, 
which were labelled as `S' in column\,6 of Tab.~\ref{obs_spec:tab}.  
\object{[CTF93]237-1} and -2 are not spectrally resolved, thus the spectra are blended. 
No additional data are available in the archive for the remaining objects in the \textit{Spitzer} archive.

All the data were taken with the \textit{Spitzer Infrared Spectrograph} \citep[IRS;][]{houck}
under a variety of programs (namely 30706, 30859, and 50374), and observed between November 2006 and November 2008. 
All the IRS spectra were obtained with a combination of short-low (SL2/SL1, 5.2-8.7/7.4-14.5\,$\mu$m, R = 64-128), and long-low (LL, 14.0-39.0\,$\mu$m, R = 64-128) 
modules (i.\,e. ranging from 5 to 39\,$\mu$m). Source \#\,9 was observed with the LL module alone.
The total exposure time on source ranges from four to eight minutes. All the spectra were observed in IRS staring mode and extracted from the Spitzer Science Center 
S18.7.0 pipeline basic calibrated data (BCD) using the Spitzer IRS Custom Extraction (SPICE) software. This process includes bad-pixel correction, optimal 
point-spread function (PSF) aperture extraction, defringing, order matching, wavelength and flux calibration. 
The extracted spectra from the eight nod positions were averaged together (four for the SL2/SL1 and LL modules, respectively),
obtaining a final flux-calibrated spectrum, which ranges from $\sim$5 to 39\,$\mu$m.

\subsection{Imaging: SofI, Archival Spitzer imaging}
\label{photdata:sec}

Acquisition images in the $K_s$ band were taken with SofI before each NIR spectroscopic observation, with an \textit{I$_t$} of one minute each.  
The data were reduced using \textit{IRAF} packages and applying standard procedures for sky subtraction, dome flat-fielding and bad pixel removal. 

Aperture photometry was performed on the resulting images using the task \textit{phot} in \textit{IRAF}. 
The images were, then, flux calibrated using several field stars as 'standards' along with their Two-Micron All Sky Survey~\citep[2MASS,][]{2MASS} $K_s$ band photometry. 
An aperture correction was estimated on a few isolated stars using the task \textit{mkapfile} in \textit{IRAF} with a formal error of 0.02--0.03\,mag. 
As a result, we get uncertainties for the YSO photometry in the range of $\sim$0.05--0.2\,mag, depending on the number of standards and the YSO brightness.

In addition, \emph{Spitzer} basic calibrated data (BCD) have been obtained from the \textit{Spitzer Heritage Archive}, reduced with the S14.0 pipeline.

They consist of several individual frames, which map the entire L\,1641 cloud, observed with the Infrared Array Camera~\citep[IRAC;][]{Fazio} in four 
channels (at 3.6, 4.5, 5.8, and 8.0\,$\mu$m) (covering a region of $\sim$1.04\,deg$^2$) 
and with the Multiband Imaging Photometer for Spitzer~\citep[MIPS;][]{RiekeMIPS} (at 24, 70, and 160\,$\mu$m) ($\sim$1.5\,deg$^2$). 
The \textit{IRAC} images were performed in High-Dynamic Range mode with integration times of 0.4 and 10.4\,s (Spitzer program ID 43), and have been presented  
in previous papers~\citep[][]{meg,fang}. The \textit{ MIPS} images (Spitzer program ID 47) have effective integration times of 80, 40, and 8\,s at 24, 70, 
and 160\,$\mu$m, respectively, and have been presented by \citet{fang}.

A mosaicking of the individual frames in the seven different bands was performed using the Spitzer MOPEX (MOsaicker and Point source EXtractor) tool~\citep[][]{makovoz}, 
which performs background matching of individual data frames, mosaics the individual frames, and rejects outliers. 
IRAC and MIPS 24\,$\mu$m photometry for our targets was reported by \citet{meg} and  \citet{fang}, and kindly provided by these authors.

We thus performed photometry only on the \emph{MIPS} 70 and 160\,$\mu$m reduced mosaics.
We manually examined the latter for target detection and used MOPEX/APEX for the source extraction. 
We detect 26 out of 27 targets at 70\,$\mu$m, and 11 out of 27 in the 160\,$\mu$m map.

As indicated in the MIPS data handbook, we used apertures of radius 16$\arcsec$ and 32$\arcsec$ for the 70\,$\mu$m and 160\,$\mu$m photometry respectively,
a sky annulus with inner and outer radii of 18$\arcsec$ and 39$\arcsec$ at 70\,$\mu$m, and 64$\arcsec$ and 128$\arcsec$ at 160\,$\mu$m
and an aperture correction factor of 2.07 and 1.98 for the 70\,$\mu$m and 160\,$\mu$m photometry, respectively.
Uncertainties were obtained from the pixel-to-pixel noise in the sky annulus. 
This error is added in quadrature to the photometric uncertainty. 
Source identification and extraction were particularly difficult for the 160\,$\mu$m map, due to
the extreme brightness and saturation of the 160\,$\mu$m diffuse emission in several parts of the map, which
translated into very large photometric errors or into a non-detection of the targets.

\subsection{Additional photometry from the literature}

Additional NIR photometry in the $J$, $H$, and $K_s$ bands was retrieved from 2MASS (observed in November 1998), with 10\,$\sigma$ detection 
limits of 16.2, 15.3, and 14.6\,mag, respectively. 

Additional photometric data for the sample were retrieved from `the Naval Observatory Merged Astrometric Dataset'~\citep[NOMAD;][]{zac} in the optical
which includes USNO optical photometry (B, V, R, I bands). More optical data were taken 
from the Sloan Digital Sky Survey~\citep[SDSS][]{york}, which includes measurements at \textit{u'g'r'i'z'} bands at
3540, 4760, 6290, 7690, and 9250\AA.

Finally sub-MM photometry was retrieved from \cite{difrancesco} (SCUBA/JCMT, at 450 and 850\,$\mu$m), and from \cite{dent} (UKT14/JCMT, at 350, 800, 850, and 1100 \,$\mu$m).

\section{Results}
\label{results:sec}

\subsection{Imaging}
\label{photometry:sec}

\begin{table}
\begin{minipage}[t]{\columnwidth}
\caption{2MASS and SofI $K_s$ band magnitudes of the observed targets}
\label{mags:tab}
\centering
\renewcommand{\footnoterule}{}
\begin{tabular}{ccccc}
\hline \hline
ID  & $K_s(2MASS)$     & $K_s(SofI)$ &     $\Delta K_s(S-2M)$\footnote{$K_s(SofI)-K_s(2MASS)$.}  \\
    &   (mag)          &   (mag)      &   ($\Delta$mag) 	  \\
\hline\\[-5pt]
 1 & 9.68$\pm$0.02     &  9.3$\pm$0.1	&   -0.38  \\ 
 2 & 8.21$\pm$0.02     &  8.2$\pm$0.1	&   -0.01  \\ 
 3 & 5.95$\pm$0.02     &  6.10$\pm$0.05 &    0.15  \\ 
 4 & 9.52$\pm$0.03     &  9.3$\pm$0.1	&   -0.22  \\ 
 5 & 9.85$\pm$0.02     &  9.8$\pm$0.1	&   -0.05  \\ 
 6 & 11.79$\pm$0.07    & 11.6$\pm$0.2	&   -0.19   \\ 
 7 & 7.98$\pm$0.02     &  7.6$\pm$0.06  &   -0.38  \\ 
 8 & 5.15$\pm$0.01     &  5.7$\pm$0.05  &    0.55    \\ 
 9 & 8.03$\pm$0.03     &  8.00$\pm$0.06 &   -0.03  \\ 
10 & 10.54$\pm$0.02    & 10.5$\pm$0.1	&   -0.04   \\ 
11 & 8.17$\pm$0.03     &  8.1$\pm$0.05  &   -0.06   \\ 
12 & 11.1$\pm$0.02     & 11.2$\pm$0.2	&    0.1     \\ 
13 & 13.47$\pm$0.06    & 13.5$\pm$0.2	&    0.03    \\ 
14 & 10.49$\pm$0.02    & 10.3$\pm$0.1	&   -0.19   \\ 
15 & 10.22$\pm$0.02    & 10.2$\pm$0.1	&   -0.02   \\ 
16 & 7.93$\pm$0.05     &  7.90$\pm$0.05 &   -0.03  \\ 
17 & 10.53$\pm$0.02    & 10.6$\pm$0.1	&    0.07   \\ 
18 & 9.29$\pm$0.02     &  9.20$\pm$0.07 &   -0.09  \\ 
19 & 12.27$\pm$0.03    & 12.4$\pm$0.2	&    0.13   \\ 
20 & 10.56$\pm$0.02    & 10.1$\pm$0.1	&   -0.41    \\ 
21 & 11.86$\pm$0.03    & 11.8$\pm$0.2	&   -0.05	   \\ 
22 & 11.52$\pm$0.02    & 11.6$\pm$0.2	&    0.08	   \\ 
23 & 10.57$\pm$0.04    & 10.1$\pm$0.1	&   -0.47	   \\ 
24 & 11.76$\pm$0.03    & 11.8$\pm$0.1	&    0.04  \\
25 & 11.06$\pm$0.03    &  7.98$\pm$0.05 &   -3.08   \\
26 & 9.09 $\pm$0.02    &  9.2$\pm$0.1	&    0.11   \\ 
27 & 10.36$\pm$0.02    & 10.5$\pm$0.1	&    0.14    \\ 
\hline
\end{tabular}
\end{minipage}
\end{table}

Table~\ref{mags:tab} reports the 2MASS, and SofI $K_s$ band photometry for the sample, as well as the YSO variability
between the SofI and 2MASS observations (corresponding to about ten years).
From a quick inspection of Column\,4 in Tab.~\ref{mags:tab}, we first note the large $K_s$ variation ($\sim$3.1\,mag) of source
\#\,25 (namely \object{[CTF93]216-2}), whose outburst was recently reported by our team~\citep[][]{caratti11}. 
Five more objects (namely \#\,1, 7, 8, 20, 22) show some degree of variability (0.3$\lesssim \Delta K_s \lesssim$0.5\,mag),
whereas the remaining twenty objects do not indicate any, inside their photometric error bars.

Tables~\ref{photometry:tab} and \ref{photometry2:tab} (reported in Appendix~\ref{appendixA:sec}) present the complete collected photometry for our targets from 0.4 to 1100\,$\mu$m.
We do not consider and report the available IRAS measurements in this table, due to the crowding of the fields at mid-IR wavelengths, 
and the availability of the Spitzer data.

\begin{figure}
 \centering
\includegraphics [width=9cm] {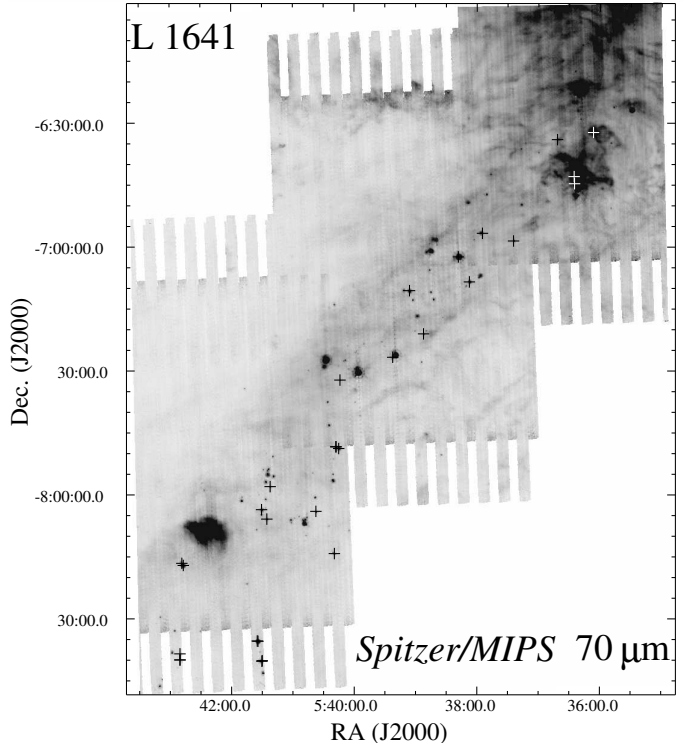}\\
  \caption{\textit{Spitzer/MIPS} 70\,$\mu$m map of L\,1641. Crosses indicate the sample targets. 
\label{spitz70:fig}}
\end{figure}

In Fig.~\ref{spitz70:fig} we report the \emph{Spitzer-MIPS} map at 70\,$\mu$m, indicating the location of the studied objects.

\begin{figure*}[!h]
 \centering
\includegraphics [width=19cm] {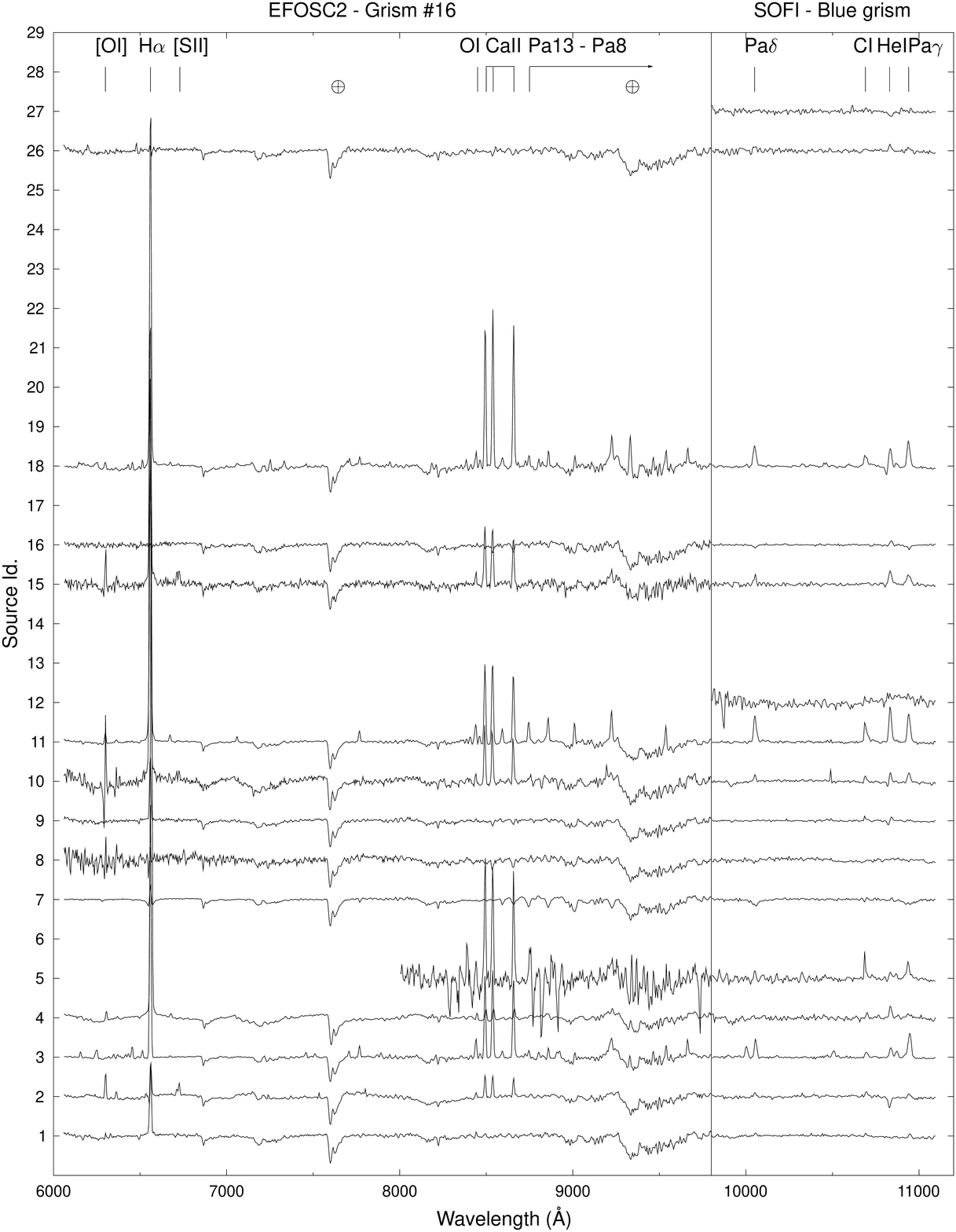}\\
  \caption{Continuum-normalised spectra, EFOSC and SofI Z band. Labels indicate the most prominent spectral features.
\label{spec1:fig}}
\end{figure*}

\begin{figure*}[!h]
 \centering
\includegraphics [width=19cm] {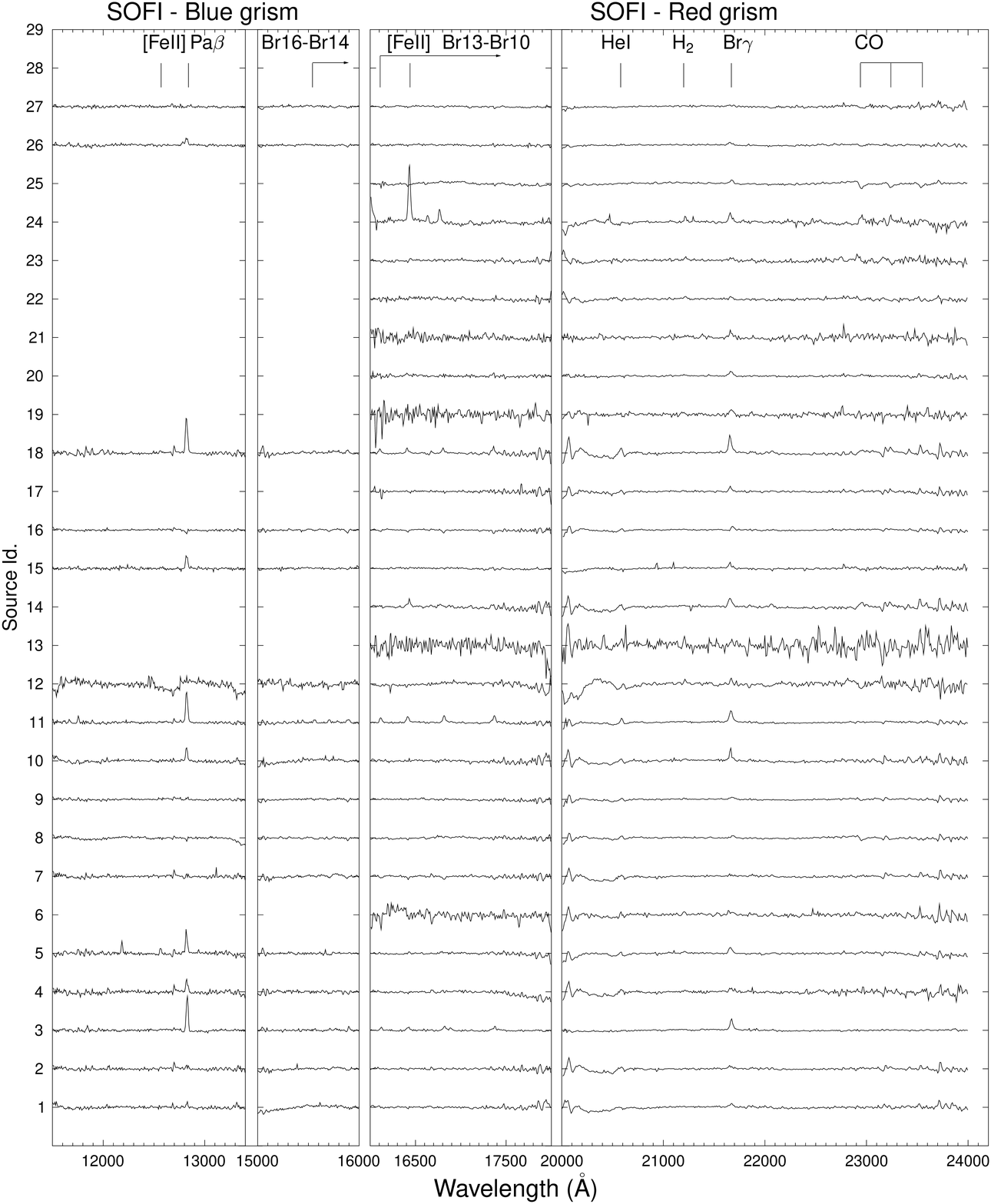}\\
  \caption{Continuum-normalised spectra, SofI J, H, and K bands. Labels indicate the most prominent spectral features.
\label{spec2:fig}}
\end{figure*}

\subsection{Optical and NIR spectra}
\label{OptIRspec:sec}


Optical and NIR continuum-normalised spectra of the sample are shown in Figures~\ref{spec1:fig} and \ref{spec2:fig}.
The most prominent features observed in the spectra are emission lines, which have been indicated in the figures.
The fluxes of the main lines, used for the analysis in this paper, are reported in Tables~\ref{lines:tab}
and \ref{lines2:tab}, along with the statistics on the detections. A complete list of all the detected lines in each spectrum, including fluxes,
full width half maximum ($FWHM$), and equivalent widths ($W_\lambda$), is given in Appendix~\ref{appendixA:sec} (Tables~\ref{tab:ctf29}--\ref{tab:ctf245B1}).
The equivalent widths and line fluxes were calculated by integrating across the line, after subtracting the continuum, which 
was estimated by interpolating 
between two line-free adjacent regions.

The detected lines are circumstellar features and are mostly associated with YSO accretion activity or inner winds, as, e.\,g., 
\ion{H}{i}, \ion{Ca}{ii}, \ion{He}{i}~\citep[see, e.\,g.,][]{muzerolle98,natta04,edwards,antoniucci}, or ejection activity, 
as, e.\,g., [\ion{O}{i}], [\ion{S}{ii}], [\ion{Fe}{ii}], H$_2$~\citep[see, e.\,g.,][]{hartigan95,nisini02,caratti06}.
A few YSOs, namely \#\,2, \#\,3, \#\,11, and \#\,18, also show permitted ionic emission lines (e.\,g., \ion{O}{i}, \ion{Fe}{i}, 
\ion{Fe}{ii}, \ion{Na}{i}, \ion{Mg}{ii}, \ion{C}{i}), which are characteristic of active young stars~\citep[see, e.\,g.,][]{hamann92-1, hamann92-2,kelly, hernandez}, 
and usually associated with chromospheric activity. In particular, the \ion{Fe}{i}, and \ion{Fe}{ii} emission lines have been observed in several 
young eruptive stars during their outburst activity~\citep[e.\,g. \object{V1647 Ori}, see, e.\,g.,][]{fedele}, or in YSOs
showing high mass-accretion rates~\citep[e.\,g.,][]{rossi}.

The most commonly detected features are from \ion{H}{i} recombination lines, H$\alpha$ in the optical spectra (11 out of 14, i.\,e. 79\% detection rate),
Paschen lines between 0.8 and 1.3\,$\mu$m (e.\,g. Pa$\beta$ has a 69\% detection rate - 10 out of 16), and Brackett lines between 1.5 and 2.2\,$\mu$m (Br$\gamma$
has a 96\% of detection rate - 26 out of 27).
Other prominent bright emission lines are the \ion{Ca}{ii} triplet between 0.85 and 0.87\,$\mu$m (detected in 57\% of the optical spectra), 
and the \ion{He}{i} line at 1.08\,$\mu$m (56\% detection rate), 
which notably displays a clear P Cygni profile in four out of nine detections (namely source \#\,9, 10, 15, 18), usually indicative of 
inner winds in accretion disc systems~\citep[see, e.\,g.,][]{edwards}, with radial velocities larger than 300\,km\,s$^{-1}$.
Source \#\,2, namely \object{[CTF93]50}, only shows the blue-shifted \ion{He}{i} line in absorption, and no indication
of emission. \citet[][]{connelley} suggest that this particular feature may be indicative of FU Orionis like stars, and
might have a disc wind origin. According to \citet[][]{fang}, the optical spectrum of \object{[CTF93]50} (their source \#\,105)
resembles those of FU Ori-like YSOs, however 
the typical absorption features of the CO band-head lines longward of 2.29\,$\mu$m, also characteristic of FU-Ori objects, are not detected
in our NIR spectrum.

Lines tracing the jet activity have also been identified in our sample, but with a detection rate lower than that of the accretion lines.
Namely the [\ion{O}{i}] line at 0.63\,$\mu$m has been identified in 43\% of the optical spectra, the [\ion{S}{ii}] doublet at 0.68\,$\mu$m in 36\%, 
the [\ion{Fe}{ii}] 1.64\,$\mu$m  and H$_2$ 2.12\,$\mu$m lines in 19\% and 32\% of the NIR spectra, respectively.
In total, jet-tracer emission lines have been observed in 14 out of 27 spectra, i.\,e. in 52\% of the sample.

Our low resolution spectra mostly lack absorption lines on the photospheric continuum, also indicating the
presence of veiling. Source \#\,7 and \#\,16 are an exception, and clearly show \ion{H}{i} absorption features at optical and NIR wavelengths, 
indicating both early spectral types and lower veiling\footnote{For those sources where absorption features such as \ion{H}{i} (spectral types F or earlier), \ion{Na}{i} and/or \ion{Ca}{i} (spectral types K or M) have been detected,
we usually get $r_K \le 1$. For the remaining sources, we can infer lower limits for the veiling, on the basis of non-detection of \ion{Na}{i} or \ion{Ca}{i} lines in the K band
\citep[see, e.\,g.,][]{antoniucci}. Assuming M0 as average spectral type and typical intrinsic equivalent widths $\sim$3.3\,\AA, we get $r_k > 1$.}, likely indicating a more evolved stage.
On the other hand, broad-band molecular absorption features, typical of cool objects, are clearly visible in both optical (VO and TiO bands) and NIR (H$_2$O bands) 
spectra.
CO overtone bands have been detected in our spectra, both in absorption (sources \#\,8, \#\,11, \#\,18, \#\,24) and emission (\#\,14 and \#\,25). 
The CO bandheads in emission are usually believed to originate in the inner gaseous disc~\citep[see e.\,g.][]{carr,antoniucci},
and have been observed in YSOs that display strong jets and Herbig-Haro (HH) objects~\citep[see, e.\,g.,][]{davis11}, whereas the CO bandheads in absorption
may originate in cooler regions, like the YSO envelope~\citep[][]{davies} or the photosphere of cool YSOs~\citep[see e.\,g.][]{rayner}.

\begin{landscape}
\begin{table}
\begin{scriptsize}
\caption{ Main detected emission lines: accretion tracers.}
\label{lines:tab}
\begin{tabular}{l|cc|cc|cc|cc|cc|cc|cc}
\hline\hline      
Source & \multicolumn{2}{c|}{H$\alpha$}  & \multicolumn{2}{c|}{\ion{Ca}{ii} 0.854/0.866\,$\mu$m} & \multicolumn{2}{c|}{\ion{He}{i} 1.083\,$\mu$m}& \multicolumn{2}{c|}{Pa$\delta$}  & \multicolumn{2}{c|}{Pa$\gamma$}  & \multicolumn{2}{c|}{Pa$\beta$} 
 & \multicolumn{2}{c|}{Br$\gamma$}  \\
ID & (F $\pm$ $\Delta$ F)10$^{-14}$ & EW & (F $\pm$ $\Delta$ F)10$^{-14}$ & EW & (F $\pm$ $\Delta$ F)10$^{-14}$ & EW & (F $\pm$ $\Delta$ F)10$^{-14}$ & EW & (F $\pm$ $\Delta$ F)10$^{-14}$ & EW & (F $\pm$ $\Delta$ F)10$^{-14}$ & EW  & (F $\pm$ $\Delta$ F)10$^{-14}$ & EW \\
    & erg s$^{-1}$ cm$^{-2}$ & \AA & erg s$^{-1}$ cm$^{-2}$ & \AA & erg s$^{-1}$ cm$^{-2}$ & \AA & erg s$^{-1}$ cm$^{-2}$ & \AA & erg s$^{-1}$ 
cm$^{-2}$ & \AA & erg s$^{-1}$ cm$^{-2}$ & \AA & erg s$^{-1}$ cm$^{-2}$ & \AA \\
\hline
1      & 13$\pm$1 &  -11.0  & $\cdots$ & $\cdots$  & $\cdots$ & $\cdots$ &  0.8$\pm$0.3\tablefootmark{a} &  -2.0  &  1.1$\pm$0.3 &  -2.8 &  1.8$\pm$0.4 &   -4.0  &  3.0$\pm$0.5 &  -5.0 \\
2      &  0.99$\pm$0.04\tablefootmark{b} &  -7 & 1.8$\pm$0.2/1.6$\pm$0.2 & -6.4/-5.7 & $\cdots$ & $\cdots$ &  1.1$\pm$0.2  & -2.5 &  2.0$\pm$0.4  & -3.3 &   4.5$\pm$0.4 & -4.4 & 7.4$\pm$0.8 & -3.3 \\
3      & 1330$\pm$1  & -73.2 & 344$\pm$5/304$\pm$5  & -26.6/-23.7 & 95$\pm$8 &   -5.1 & 152$\pm$8   &  -8 & 283$\pm$8   &  -16.2  & 386$\pm$8   &  -23.2 &  181$\pm$8   &  -13.0\\
4      &  82.3$\pm$0.2  & -143.1 & 2.3$\pm$0.4/3.1$\pm$0.4 & -4/-5.5 & 9$\pm$2 &  -8.5 &  6$\pm$2 &  -5.7  &  8$\pm$2 &  -8.1  &  10$\pm$1  &  -9.2 &  3.3$\pm$0.8 &  -4.5\\
5      &  0.13$\pm$0.05\tablefootmark{a} &   $\cdots$ &  0.41$\pm$0.03/0.38$\pm$0.03 & -19.0/-17.5 & 0.33$\pm$0.08  & -3.7   & $\cdots$ & $\cdots$ &  0.89$\pm$0.08 & -9.5 & 2.8$\pm$0.1 & -12.7 &  3.8$\pm$0.4  & -7.6 \\
6      & $\cdots$ & $\cdots$  & $\cdots$ & $\cdots$ & $\cdots$   & $\cdots$ & $\cdots$ & $\cdots$  & $\cdots$ & $\cdots$ &   $\cdots$ & $\cdots$ & 0.4$\pm$0.1 & -6.5 \\
7      & 18$\pm$1 & -6.3  & $\cdots$ & $\cdots$ & $\cdots$ & $\cdots$   & $\cdots$ & $\cdots$ & $\cdots$ & $\cdots$ &   $\cdots$ & $\cdots$ & 12$\pm$3 & -3.1 \\
8      & $\cdots$ & $\cdots$  & $\cdots$ & $\cdots$ & $\cdots$ & $\cdots$   & $\cdots$ & $\cdots$ & $\cdots$ & $\cdots$ &   $\cdots$ & $\cdots$ & 140$\pm$10 & -3.6 \\
9      &  13.4$\pm$0.2  & -17 & $\cdots$ & $\cdots$ &  3.8$\pm$0.8\tablefootmark{b} & -1.9  & $\cdots$ & $\cdots$ & $\cdots$ & $\cdots$ &  42$\pm$1 &  -1.6  & 10$\pm$1 & -3.8   \\
10     &  3.05$\pm$0.03 & -96.1 &  1.2$\pm$0.1/1.1$\pm$0.1  & -18.6/-14.1 & 0.89$\pm$0.08\tablefootmark{b}  & -4.5 &  0.5$\pm$0.1  & -3.3 &  1.3$\pm$0.2  & -6.1 &  2.2$\pm$0.1 & -8.5 &  2.2$\pm$0.4 &  -8.9 \\
11     & 69.9$\pm$0.1 & -88.2 & 26.4$\pm$0.4/23.0$\pm$0.4  & -27.9/-24.4 &  28.8$\pm$0.7  & -17.5 &  19$\pm$2 & -14.7 &  27.0$\pm$0.8 & -16.1 &  45$\pm$1   &  -21.2 & 30$\pm$1 &  -13.1  \\
12     & $\cdots$ & $\cdots$  & $\cdots$ & $\cdots$ & $\cdots$  & $\cdots$ & $\cdots$ & $\cdots$  & $\cdots$ & $\cdots$ &   $\cdots$ & $\cdots$ & 0.6$\pm$0.1 &  -4.3 \\
13     & $\cdots$ & $\cdots$  & $\cdots$ & $\cdots$ & $\cdots$   & $\cdots$ & $\cdots$& $\cdots$  & $\cdots$ & $\cdots$ &   $\cdots$ & $\cdots$ &  0.2$\pm$0.1\tablefootmark{a}  &  $\cdots$  \\
14     & $\cdots$ & $\cdots$  & $\cdots$ & $\cdots$ & $\cdots$   & $\cdots$ & $\cdots$& $\cdots$  & $\cdots$ & $\cdots$ &   $\cdots$ & $\cdots$ &  3.0$\pm$0.3  &  -11.1  \\
15     & 5.18$\pm$0.05 & -95.6 & 1.35$\pm$0.4/1.06$\pm$0.6 &  -21.1/-16.5 &  1.1$\pm$0.1\tablefootmark{b} &  -8.7 & 0.5$\pm$0.1  & -5.5  & 1.2$\pm$0.1  & -9.5 & 2.0$\pm$0.2 &  -10.0 &  2.5$\pm$0.4 &  -7.2 \\
16     & $\cdots$ & $\cdots$  & $\cdots$ & $\cdots$ & $\cdots$   & $\cdots$ & $\cdots$& $\cdots$  & $\cdots$ & $\cdots$ &   $\cdots$ & $\cdots$ & 7$\pm$1 & -2.5 \\
17     & $\cdots$ & $\cdots$  & $\cdots$ & $\cdots$ & $\cdots$   & $\cdots$ & $\cdots$& $\cdots$  & $\cdots$ & $\cdots$ &   $\cdots$ & $\cdots$ & 1.8$\pm$0.3 & -6.9 \\
18     &  403.0$\pm$0.5 &-102.9 & 42.2$\pm$0.6/35.8$\pm$0.6 & -44/-36 & 21.5$\pm$1\tablefootmark{b}  & -9.7 & 29$\pm$1 & -13.1  & 38$\pm$1 & -17.6  &  45$\pm$1 & -25.2 &  14.1$\pm$0.5  & -16.5	\\
19     & $\cdots$ & $\cdots$  & $\cdots$ & $\cdots$ & $\cdots$   & $\cdots$ & $\cdots$& $\cdots$  & $\cdots$ & $\cdots$ &   $\cdots$ & $\cdots$ & 0.3$\pm$0.1 &  -9.8   \\
20     & $\cdots$ & $\cdots$  & $\cdots$ & $\cdots$ & $\cdots$   & $\cdots$ & $\cdots$& $\cdots$  & $\cdots$ & $\cdots$ &   $\cdots$ & $\cdots$ & 2.7$\pm$0.4 & -7.1 \\
21     & $\cdots$ & $\cdots$  & $\cdots$ & $\cdots$ & $\cdots$   & $\cdots$ & $\cdots$& $\cdots$  & $\cdots$ & $\cdots$ &  $\cdots$ & $\cdots$ & 0.7$\pm$0.1 &  -7.3	\\
22     & $\cdots$ & $\cdots$  & $\cdots$ & $\cdots$ & $\cdots$  & $\cdots$ & $\cdots$ & $\cdots$  & $\cdots$ & $\cdots$ &  $\cdots$ & $\cdots$ & $<$0.1 & $\cdots$ \\
23     & $\cdots$ & $\cdots$  & $\cdots$ & $\cdots$ & $\cdots$   & $\cdots$ & $\cdots$& $\cdots$  & $\cdots$ & $\cdots$ &  $\cdots$ & $\cdots$ & 3.8$\pm$0.9 & -5.0 \\
24     & $\cdots$ & $\cdots$  & $\cdots$ & $\cdots$ & $\cdots$  & $\cdots$ & $\cdots$ & $\cdots$  & $\cdots$ & $\cdots$ &  $\cdots$ & $\cdots$ & 1.74$\pm$0.08 & -12.4 \\
25     & $\cdots$ & $\cdots$  & $\cdots$ & $\cdots$ & $\cdots$  & $\cdots$ & $\cdots$ & $\cdots$  & $\cdots$ & $\cdots$ &  $\cdots$ & $\cdots$ & 10$\pm$2 & -3.6 \\
26     & $\cdots$ & $\cdots$   & $\cdots$ & $\cdots$  & 0.40$\pm$0.09  &  -3.9 &  0.2$\pm$0.1\tablefootmark{a}  &  -3.8 &  0.4$\pm$0.1  &  -4.0 &  1.5$\pm$0.2  & -5.5 &  4.1$\pm$0.6 &	-5.2  \\
27     & $\cdots$ & $\cdots$  & $\cdots$ & $\cdots$ & $\cdots$   & $\cdots$ & $\cdots$ & $\cdots$  & $\cdots$ & $\cdots$ &  $\cdots$ & $\cdots$ &  1.6$\pm$0.8\tablefootmark{a} &  -11.1   \\
\hline\
Detections & \multicolumn{2}{c|}{11/14 - 79\%}  & \multicolumn{2}{c|}{8/14 - 57\%} & \multicolumn{2}{c|}{9/16 - 56\%} & \multicolumn{2}{c|}{9/16 - 56\%}  & \multicolumn{2}{c|}{10/16 - 62\%}  & \multicolumn{2}{c|}{11/16 - 69\%} 
 & \multicolumn{2}{c}{26/27 - 96\%}  \\
\hline
\end{tabular}
%
\caption{Main detected emission lines: ejection tracers. }
\label{lines2:tab}
\begin{tabular}{l|cc|cc|cc|cc}
\hline\hline      
Source & \multicolumn{2}{c|}{[\ion{O}{i}] 0.630\,$\mu$m}  & \multicolumn{2}{c|}{[\ion{S}{ii}] 0.672/3\,$\mu$m} & \multicolumn{2}{c|}{[\ion{Fe}{ii}] 1.644\,$\mu$m}&\multicolumn{2}{c|}{H$_2$ 2.122\,$\mu$m} \\
ID & (F $\pm$ $\Delta$ F)10$^{-14}$ & EW & (F $\pm$ $\Delta$ F)10$^{-14}$ & EW & (F $\pm$ $\Delta$ F)10$^{-14}$ & EW & (F $\pm$ $\Delta$ F)10$^{-14}$ & EW   \\
 & erg s$^{-1}$ cm$^{-2}$ & \AA & erg s-2.3$^{-1}$ cm$^{-2}$ & \AA & erg s$^{-1}$ cm$^{-2}$ & \AA & erg s$^{-1}$ cm$^{-2}$ & \AA  \\
\hline\\[-5pt]
2    & 0.77$\pm$0.02 &  -7.1 & 0.33$\pm$0.03/0.55$\pm$0.03 & -2.1/-3.4 & 2.1$\pm$0.8\tablefootmark{a}	& -1.1 & $\cdots$ & $\cdots$\\
4    & 1.3$\pm$0.1 & -2.7   & $\cdots$ & $\cdots$  & $\cdots$ & $\cdots$  & $\cdots$ & $\cdots$  \\
5    & $\cdots$ & $\cdots$  & $\cdots$ & $\cdots$  &  2.9$\pm$0.2 & -1.9 &  1.4$\pm$0.4  & -2.8 \\ 
6    & $\cdots$ & $\cdots$  & $\cdots$ & $\cdots$  & $\cdots$ & $\cdots$  &  0.2$\pm$0.1\tablefootmark{a} & -5.4  \\ 
9    & $\cdots$ & $\cdots$  & $\cdots$ & $\cdots$  & $\cdots$ & $\cdots$  &  3$\pm$1   &  -1.3   \\
10   & 0.26$\pm$0.02\tablefootmark{b}  & -17.2 &  0.07$\pm$0.03$^a$/0.09$\pm$0.03 & -2.6/-3.1 & $\cdots$ & $\cdots$   &  0.9$\pm$0.4\tablefootmark{a} & -3.5   \\
11   & 2.0$\pm$0.1   &   -3.1 &  0.4$\pm$0.1 &  -0.5 & $\cdots$ & $\cdots$  & $\cdots$ & $\cdots$ \\
12   & $\cdots$ & $\cdots$  & $\cdots$ & $\cdots$  & $\cdots$ & $\cdots$  & 0.7$\pm$0.1 & -4.3 \\
14   & $\cdots$ & $\cdots$  & $\cdots$ & $\cdots$  &  0.57$\pm$0.07  & -6.6 &  $\cdots$ & $\cdots$  \\
15   & 0.54$\pm$0.05  &  -12.6 &   0.26$\pm$0.04/0.32$\pm$0.04 & -4.5/-5.5 &  $\cdots$ & $\cdots$  &  0.8$\pm$0.2 & -2.5   \\
18   &  9.1$\pm$0.4 & -2.6  &	 1.5$\pm$0.4/3.0$\pm$0.4 & -0.4/-0.8 &  4$\pm$1  & -3.2 & $\cdots$ & $\cdots$  \\
22   &   $\cdots$ & $\cdots$  &  $\cdots$ & $\cdots$  &   $\cdots$ & $\cdots$  & 0.2$\pm$0.1\tablefootmark{a} & -2.8 \\
24   &   $\cdots$ & $\cdots$  &  $\cdots$ & $\cdots$  & 2.48$\pm$0.05 & -45.8 & 0.7 $\pm$0.06 & -4.9 \\
\hline
Detections & \multicolumn{2}{c|}{6/14 - 43\%}  & \multicolumn{2}{c|}{5/14 - 36\%} & \multicolumn{2}{c|}{5/27 - 19\%} &\multicolumn{2}{c}{8/27 - 30\%}\\
\hline
\end{tabular}
\tablefoot{
\tablefoottext{a} { S/N ratio below 3$\sigma$.}
\tablefoottext{b} { Line with P Cygni profile.}}
\end{scriptsize}
\end{table}

\end{landscape}

\subsection{Spitzer IRS spectra}
\label{MIRspec:sec}

\begin{figure*}
 \centering
\includegraphics [width=6cm] {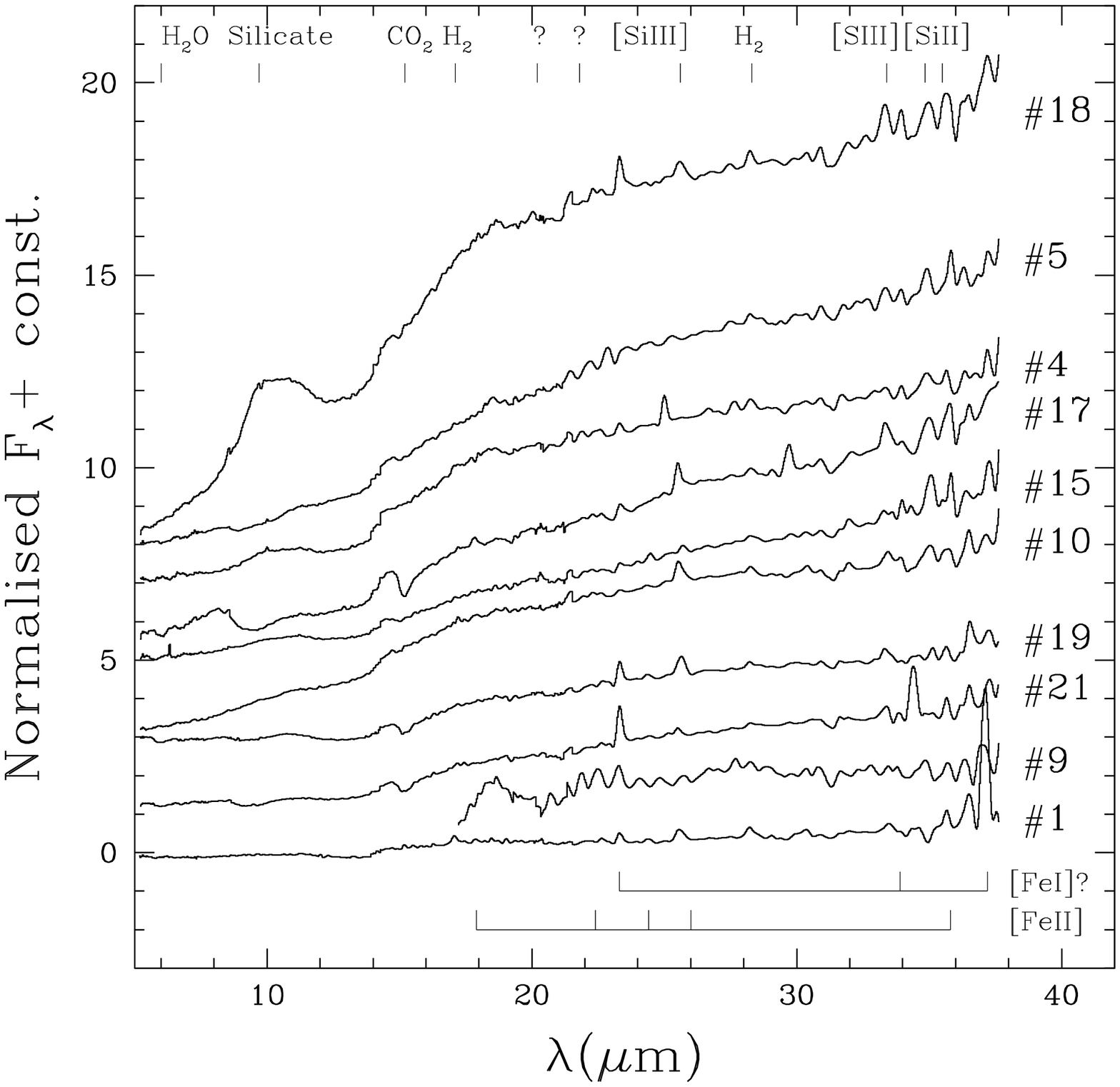}
\includegraphics [width=6cm] {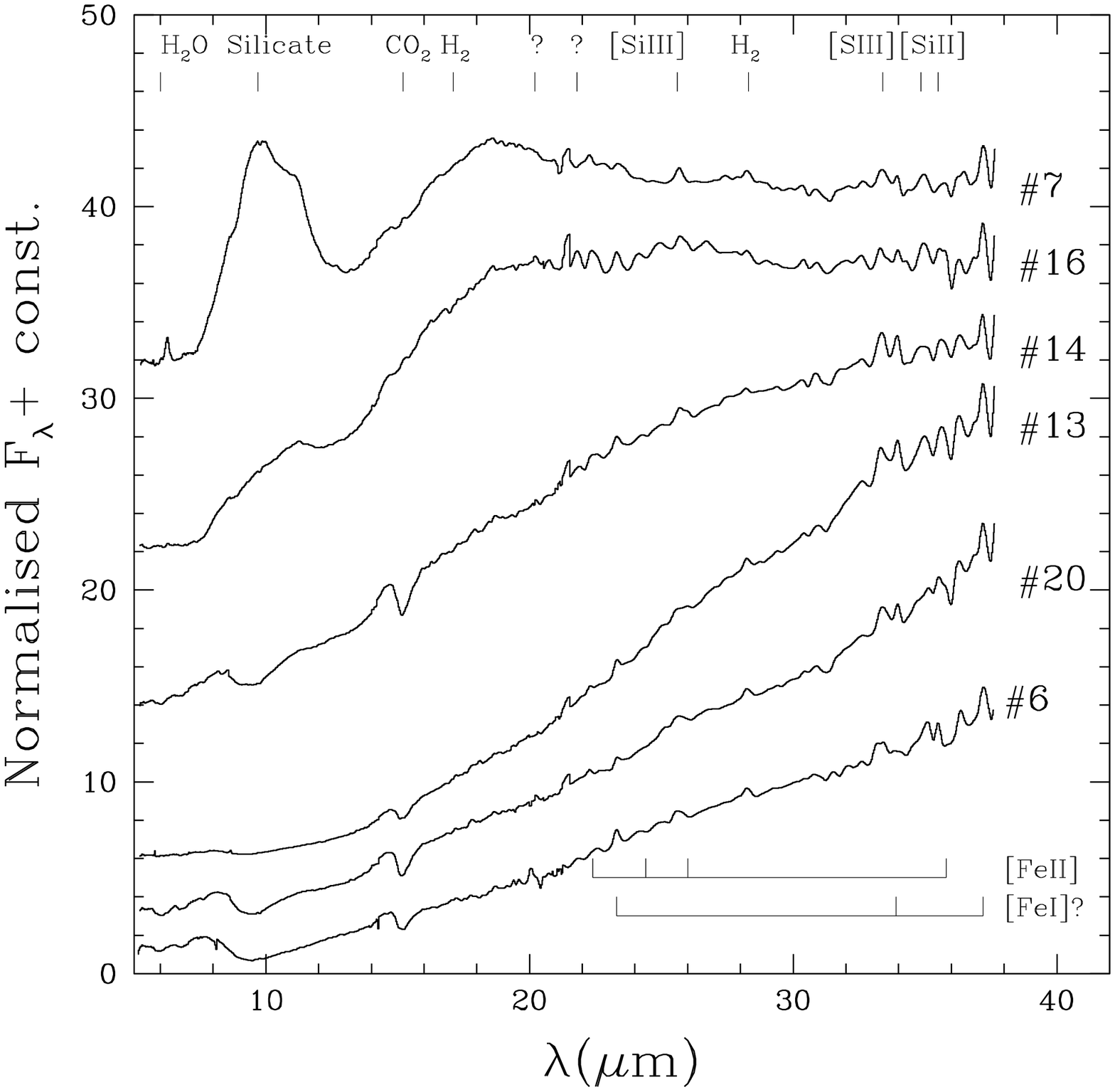}
\includegraphics [width=6cm] {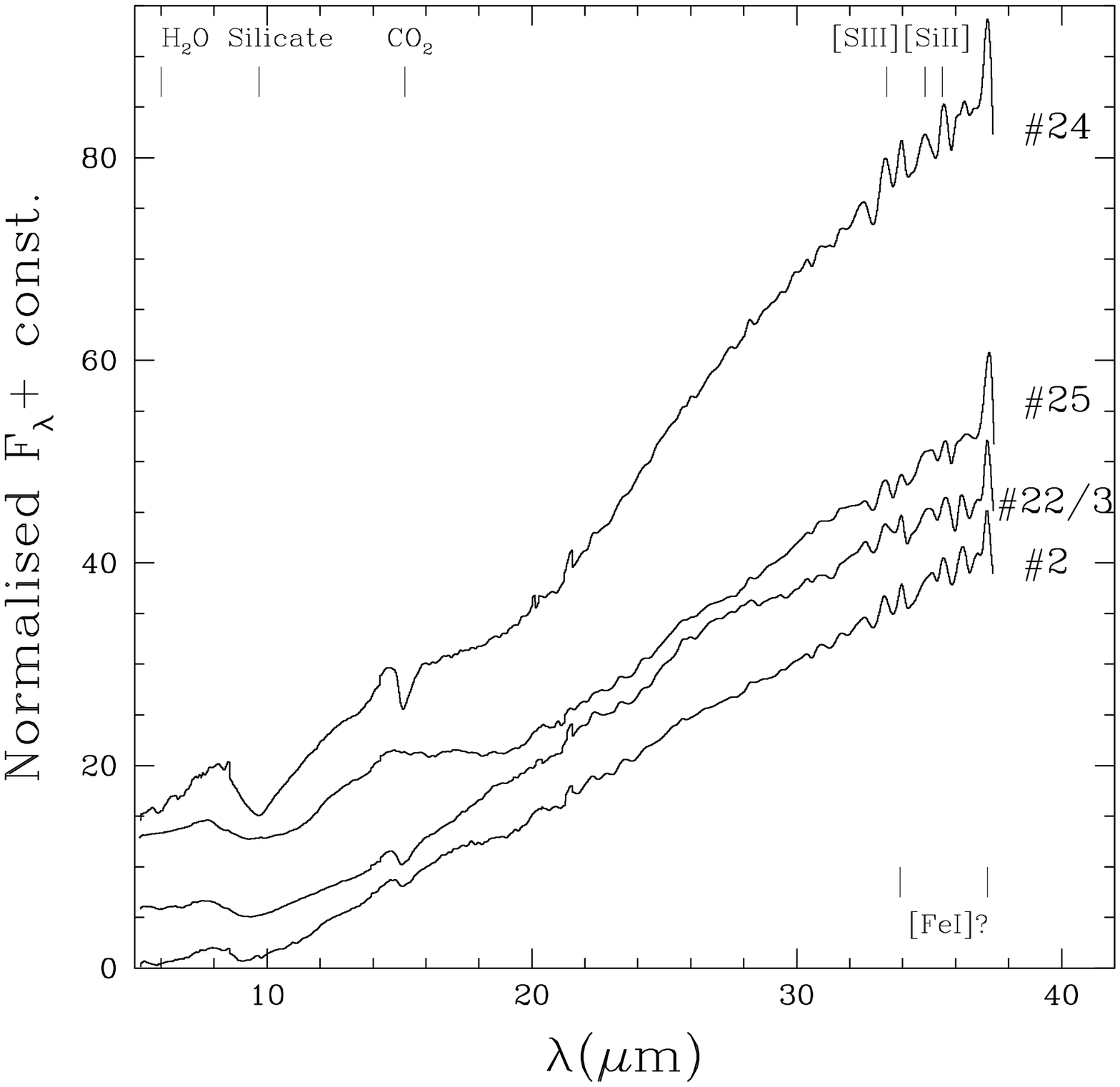}
  \caption{Spitzer IRS spectra. Flux densities have been shifted and normalised by a constant value
for a better display. Labels indicate the detected spectral features.
\label{spitzer_spec:fig}}
\end{figure*}

Our IRS Spitzer spectra are shown in Figure~\ref{spitzer_spec:fig}. 
The flux densities for the Y-axes have been scaled
and shifted for a better display.
Moreover, the three panels have a different Y-axis range to properly fit the different steepness of the spectra. 
Such a variety already shows that the SED spectral index of our YSOs has a spread larger than originally expected (see Sect.~\ref{sample:sec}), 
i.\,e. $\alpha > -1$, instead of $\alpha > -0.4$, as we will see later in Sect.~\ref{phot:sec}.

There are a few emission features detected in the Spitzer spectra, mostly H$_2$ pure rotational lines (at 17.0 and 28.2\,$\mu$m) 
(detected in 50\%, i.\,e. 10/20), [\ion{Fe}{ii}] (at 17.9, 22.9, 24.5, and 35.8\,$\mu$m) (detected in 40\%, i.\,e. 8/20), 
[\ion{Si}{ii}] and [\ion{Si}{iii}] (at 34.8 and 35.3\,$\mu$m, and at 25.6\,$\mu$m) (detected in 50\% and 55\%, respectively, i.\,e. 10/20 and 11/20),
 and [\ion{S}{iii}] (at 33.5\,$\mu$m) (detected in 65\%, i.\,e. 13/20). There 
are also a few lines which we tentatively assign to [\ion{Fe}{i}] (at 24, 28.5, and 36.5\,$\mu$m) (detected in 30\%, i.\,e. 6/20). 
These lines are quite common in Mid-IR YSO spectra~\citep[see, e.\,g.,][]{forbich,lahuis10,ba-s},
and mostly originate from both dissociative (e.\,g. [\ion{S}{iii}], [\ion{Si}{ii}], [\ion{Si}{iii}], [\ion{Fe}{ii}], [\ion{Fe}{i}]) 
and non-dissociative (H$_2$) shocks along the YSOs jets~\citep[][]{lahuis10}, whereas a few lines may also originate from the discs, 
as H$_2$, [\ion{Si}{ii}], [\ion{Si}{iii}], [\ion{Fe}{i}]~\citep[see, e.\,g.,][]{lahuis}. 
A complete list of all the detected lines is given in Appendix~\ref{appendixA:sec} (Tables~\ref{tab:ctf29}-\ref{tab:ctf245B1}).

Several YSOs in our sample show prominent ice absorption features at 5--8\,$\mu$m (H$_2$O) and 15.2\,$\mu$m (CO$_2$) (79\% of the sample)
as well as the amorphous silicate absorption feature at 9.7\,$\mu$m (observed in all the IRS Spitzer spectra). 
This last feature is detected in absorption in all sources except \#7 and \#18, where it is observed in emission, 
usually indicating more evolved objects~\citep[see, e.\,g.,][]{watson}. 
These broadband features are usually connected to dust and ice mantels on dust in the interstellar medium and in the YSO envelope,
and their optical depths have been used in several works to compute extinction towards the protostellar photospheres~\citep[see, e.\,g.,][]{alexander,chiar}.


\subsection{Derived stellar parameters} 

\begin{figure}
 \centering
\includegraphics [width=8cm] {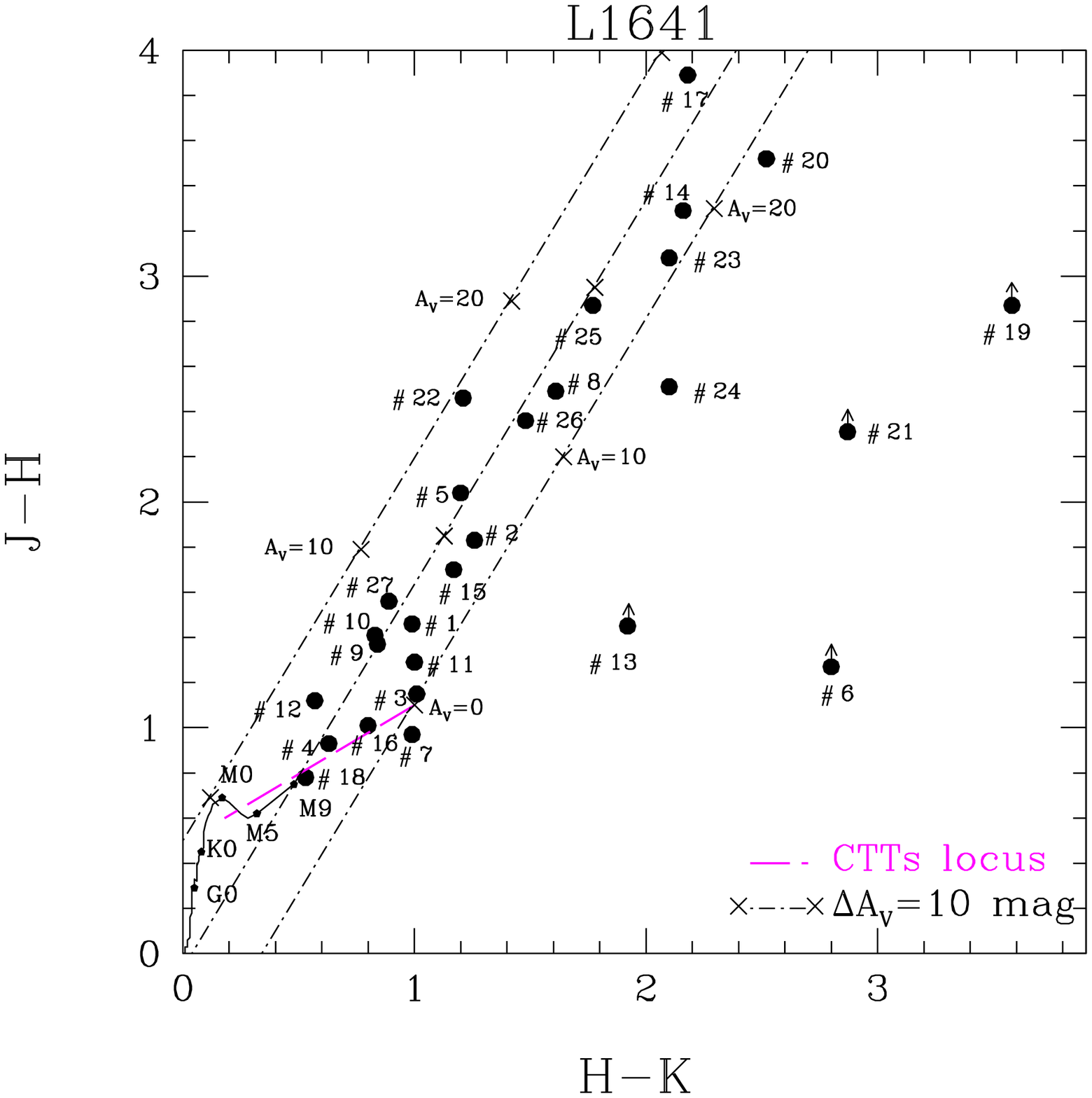}\\
\includegraphics [width=8cm] {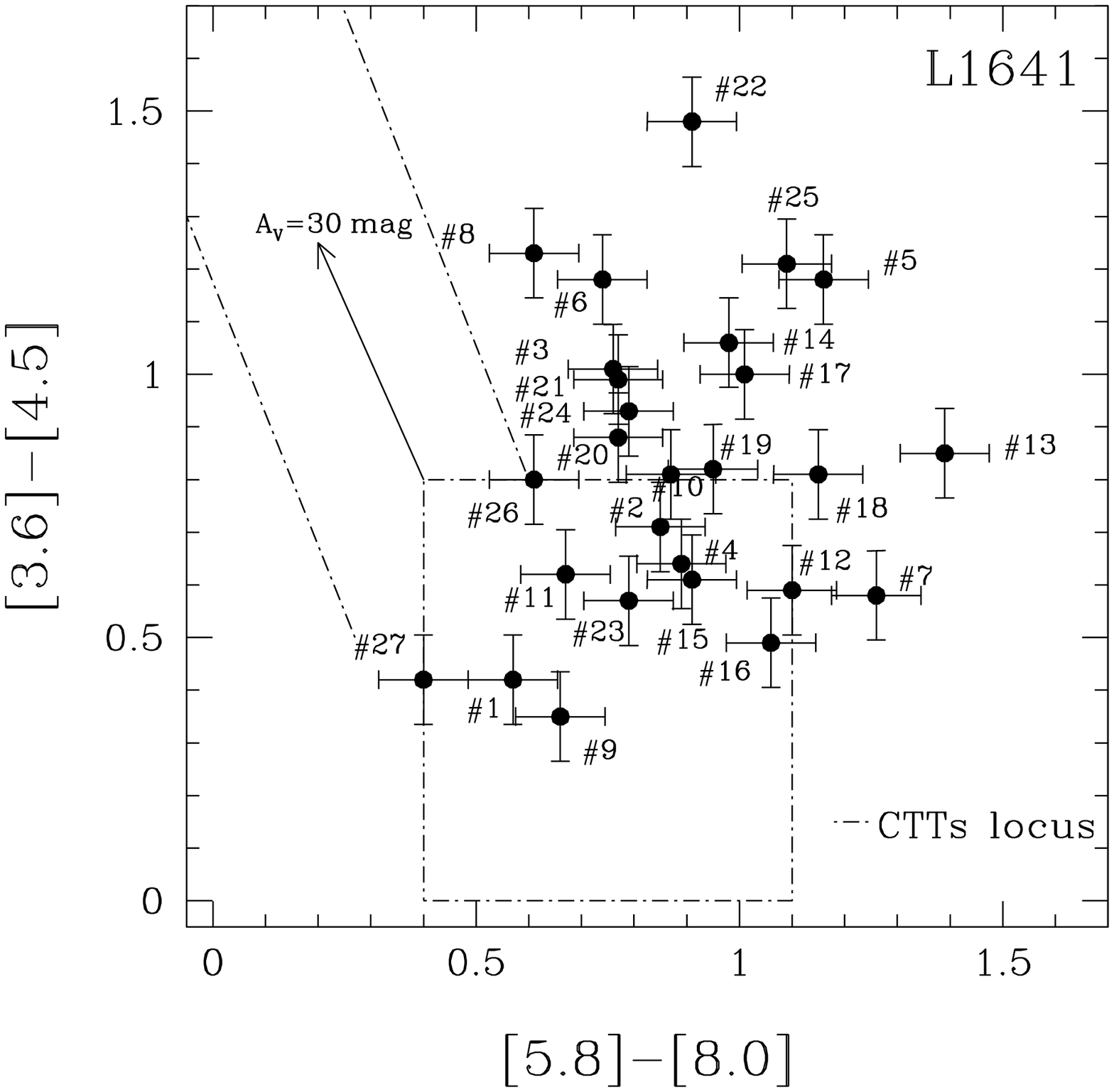}\\
 \caption{{\bf Top panel:} $JH$ vs. $HK$ diagram for our L1641 targets. 
Dwarf MS and classical T Tauri loci~\citep[][]{meyer} are indicated by a continuous black line and a dashed magenta line, respectively.
Black dotted dashed lines show the reddening vectors.
{\bf Bottom panel:} IRAC colour-colour diagram for the L1641 targets. The black dotted dashed rectangle delineates the
domain of Class\,II YSOs~\citep[][]{allen}, two parallel lines indicate the locus of reddened Class\,II sources~\citep{meg04},
and an average reddening vector for $A_\mathrm{V}$=30\,mag is plotted, as well.
\label{colors:fig}}
\end{figure}

\subsubsection{Reddening}
\label{reddening:sec}

To obtain correct estimates of important stellar parameters, such as the accretion luminosity ($L_{acc}$), bolometric and stellar luminosities 
($L_{bol}$ and $L_*$), mass accretion and mass ejection rates ($\dot{M}_{acc}$, and  $\dot{M}_{out}$), 
the observed line fluxes and photometry must be properly dereddened.
Thus, deriving an accurate value of the visual extinction ($A_\mathrm{V}$) towards the stellar photosphere and its circumstellar region is a fundamental task. 
There are several independent methods to compute $A_\mathrm{V}$, and they are successfully applied to CTTs, that are usually less embedded. Determining 
the extinction towards more embedded protostars is particularly difficult, however, due to 
the high extinction and/or scattered light from the outflow cavities and reflection nebulae surrounding the 
YSOs~\citep[see, e.\,g.,][]{whitney03-1,beck,connelley}.

In this work, we have combined our multi-wavelength spectroscopy and photometry to obtain independent estimates of the visual extinction (see Table~\ref{av:tab}), namely:

\textit{i)} Line ratios of transitions arising from the same upper level can be used to evaluate $A_{\rm V}$.
Indeed the observed ratio depends only on the differential extinction, if the emission is optically thin, and once the theoretical value,
which depends on the Einstein coefficients and frequencies of the transitions, is known.

We use the Br$\gamma$ to Pa$\delta$ ratio, assuming that the observed emission arises from optically thin gas.
Results are reported in column 2 of Tab.~\ref{av:tab}.
[\ion{Fe}{ii}] line ratios are also used, in particular 1.644/1.257\,$\mu$m (source \#2, \#5, and \#18), 1.644/1.321\,$\mu$m (source \#5) 
or 1.71/1.60\,$\mu$m (source \#24), adopting the transition probabilities of \citet{NS}. Results are reported in column 3 of Tab.~\ref{av:tab}.

\textit{ii)} The optical depth of the 9.7\,$\mu$m amorphous silicate absorption feature ($\tau_{9.7}$) and optical depth of the 
CO$_2$ ice feature at 15.2\,$\mu$m ($\tau_{15.2}$) can also be used to estimate the extinction.

To measure the optical depths, the following steps were followed. 
First we determined the stellar continuum by fitting a low-order polynomial on a log-log scale to the spectral segments between
5 and 35\,$\mu$m, which are not affected by absorption features~\citep[see e.\,g.][]{boogert}. Then, we applied to derive 
$\tau_\lambda$, the following relationship~\citep[see, e.\,g.,][]{alexander,chiar}:

\begin{equation}
\label{tau:eq}
\tau_{\lambda} = -ln(F_{source}(\lambda)/F_{cont}(\lambda))
\end{equation}
where $F_{source}$ and $F_{continuum}$ are the flux densities of the source
and of the continuum fit at $\lambda$, respectively.
$\tau_\lambda$(CO$_2$)$_{ice}$ can be converted to a column density, using the following relationship~\citep[see, e.\,g.,][]{alexander}:

\begin{equation}
\label{NCO2:eq}
N(CO_2) =  \frac{\int{ \tau_\lambda d\lambda}}{\lambda^2_{peak} A}
\end{equation}

where $\int{ \tau_\lambda d\lambda}$ is the equivalent width, $\lambda_{peak}$ is the wavelength of the feature peak optical depth,
and $A$ is the ice band strength ($1.1\times10^{–17}$\,cm\,molecule$^{-1}$) assuming pure CO$_2$ ice~\citep[][]{gerakines, alexander}.

The $A_\mathrm{V}$ value for the silicates is obtained from~\citet{R&L} as $A_\mathrm{V}$= $\tau_{9.7}\times18$,
and from the CO$_2$ column density using the relationships of ~\cite{bergin}. Results from the CO$_2$ and silicate measurements
are reported in Columns 4 and 5 of Tab.~\ref{av:tab}, respectively. 

\textit{iii)} Colour-colour diagrams from $J-$, $H-$, and $K-$band $2MASS$ photometry. Fig.~\ref{colors:fig} (top panel) shows
the $J-H$ vs $H-K$ colours of our targets, along with the dwarf-MS and CTTs loci~\citep{meyer}, and the reddening vectors.
Many of the sources are located along the reddening band extending from the CTTs locus, while there are a few objects 
positioned on the right side of the strip, indicating less evolved objects. Arrows on these objects indicate lower limits, 
because their $J-$band magnitude is an upper limit 
or has a S/N ratio $\le$3. Derived $A_\mathrm{V}$ values are reported in column 6 of Tab.~\ref{av:tab}.

\textit{iv)} Self-consistent method \citep[SCM, see Sect.~\ref{Sparam:sec} and ][for a detailed description]{antoniucci}.
This method relies on the fact that there are two ways of determining the source stellar luminosity $L_*$, both depending on
the actual extinction value:
i) from the bolometric luminosity $L_{bol}$ (see Sect.~\ref{phot:sec}) and the 
accretion luminosity $L_{acc}$ derived from the de-reddened Pa$\beta$ (and/or Br$\gamma$) flux 
(Sect.~\ref{acc:sec}), assuming that $L_* = L_{bol}-L_{acc}$;
ii) from the observed $K$ band magnitude of the source, considering the distance, the spectral type, and an estimate of the veiling in the 
relevant band (see Sect.~\ref{SpT:sec}). 
An average veiling value of $r_K$ = 1 for the whole sample is assumed. Because veiling dependence on magnitude is logarithmic, while
extinction dependence is linear, the computation is less sensitive to veiling variations. For example, assuming $r_K$ = 5, instead of one, would increase $A_\mathrm{K}$ of $\sim$0.48, i.\,e.
$\sim$5\,mag in the $V$ band.

The best estimate of the extinction will be the one for which we obtain 
the same value of $L_*$ from both computations.

For comparison, additional extinction estimates retrieved from the literature~\citep{strom,fang,connelley} are also reported in column 8 of
Tab.~\ref{av:tab}.

The inferred $A_\mathrm{V}$ values range from 0 to $\sim$30\,mag, although the different methods adopted here usually produce
a wide range of results for each object. Indeed this is a well known problem for Class I sources~\citep[see e.\,g.][]{beck,prato,davis11}, and it is due to 
several reasons, 
as e.\,g.:\textit{ i)} the considered lines trace different regions of the YSO (e.\,g., the circumstellar region or the jet, 
the \ion{H}{i} or the [\ion{Fe}{ii}] lines, respectively), 
possibly with different extinctions;
\textit{ ii)} scattering is not taken into account by any of the above methods, although those based on the [\ion{Fe}{ii}] line ratios or 
on MIR optical depths should not be significantly affected;
\textit{ iii)} a standard ISM law~\citep[][]{R&L} has been assumed to correct for the differential extinction, 
without considering that it can vary, depending on the size and properties of the grains in the disc, envelope, or jet~\citep[see, e.\,g.][]{cardelli}.

Despite these limitations, Tab.~\ref{av:tab} shows a relatively good agreement among the different methods. 
In particular, the cross correlation coefficient between different sets of $A_\mathrm{V}$ ranges between $r=0.95$ (Si vs SCM) with an rms
of $\sim$3\,mag, to $r=0.66$ (CC vs SCM) with an rms of $\sim$6\,mag. The average value is $r=0.86$.
We then adopt an average $A_\mathrm{V}$ value for each object (column 9 of Tab.~\ref{av:tab}), after discarding those that deviate from 
the average by more than one sigma.

\begin{table*}
\caption{Visual extinction values of the sample.}
\label{av:tab}
\centering
\begin{tabular}{ccccccccc}
\hline \hline
Source	        &   $A_\mathrm{V}$(HI)  &  $A_\mathrm{V}$([\ion{Fe}{ii}])    & $A_\mathrm{V}$(CO$_2$) & $A_\mathrm{V}$(Si)	 & $A_\mathrm{V}$-CC  &   $A_\mathrm{V}$-SCM  & $A_\mathrm{V}$ pub  &   $A_\mathrm{V}$-adp   \\
ID              &   (mag)      &  (mag)      &   (mag)     & (mag)         & (mag)       & (mag)   & (mag)   & (mag)   \\
\hline\\[-5pt]
1	        & 8$\pm$3      & $\cdots$    & $\cdots$    & 1-2	   &  4$\pm$0.5    &  1  &  4.2$^2$         & 2     \\  
2	        & 9.5$\pm$1    &  5$\pm$5    &  11	   &   8-9	   &  8$\pm$0.5 &  8.6  & 9$^1$; 6$^2$     & 8.5     \\  
3	        & 3.7$\pm$0.5  &  $\cdots$   &  $\cdots$   & $\cdots$	   &  2$\pm$0.4 &  0   & 1.7--9$^1$  & 2    \\  
4	        &   1$\pm$0.5  & $\cdots$    & $\cdots$    & 3-6	   &  1.5$\pm$0.6  &  4  & 0.2$^1$; 5.7$^2$ & 3    	\\  
5	        &   $\cdots$   &  11$\pm$4   &    6-9	   &   5-7	   &  10.5$\pm$0.4 &  9  & 12$^1$	    & 10    \\  
6	        &   $\cdots$   &  $\cdots$   &  25-30	   &   20-22	   &  $>$14        &  24 & $\cdots$         & 24     \\  
7	        &   $\cdots$   &  $\cdots$   &  $\cdots$   &   $<$1	   &  0$\pm$0.5    &  0  & 9$^1$; 6$\pm^{(1.6)}_{(0.6)}\,^3$ & 0      \\ 
8	        &   $\cdots$   &  $\cdots$   &  $\cdots$   &  $\cdots$     &  19$\pm$0.5   &   $\cdots$   & 19$^1$           & 19       \\  
9               &   $\cdots$   &  $\cdots$   &  3	   &   $\cdots$    &  4.5$\pm$0.4  &  1   & 4.7$^1$          & 2      	\\  
10              & 8.3$\pm$1.2  &  $\cdots$   &  $\cdots$   & 5-7.2	   &  6$\pm$0.5    &  11  & $\cdots$         & 8.2    	 \\  
11	        & 4.6$\pm$0.5  &  $\cdots$   &  $\cdots$   &  $\cdots$     &  2.5$\pm$0.5    &  1.5  & 7$^1$            & 2       \\  
12	        &   $\cdots$   &  $\cdots$   &  $\cdots$   &  $\cdots$     &  4$\pm$1      &  12 & $\cdots$         & 8      	\\  
13              &   $\cdots$   &  $\cdots$   &  22-25	   &   12-18	   &  $>$8         &  36 & $\cdots$         & 25     	\\  
14              &   $\cdots$   &  $\cdots$   &  21-27	   &   22-30	   &  22$\pm$2     &  24 & 28$^1$; 25.4$^3$ & 24      \\  
15	        & 8.7$\pm$0.7  &  $\cdots$   &  9.5	   &   4.5-6.5     &  7.5$\pm$0.5  &  8  & 13$^1$; 5.8$^2$  & 8.5        \\  
16	        &   $\cdots$   &  $\cdots$   &  4.5	   &   3-6	   &  2$\pm$0.4    &    $\cdots$ & 4.4$^1$; 6.7$^2$ & 4    	\\  
17              &   $\cdots$   &  $\cdots$   &  23-26	   &   11-15	   &  28$\pm$2     &  16 & 20$^1$           & 23       \\  
18	        & 0.5$\pm$0.5  &  3$\pm$2    &  4.5	   &   1-2	   &   0$\pm$0.4   &   2 & 5$^1$; 2.6$^2$   & 2       \\  
19	        &   $\cdots$   &  $\cdots$   &  25	   &   13-15	   &  21$\pm$4     &  22 & $\cdots$         & 20     	\\  
20	        &   $\cdots$   &  $\cdots$   &  25-32	   &   14-20	   &  23$\pm$4     &  13 & 26$^1$           & 23     	\\  
21	        &   $\cdots$   &  $\cdots$   &  20-35	   &   17-20	   &  $>$16        &  21 & $>$4.5$^3$         & 22.5   	 \\  
22              &   $\cdots$   &  $\cdots$   &  $\cdots$   &  $\cdots$	   &  14$\pm$2     &  10 & $\cdots$         & 12     	 \\ 
23              &   $\cdots$   &  $\cdots$   &  21	   &  18-25	   &  19$\pm$2     &  24 & 16$^1$           & 21      \\   
24              &   $\cdots$   &  20$\pm$10  &  $\cdots$   &  25-30	   &  16$\pm$2     &  26 &  34$^1$; 13-24$^3$         & 25     	 \\  
25              &   $\cdots$   &  $\cdots$   &  26-30	   &  14-16	   &  18.5$\pm$2   &  16 & $\cdots$       & 18   	     \\  
26              & 11.3$\pm$5  &  $\cdots$   &  $\cdots$   &  $\cdots$    &  13.0$\pm$0.5   &   6   & $\cdots$         & 10      	 \\  
27              &  $\cdots$   &  $\cdots$   &  $\cdots$   &  $\cdots$      &   7.5$\pm$0.5 &  8  & 6.3$^1$          & 8      	\\  
\hline
\end{tabular}
\tablebib{$^1$~\citet{strom}; $^2$~\citet{fang}; $^3$~\citet{connelley}.}
\end{table*}
\begin{figure*}[!t]
\resizebox{\textwidth}{!}{
\includegraphics{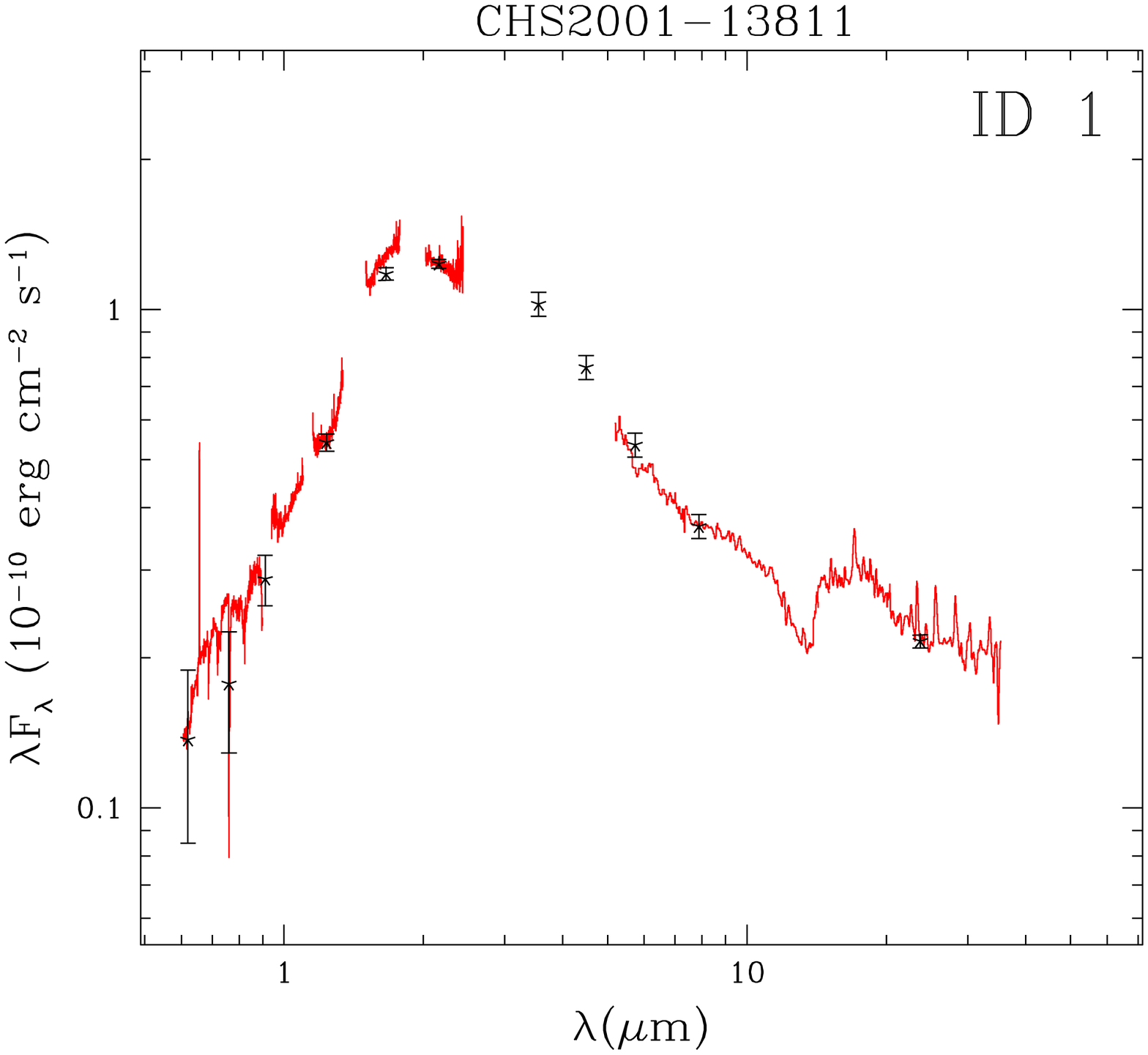} \includegraphics{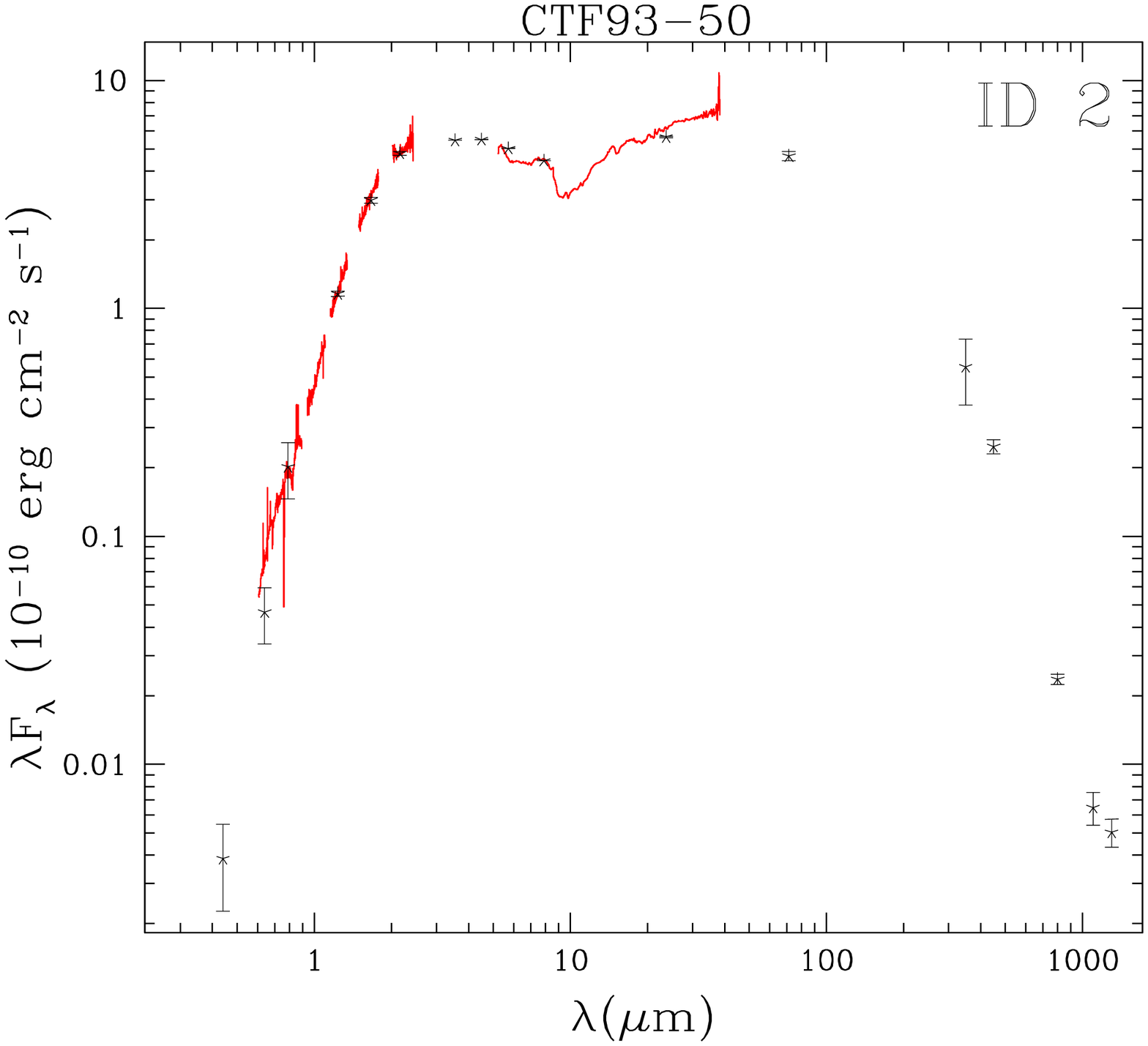} \includegraphics{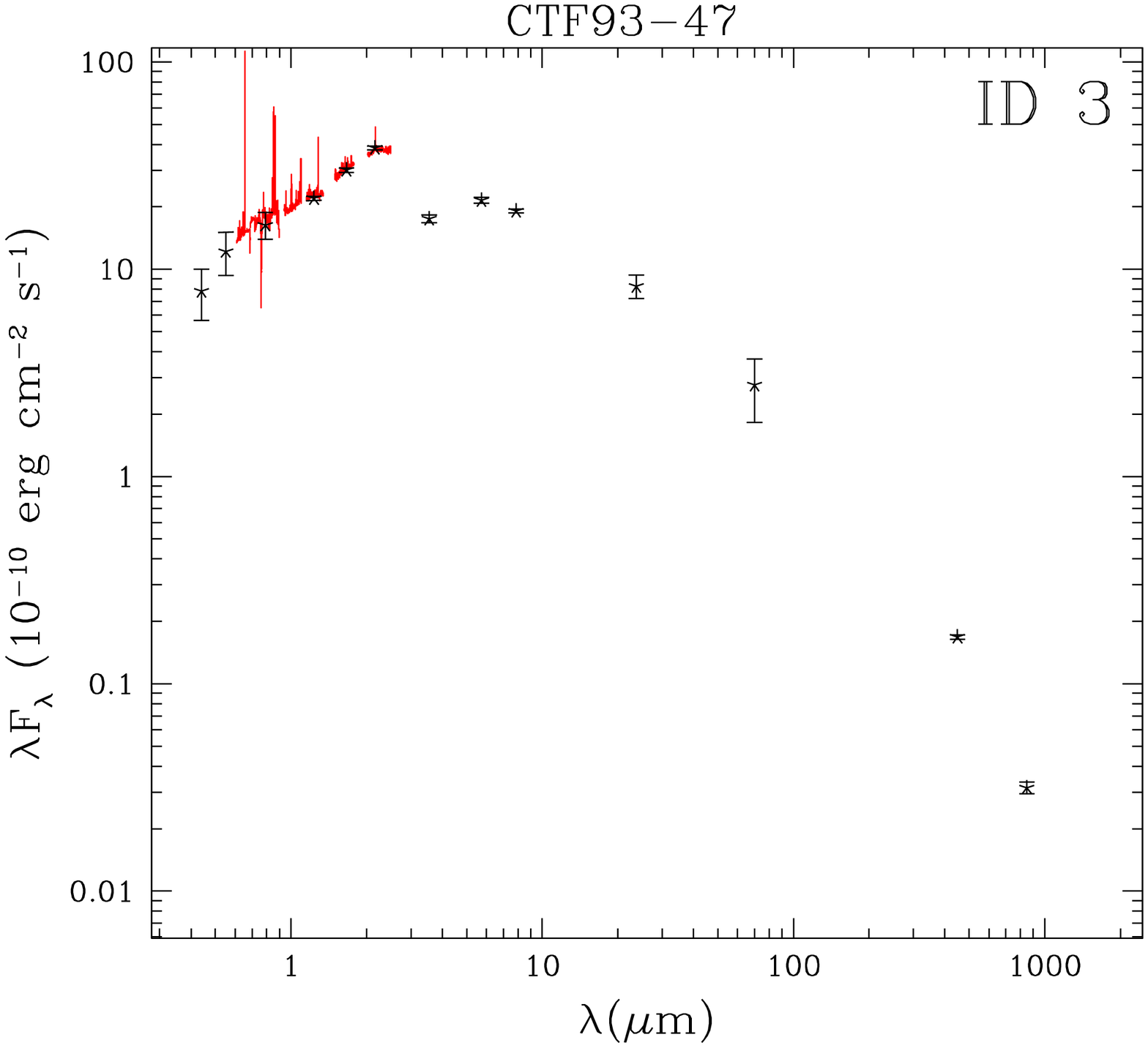} \includegraphics{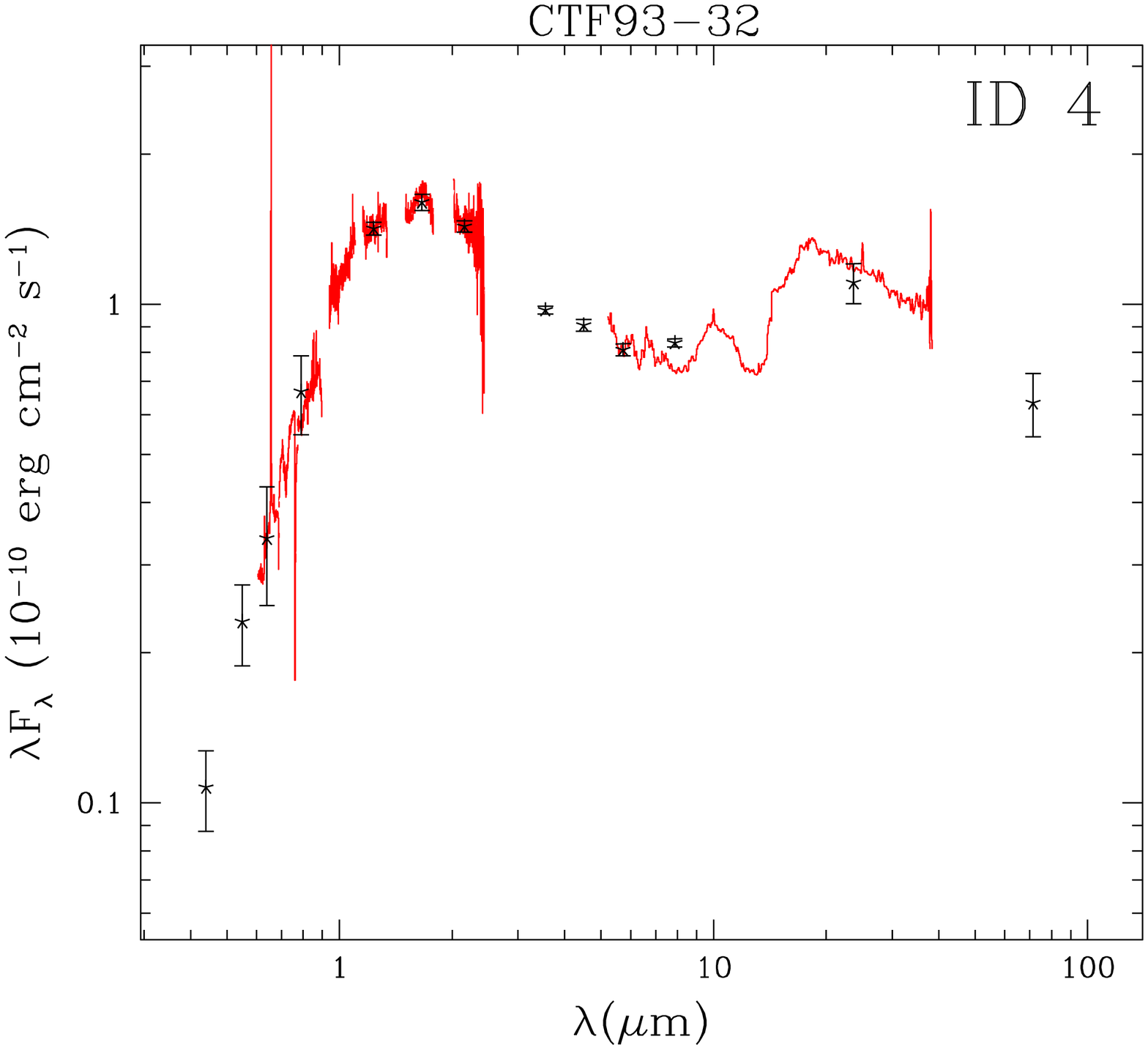}}

\resizebox{\textwidth}{!}{
\includegraphics{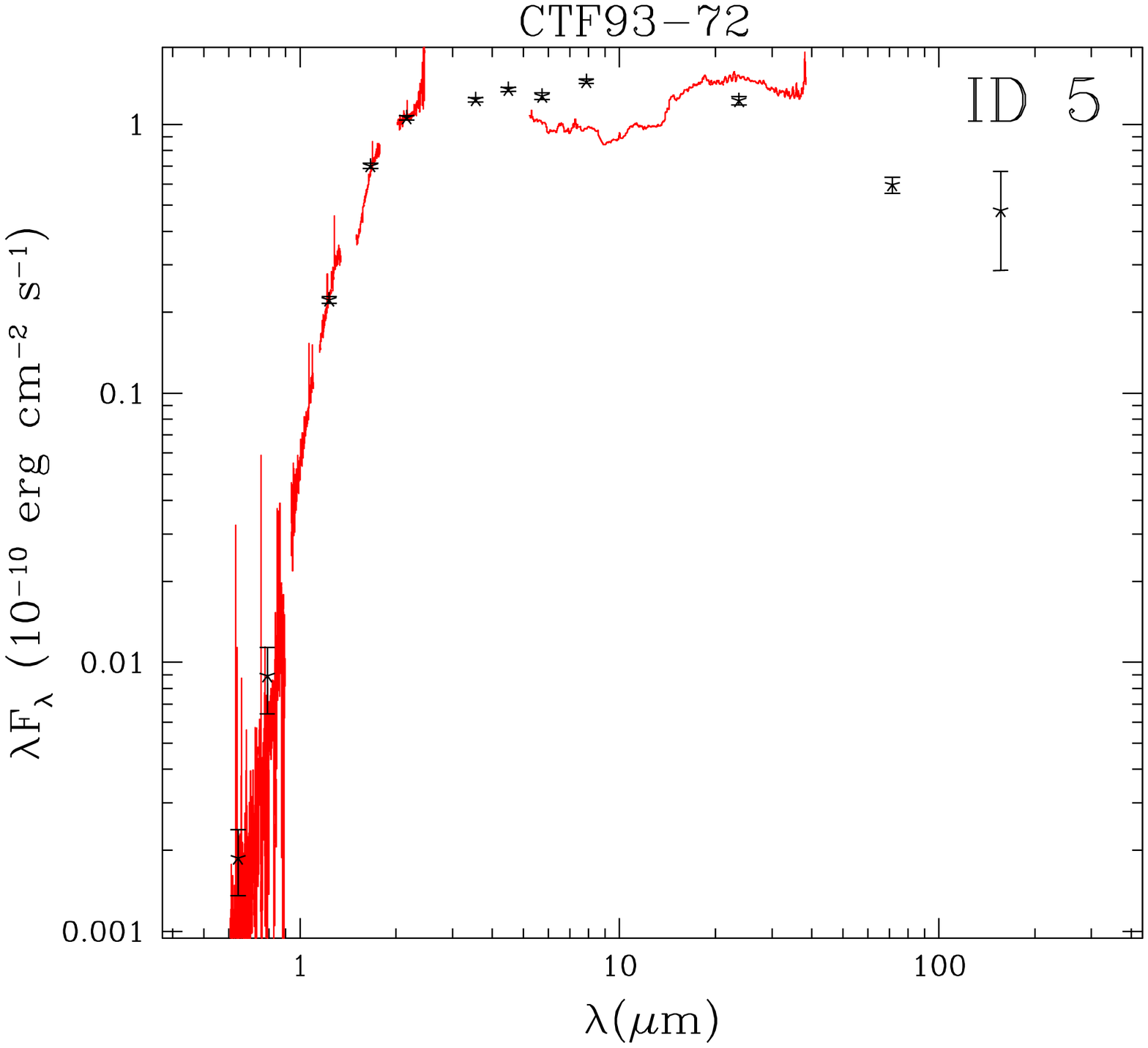} \includegraphics{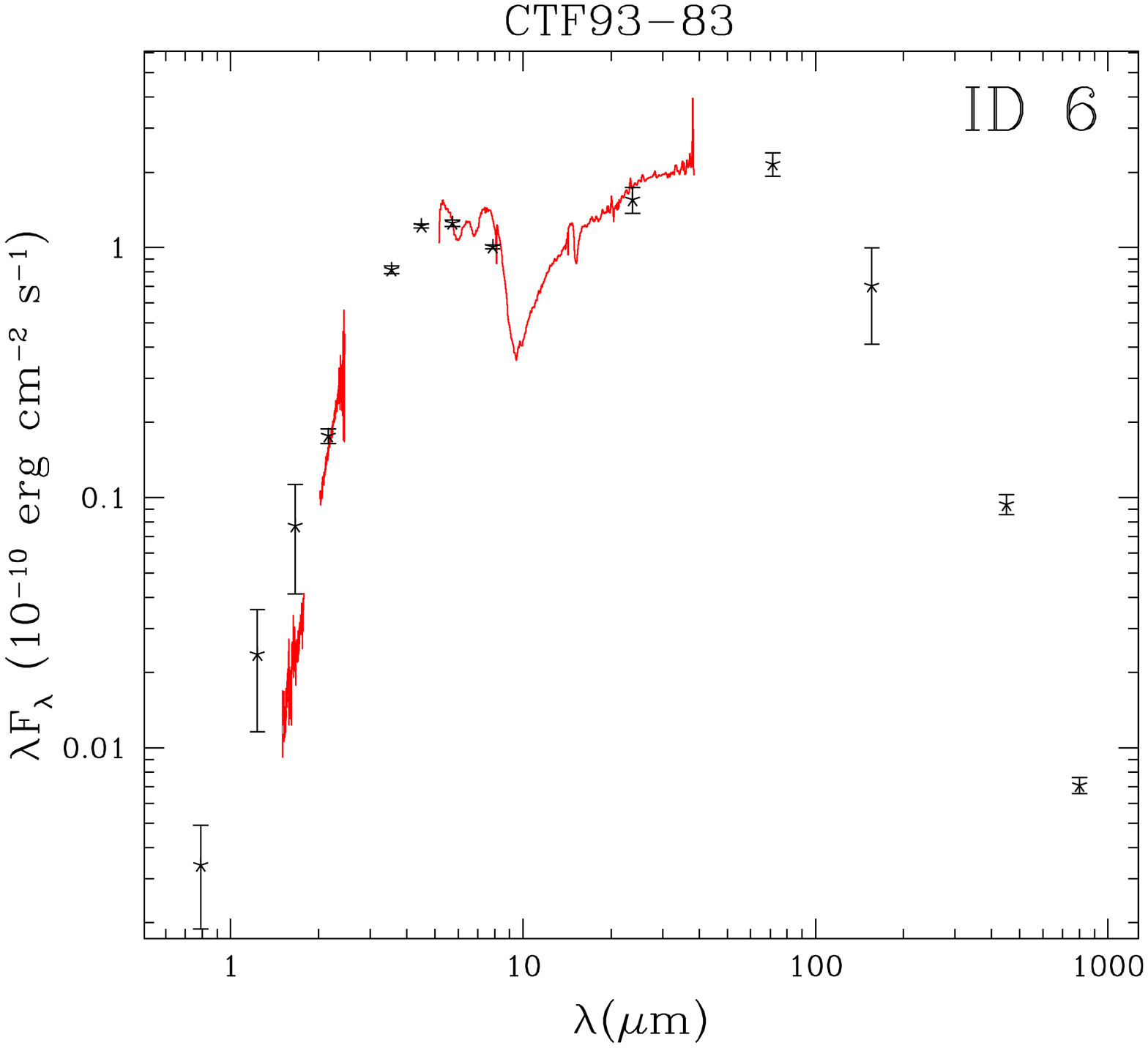} \includegraphics{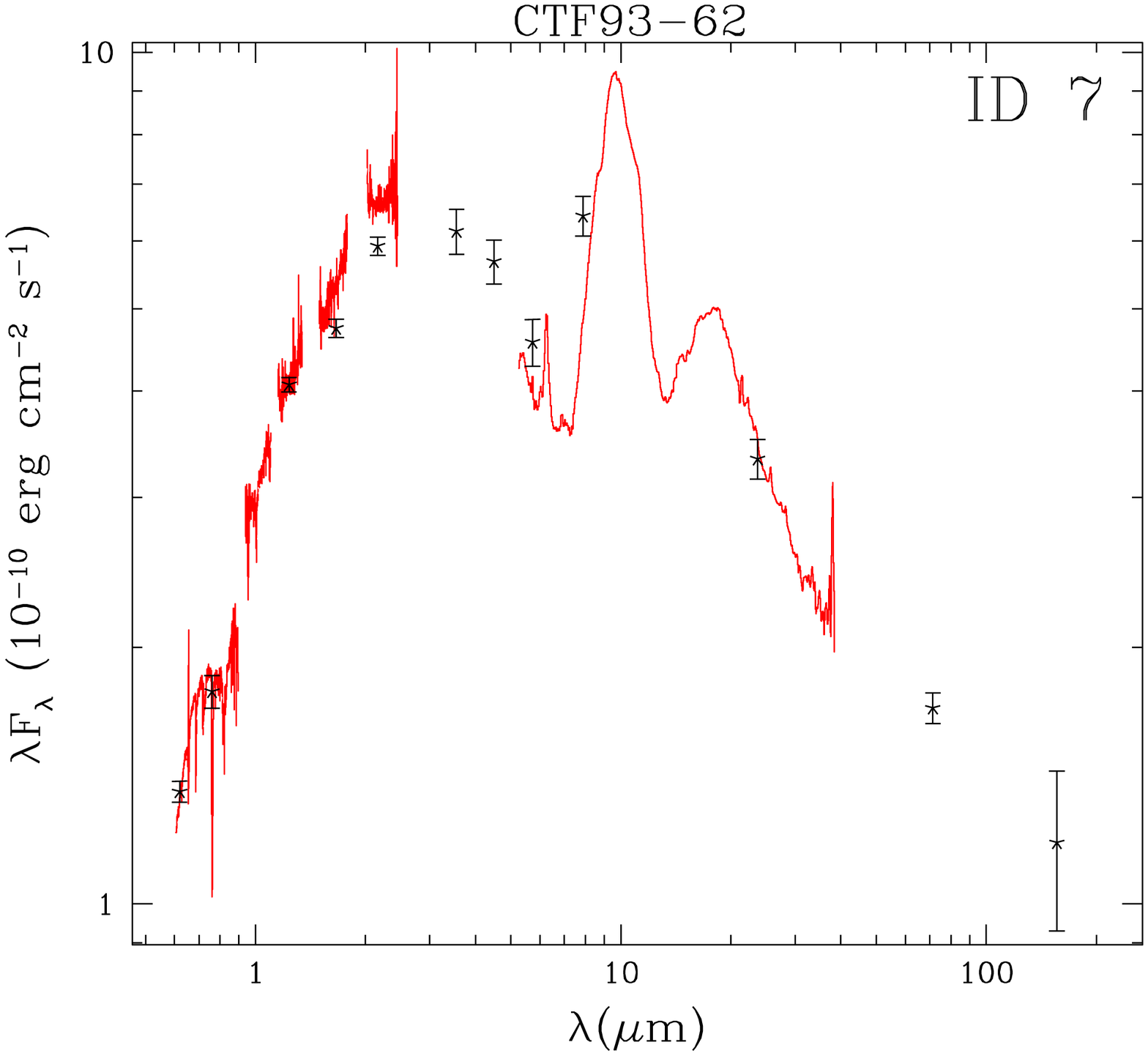} \includegraphics{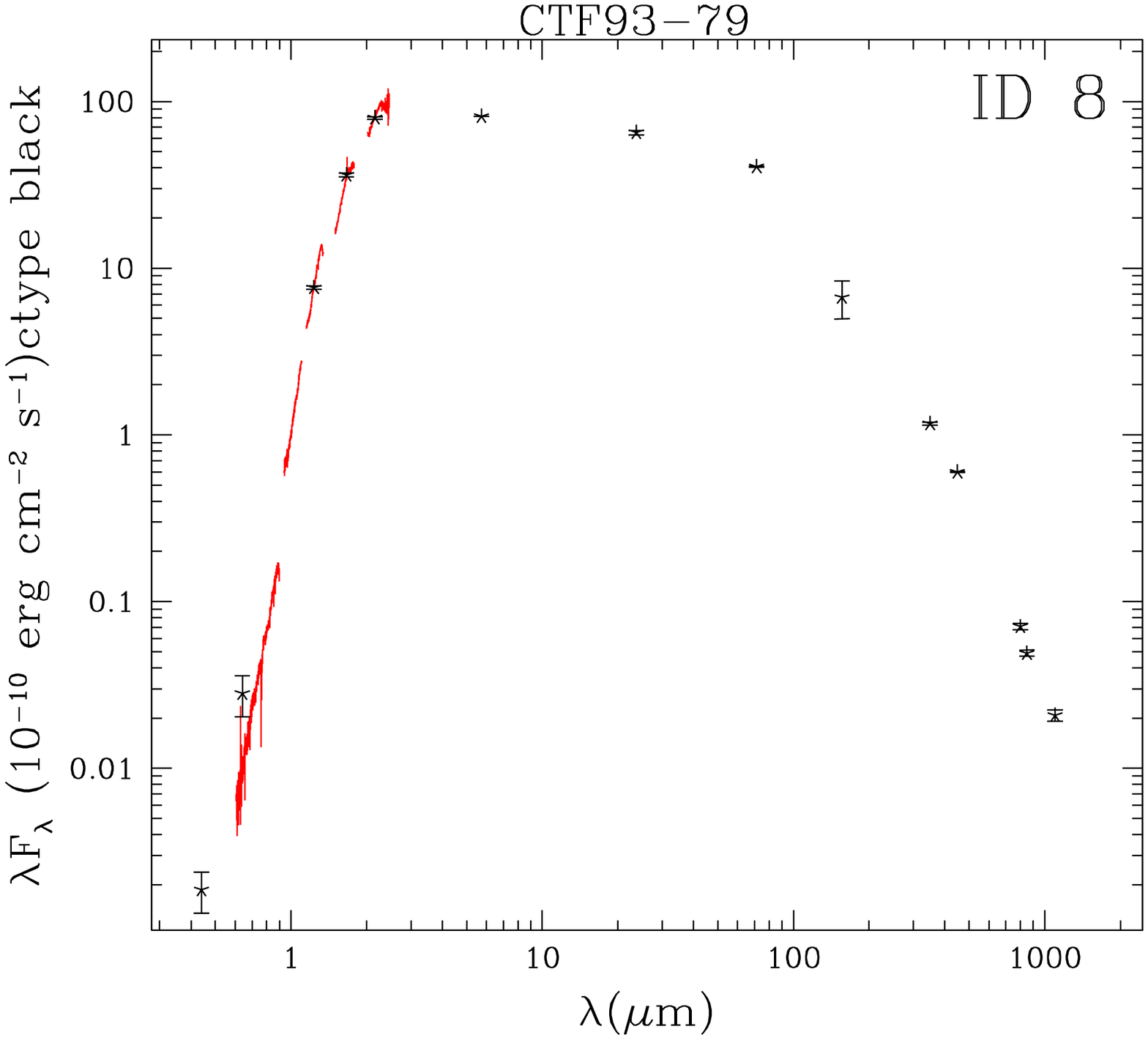}}

\resizebox{\textwidth}{!}{
\includegraphics{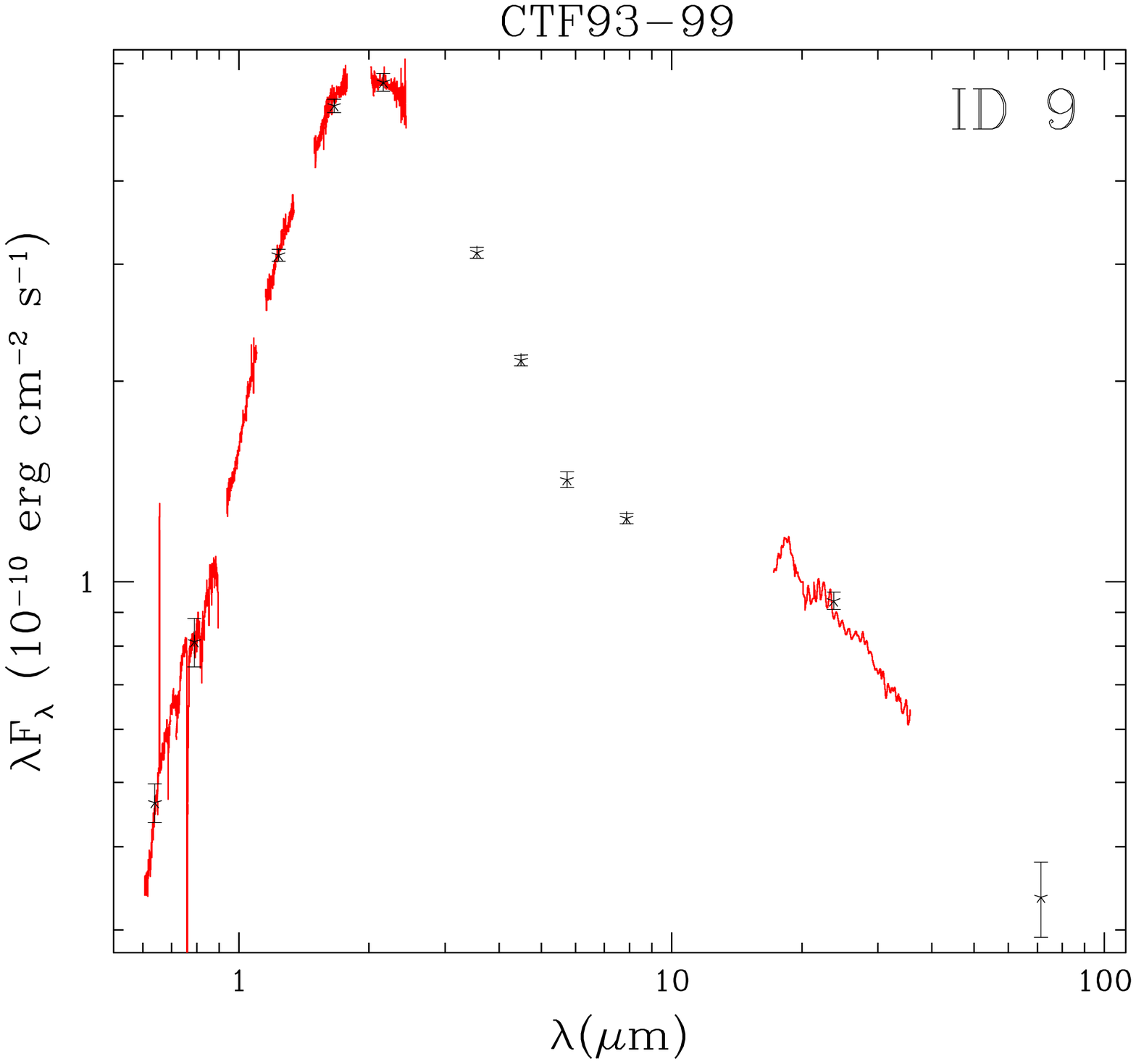} \includegraphics{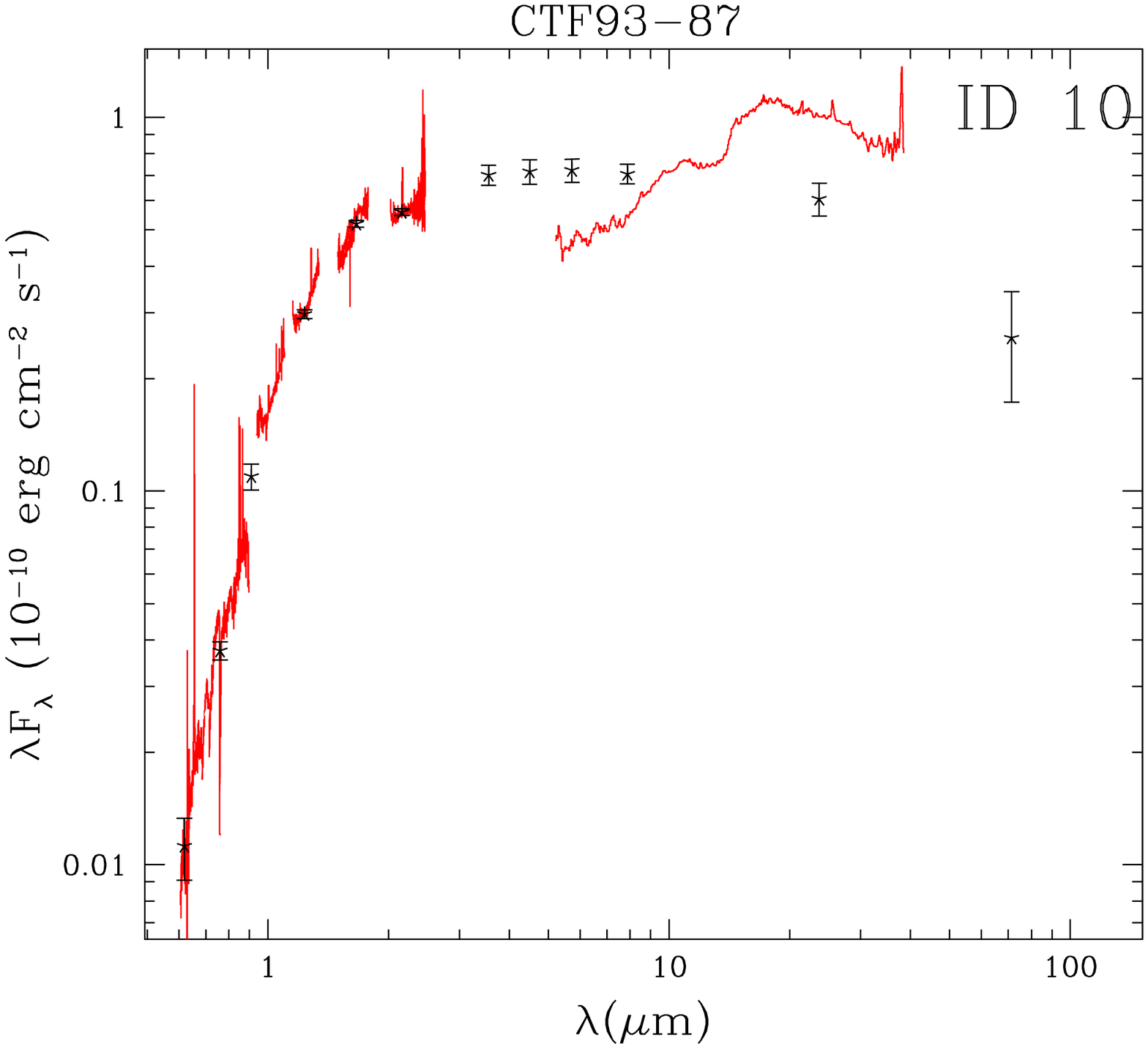} \includegraphics{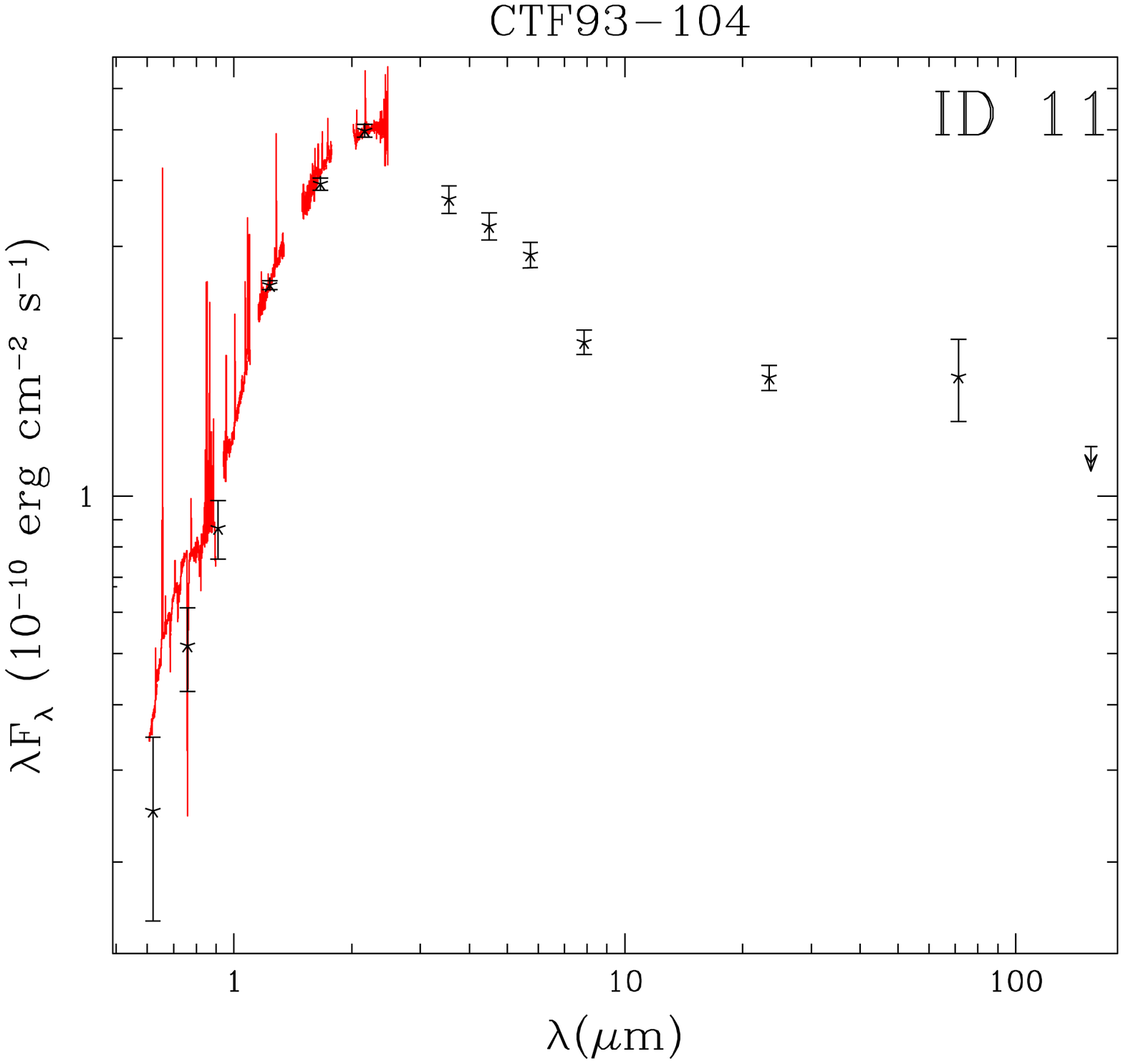} \includegraphics{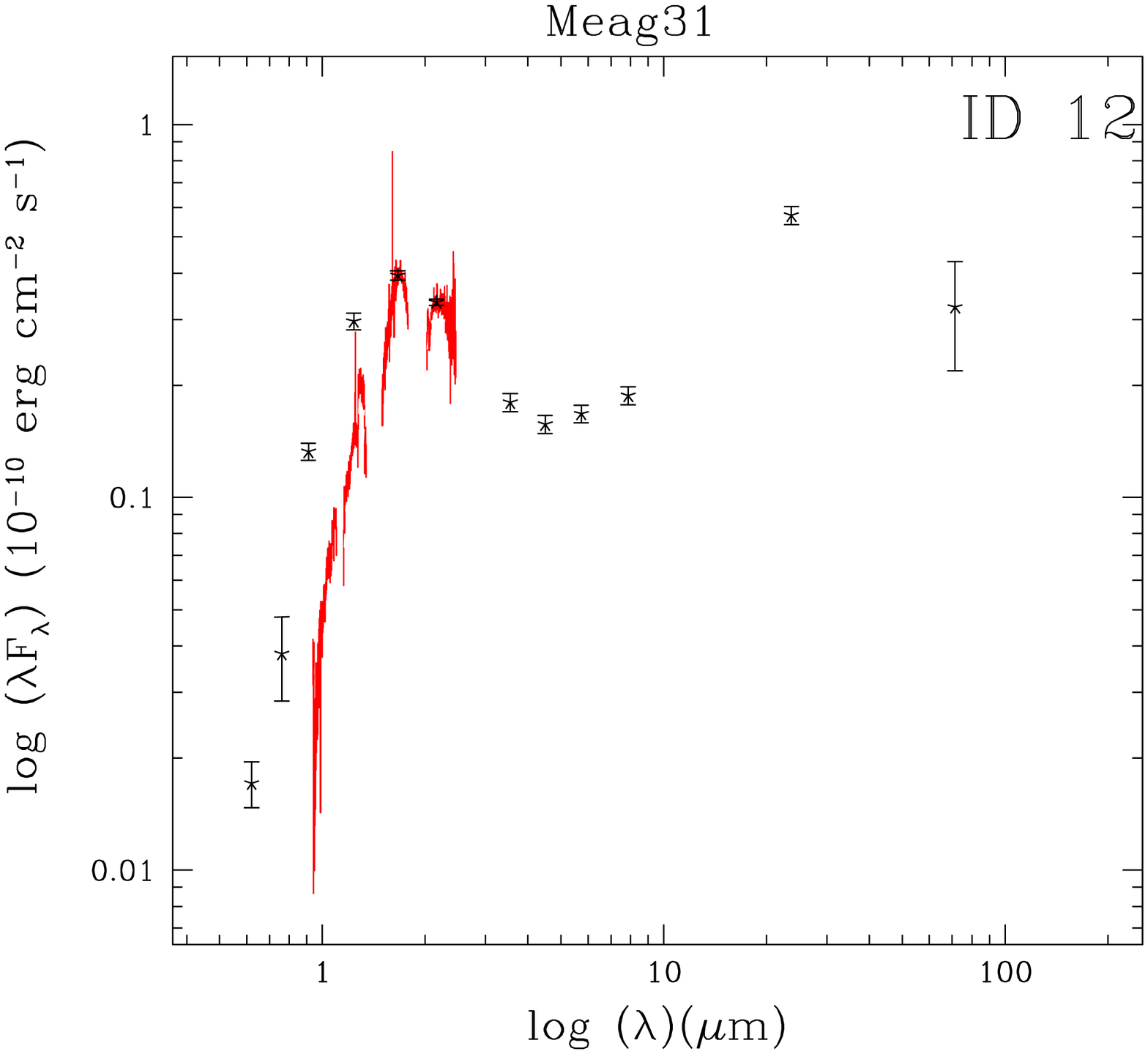}}
\resizebox{\textwidth}{!}{
\includegraphics{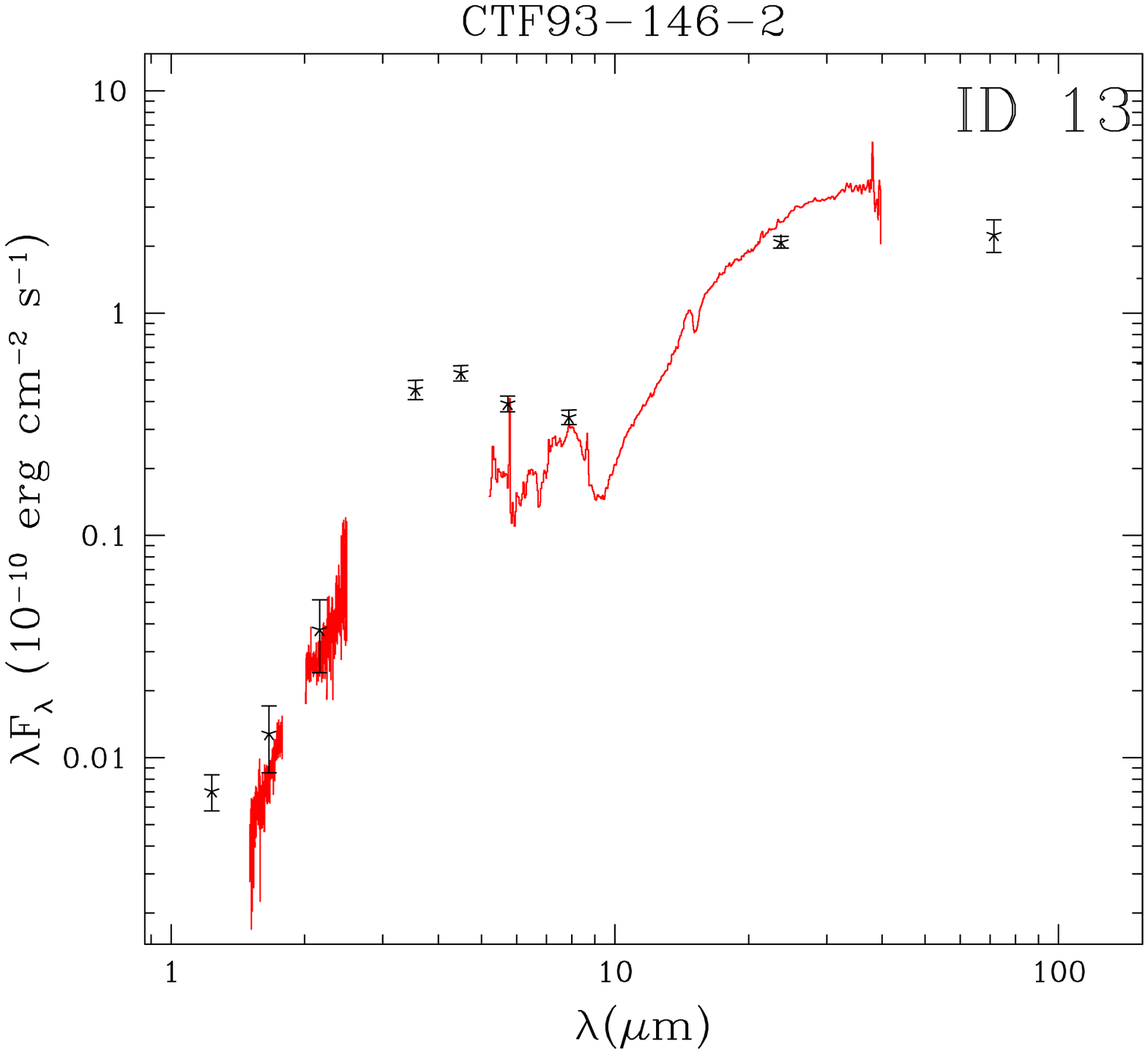} \includegraphics{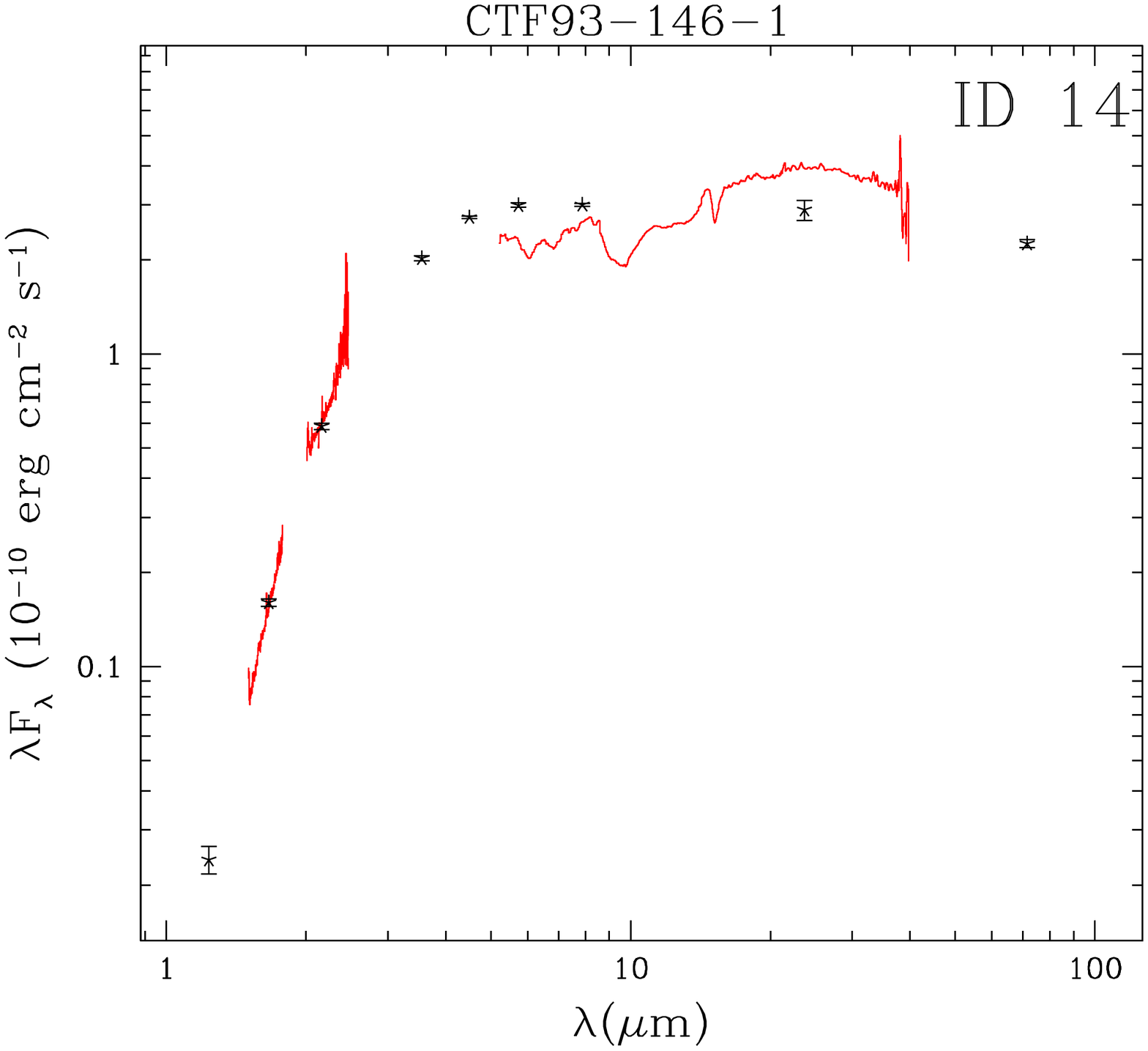} \includegraphics{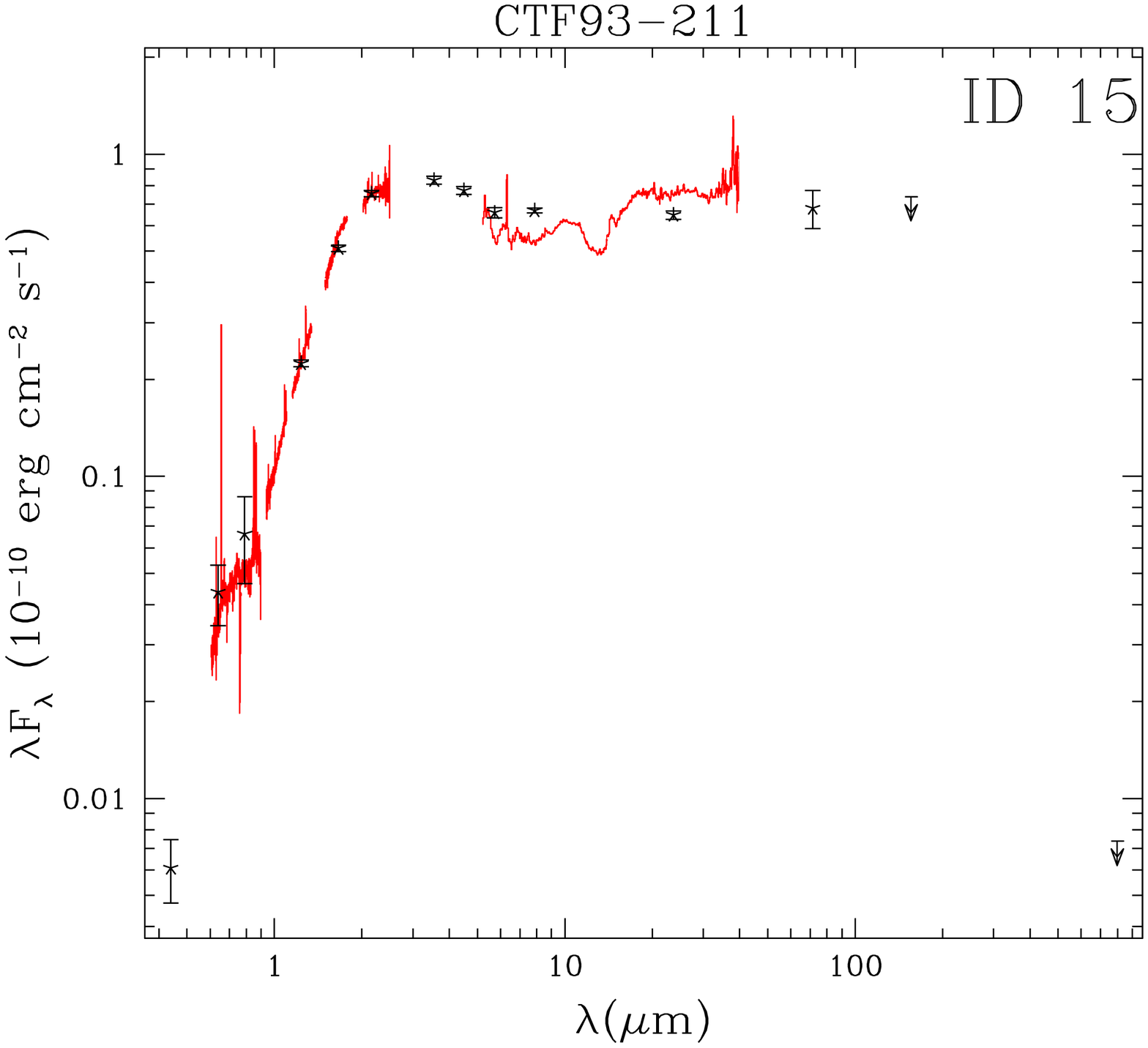} \includegraphics{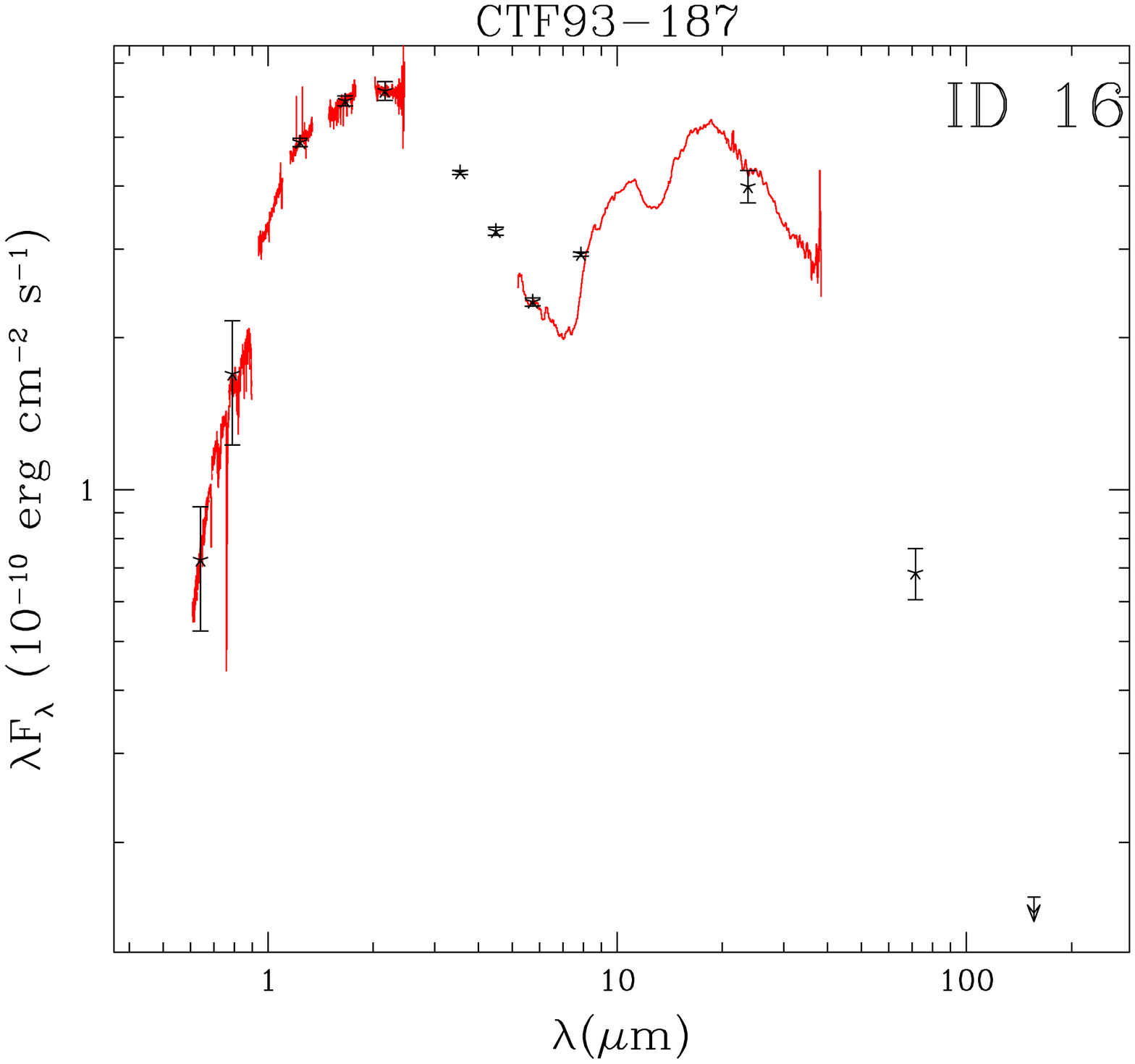}}
\resizebox{\textwidth}{!}{
\includegraphics{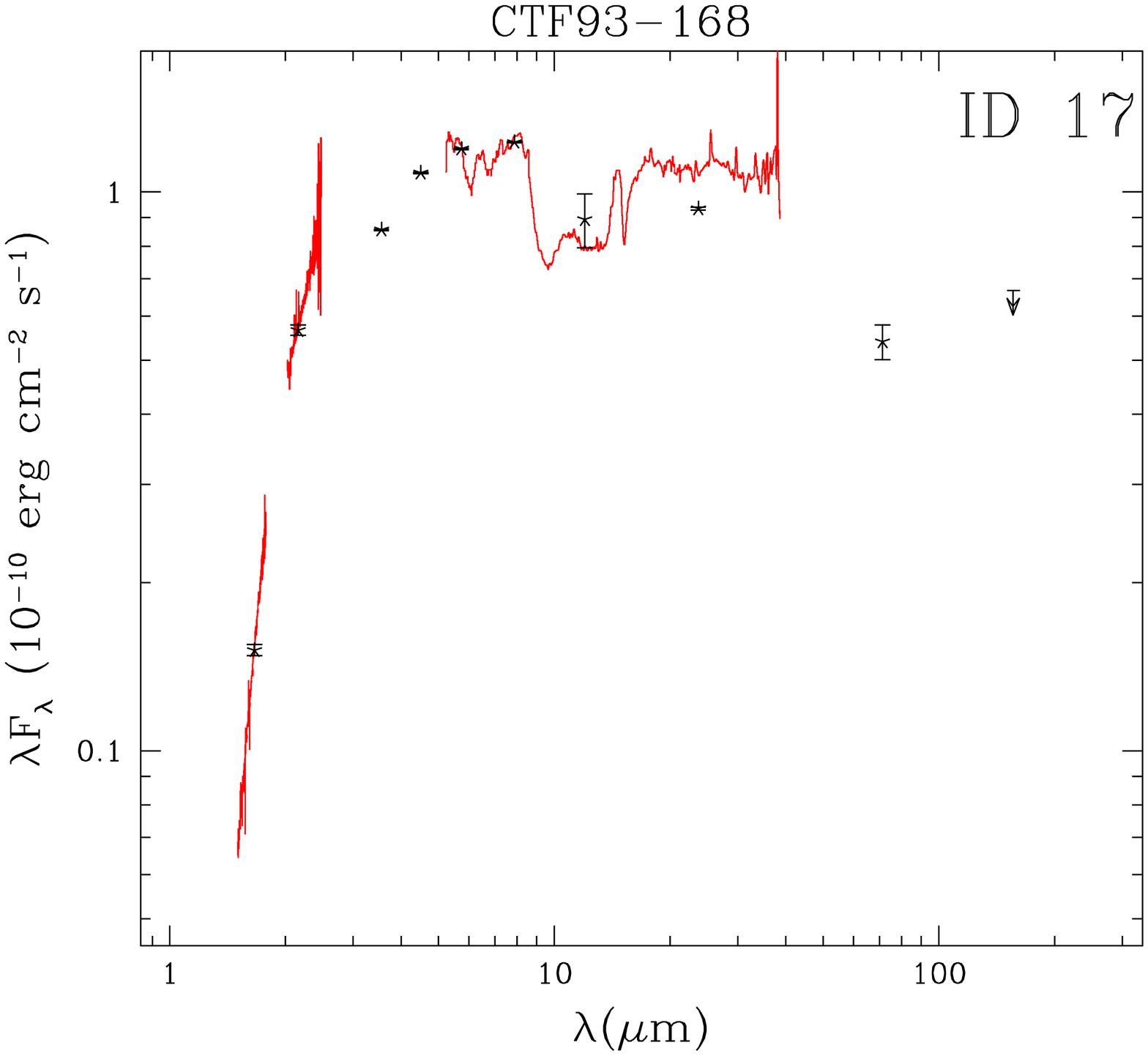}
\includegraphics{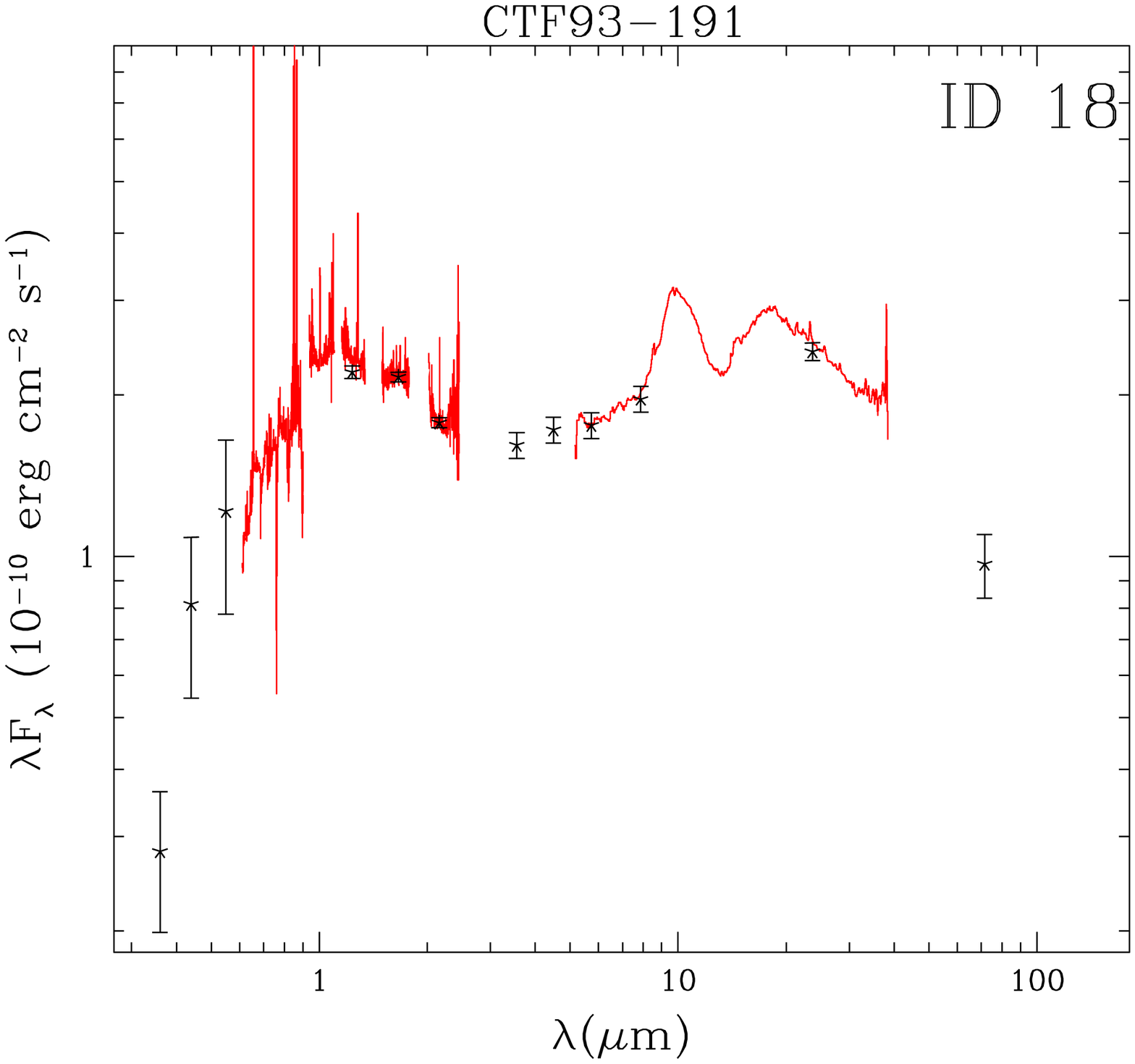}
\includegraphics{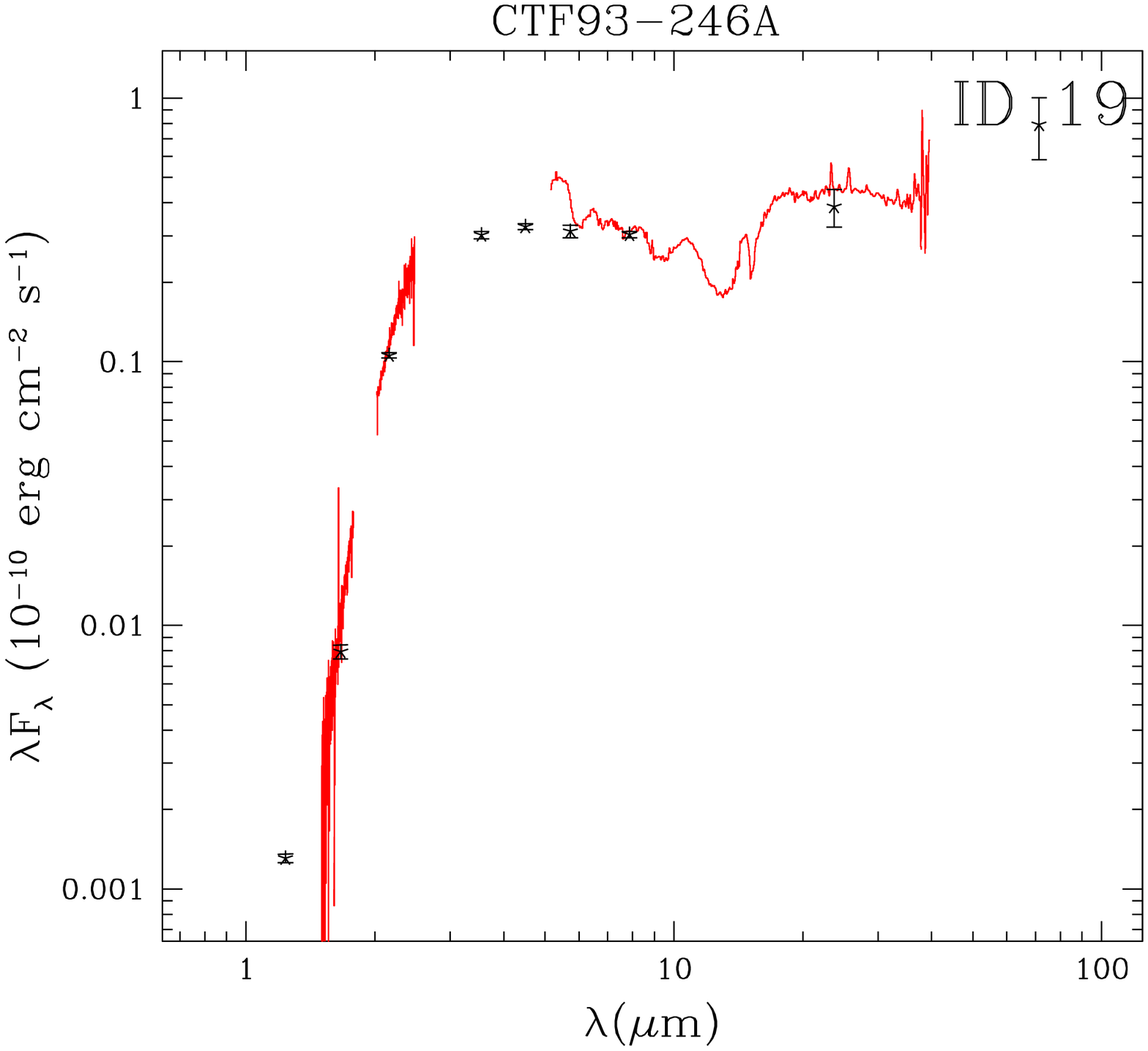}
\includegraphics{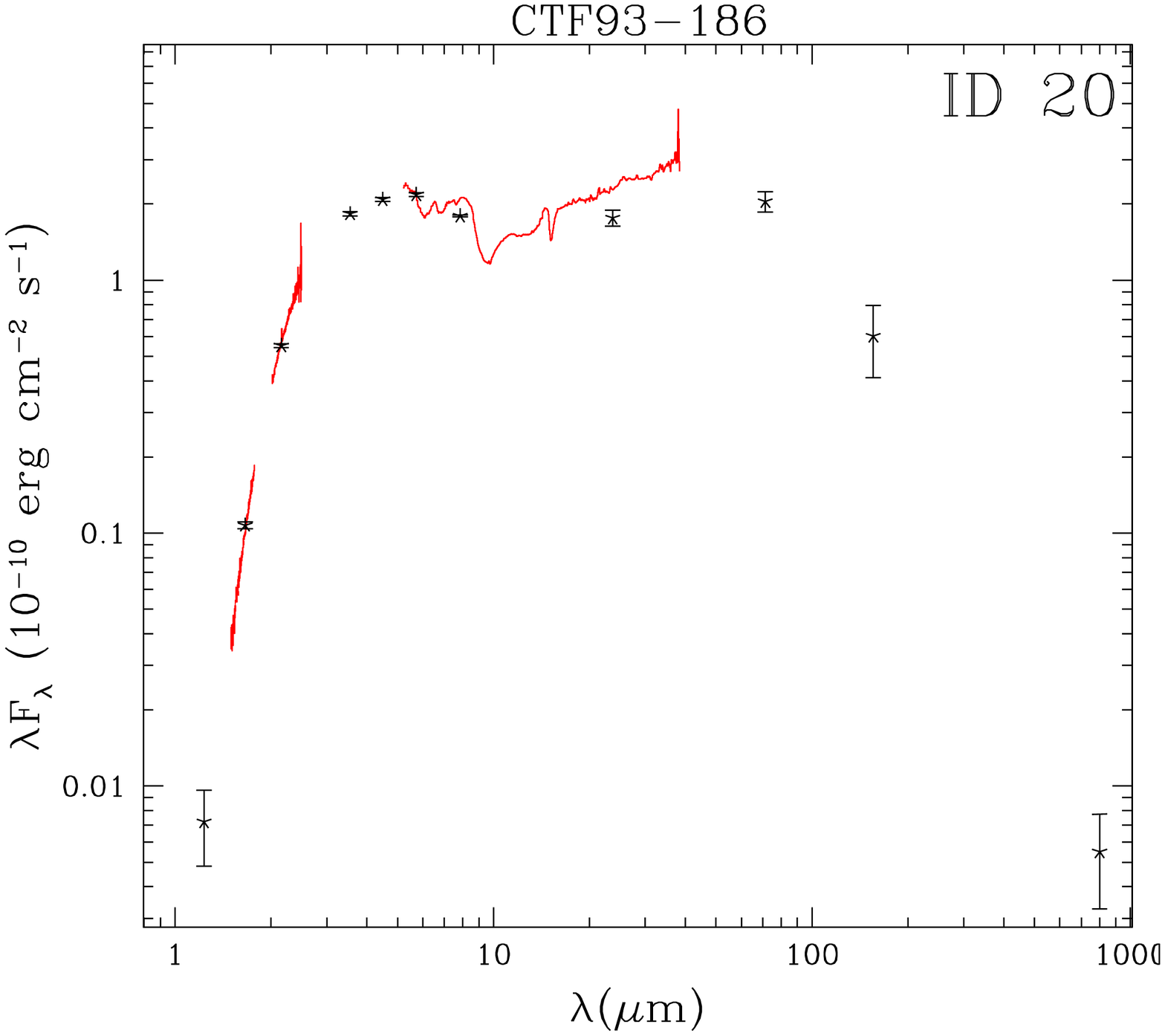}
}
\caption{Target SEDs and spectra. Photometric data with error bars are in black, spectroscopic data (ESO-NTT and Spitzer) are in red. Target IDs and names are also reported. - to be continued}
\label{SEDs:fig}
\end{figure*}

\addtocounter{figure}{-1}

\begin{figure*}[!t]
\resizebox{\textwidth}{!}{
\includegraphics{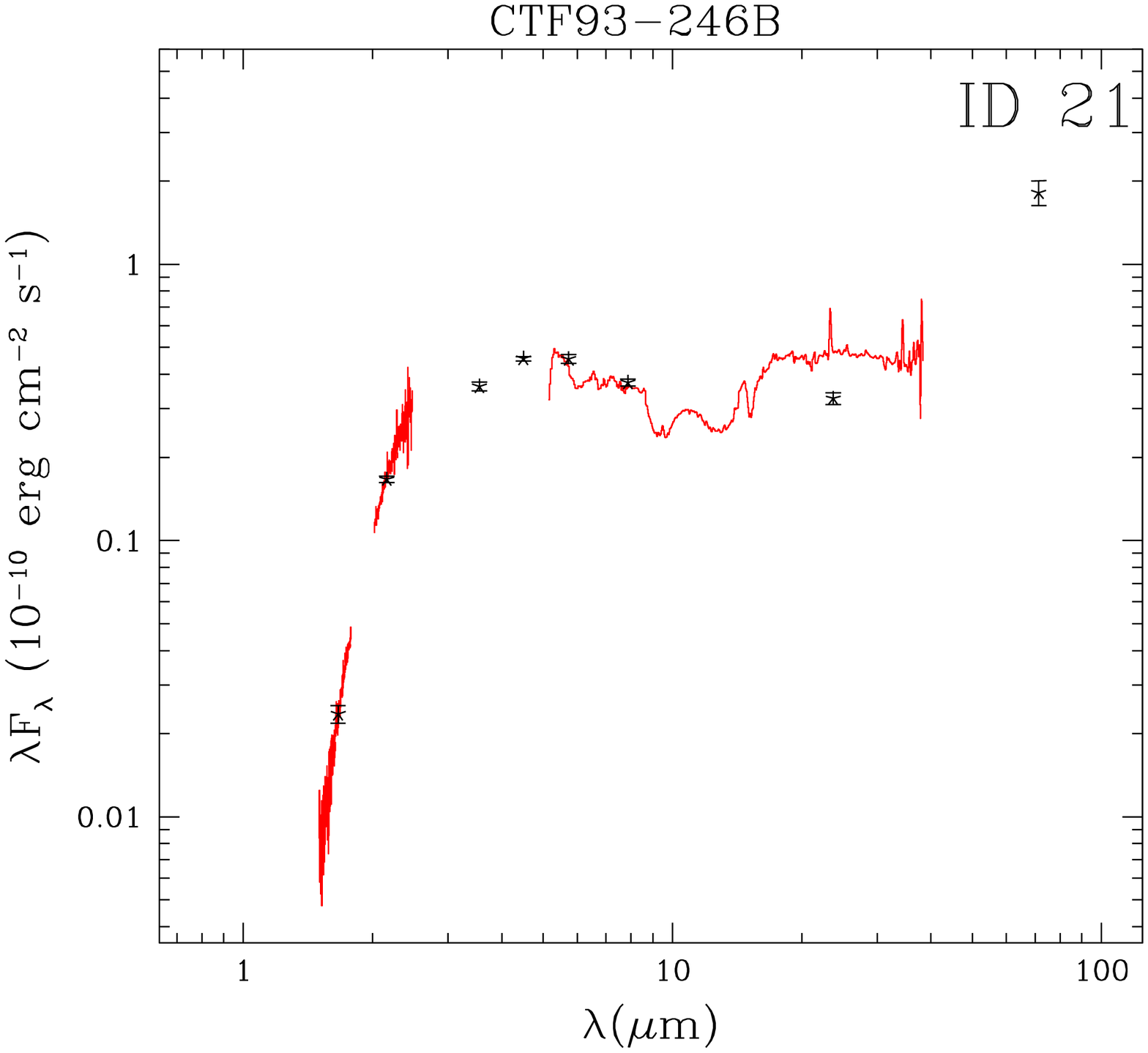} \includegraphics{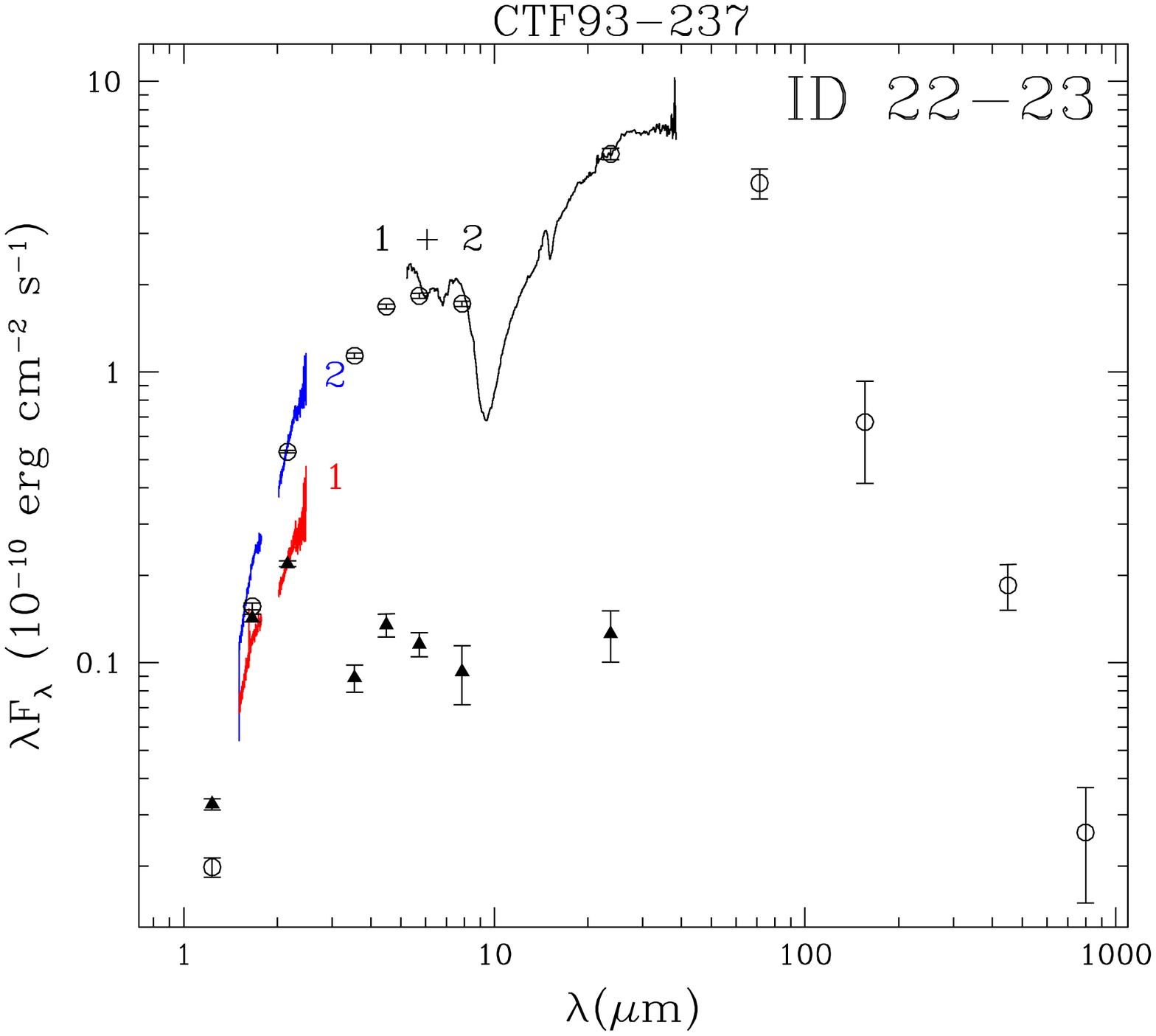} \includegraphics{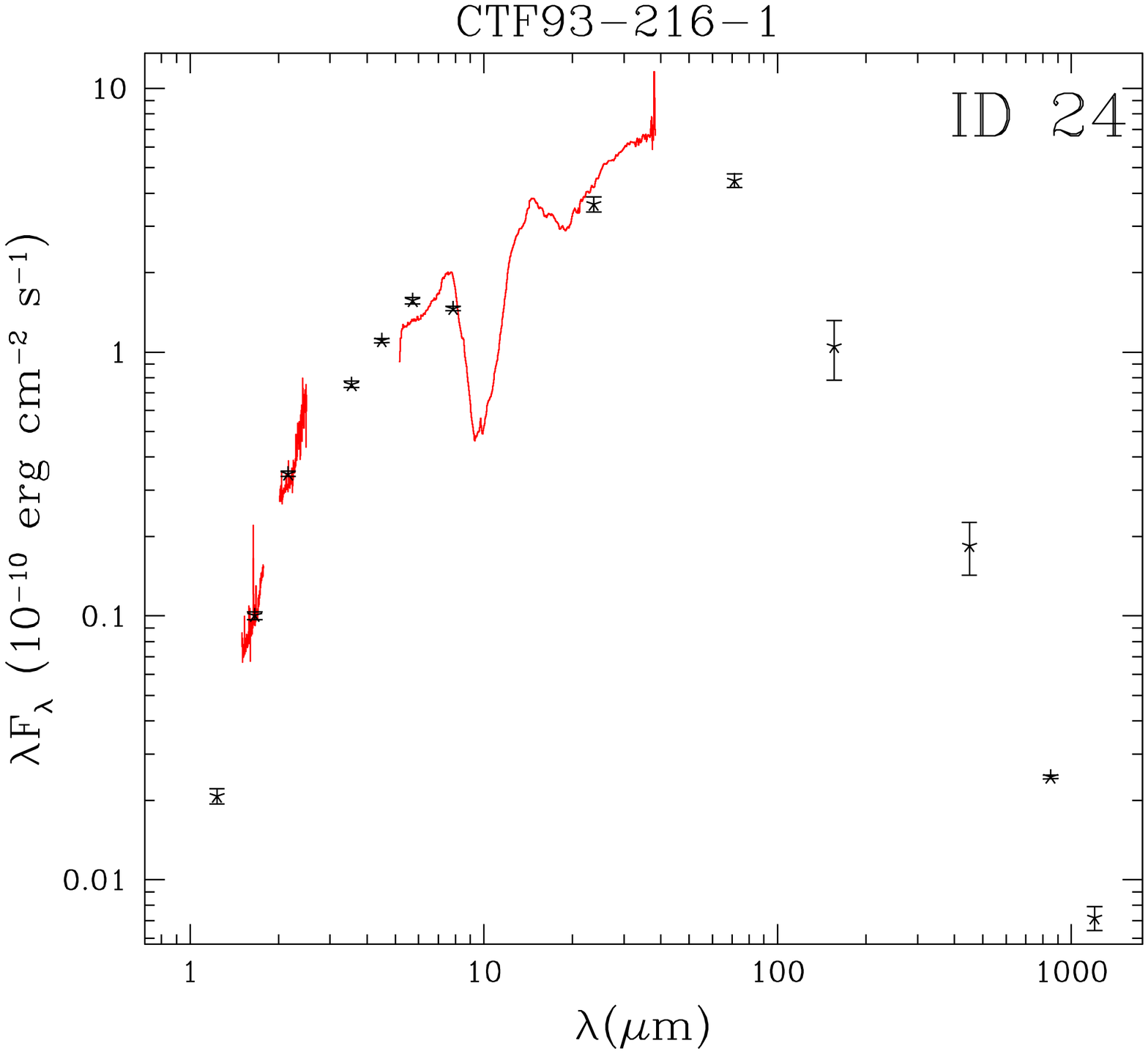} \includegraphics{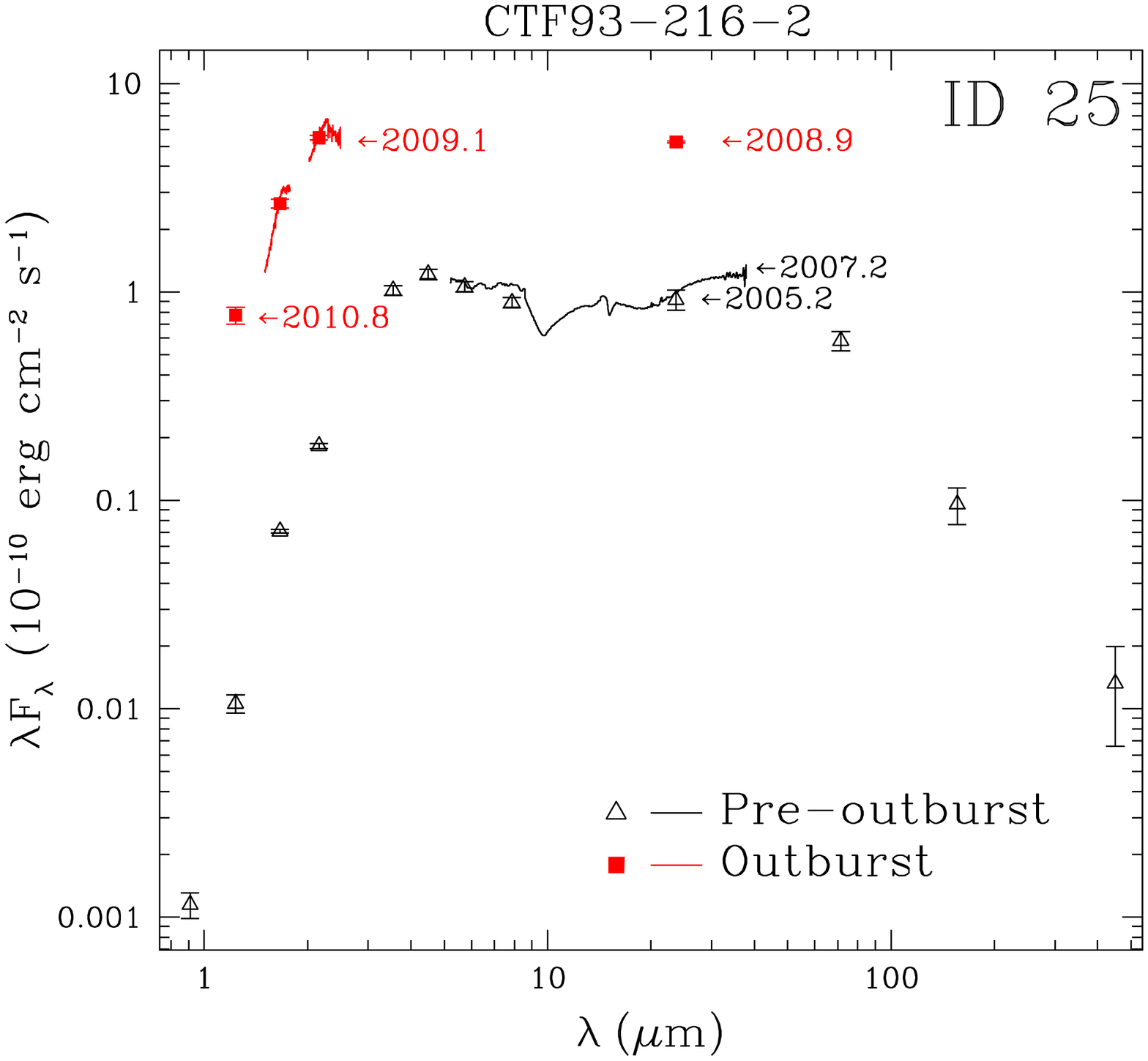}}
\resizebox{0.5\textwidth}{!}{
\includegraphics{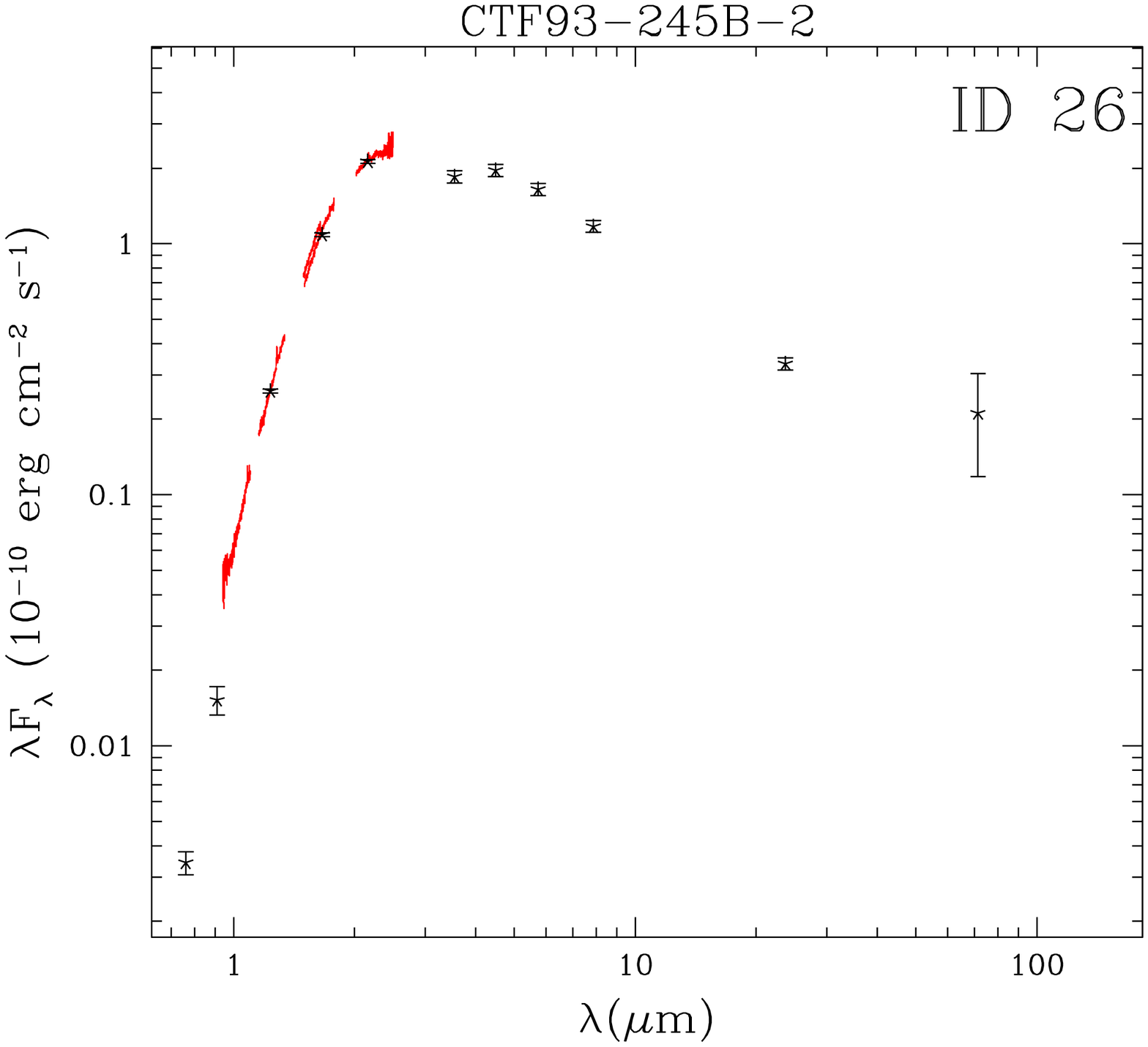} \includegraphics{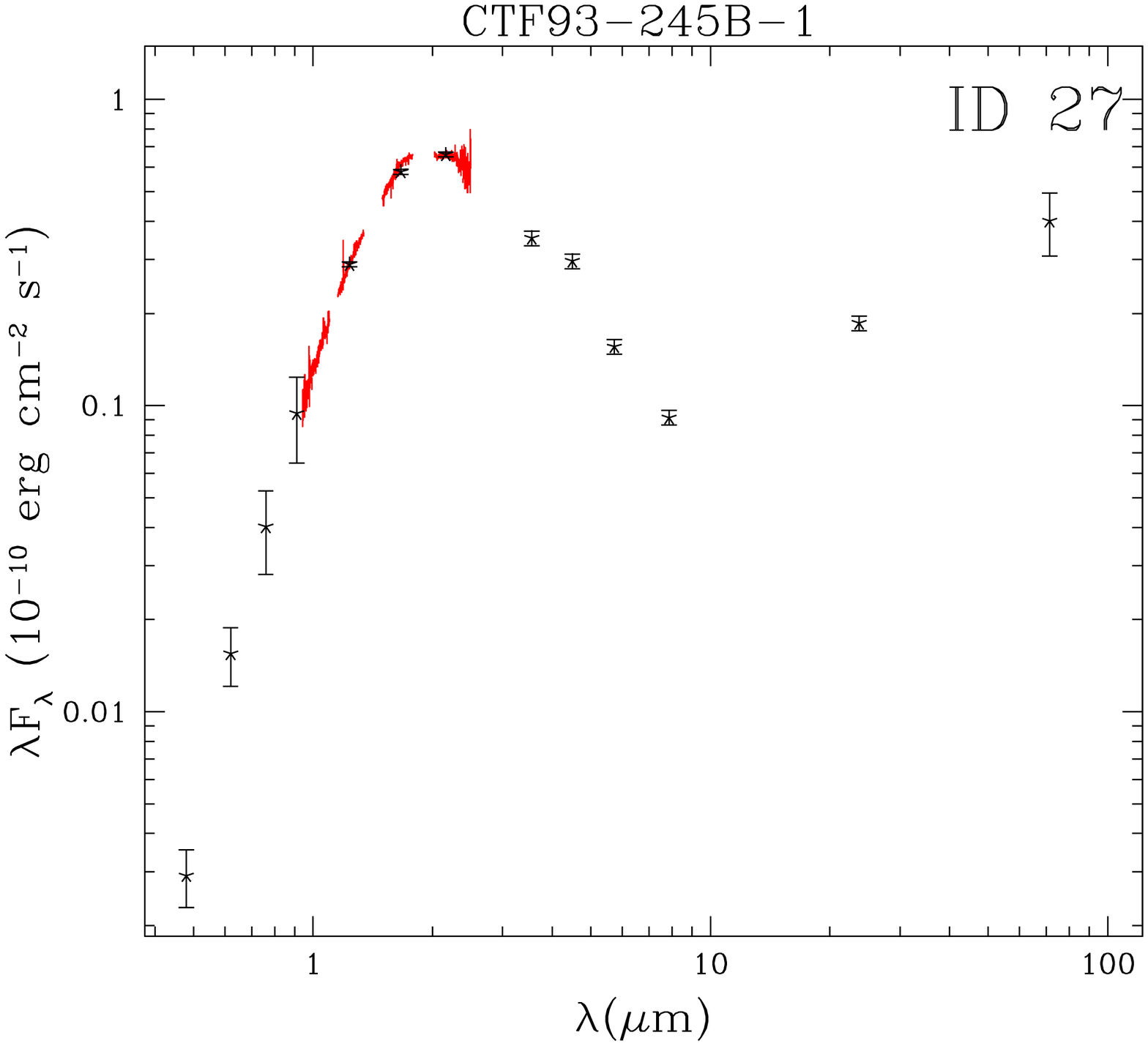}}

\caption{Target SEDs and spectra. Photometric data with error bars are in black, spectroscopic data (ESO-NTT and Spitzer) are in red. Target IDs and names are also reported. - continued}
\label{SEDsc:fig}
\end{figure*}


\subsubsection{SEDs and $L_{bol}$}
\label{phot:sec}

Both photometric and spectroscopic data of our sources are incorporated in the SEDs presented in Figure~\ref{SEDs:fig}, 
where both target IDs and names are also reported. 
In general, there is an excellent agreement between observed optical-IR spectra and photometric points, independently obtained from images.
However, there is a marginal discrepancy between the two datasets for a few objects, especially at short wavelengths (see, e.\,g., source \#\,6,
\#\,13, \#\,19), which is likely due to scattering effects, and/or intrinsic YSO variability. Mismatches between Spitzer spectra and photometry are also
observed in a few sources (see, e.\,g., \#\,5, \#\,10, \#\,14), and might be due to variability, as already observed in the literature~\citep[see, e.\,g.,][]{morales}. 
Special mention is needed of source \#\,25 (or \object{[CTF93]216-2}), which underwent a strong outburst between
2007 and 2008~\citep{caratti11}, increasing in brightness by $\sim$4.6, 4.0, 3.8, and 1.9\,mag in the $J$, $H$, $K_s$ bands and at 24\,$\mu$m.
The panel in Figure~\ref{SEDs:fig} shows both pre- and outburst SEDs. Photometric pre-outburst and outburst values are reported 
in Tab.~\ref{photometry:tab} and in \cite{caratti11}, respectively.

The YSO Class is determined from the SED slope ($\alpha$, or spectral index), 
which is calculated over all the available data points between 2.2 and 24\,$\mu$m.
The spectral index value is obtained from a linear fit to the logarithms, taking into account the uncertainties
in the flux measurements. Depending on $\alpha$, our sources were classified as Class\,I ($\alpha\geq0.3$),
Flat spectrum ($−0.3\leq\alpha\leq0.3$), or Class\,II ($−1.6\leq\alpha<−0.3$). 
$\alpha$ values of the sample range from 1.60 to -1.04, including 9 Class\,I, 11 Flat, and 7 Class\,II YSOs.
Notably, as already observed for other outbursting sources~\citep[see, e.\,g.,][and references therein]{kospal}, source \#\,25 changed its colours and its SED 
slope (from 0.69 to 0.25, pre- and outburst phase, respectively). Despite its actual $\alpha$ values, in this paper we consider it as a Class\,I source.

The source ID, their $\alpha$ values and corresponding classifications are reported in Columns 1, 2, and 3 of Table~\ref{parameters:tab}, respectively.
As already mentioned in Sect.~\ref{sample:sec}, the $\alpha$ values (and thus the YSO classification) listed in \cite{chen94} often differ
from ours, in some cases substantially as, e.\,g., for source \#9 (-0.45 vs 1.1, our and their value, respectively) or for 
source \#11 (-0.49 vs 1),
indicating that IRAS fluxes \textit{should not be} used for those objects located in crowded regions, as the resulting classification \textit{is not reliable}.

The total bolometric luminosity ($L_{bol}$) of each source was derived by integrating the observed SED (see Figure~\ref{SEDs:fig}), 
which includes data ranging from optical to mm wavelengths. The calculation was performed starting from
the first available SED data point (which varies from source to source, see also Table~\ref{photometry:tab}), using straight line interpolation
(in the $Log(\lambda)-Log(\lambda F_\lambda)$ plot) between all the SED points, also including additional points from the observed
spectra. A final correction at the longest wavelength was applied assuming that $F_\lambda$ decreases as  $\lambda^{-2}$.
A distance value of 450\,pc was assumed for the entire sample~\citep[see, e.\,g.,][]{allendavis}.
$L_{bol}$ inferred values for the sample range from $\sim$0.3 to 188\,$L_{\sun}$, and are listed in column 4 of Table~\ref{parameters:tab}.
For the outburst source \#25, both pre- and outburst $L_{bol}$ and $\alpha$ values are reported.

It worths noting that no correction for geometrical effects has been applied to the inferred bolometric luminosities, because system inclinations are not well defined.
On the other hand, SED models from \citet{whitney03-1} indicate that, due to geometrical effects, the observed bolometric luminosities of embedded objects 
may differ from their intrinsic luminosities up to a factor of two between pole-on and edge-on systems. 
Another source of uncertainty for $L_{bol}$ may be given by the foreground extinction, which cannot be correctly estimated, because the inferred $A_\mathrm{V}$
values of Sect.~\ref{reddening:sec} do not discriminate between foreground and circumstellar contribution. 
However, extinction maps can provide us with upper limits for the foreground extinction. From \citet{RF} maps, we infer an average $A_\mathrm{V}$ value of $\sim$6\,mag towards our
targets, which would modify $L_{bol}$ values between 5 and 10\%, because foreground extinction only affects the shortest wavelengths of the SED.
As a consequence, estimates of $L_{bol}$ and other stellar parameters (i.\,e., $L_{*}$, $M_{*}$, $R_{*}$, age, and $\dot{M}_{acc}$), 
that will be later derived from it, might also be affected 
by such uncertainties. 


\subsubsection{Accretion luminosity}
\label{acc:sec}

The accretion luminosity is the main indicator of YSO accretion activity.
Many different techniques exist in the literature to derive it~\citep[see, e.\,g.,][]{calvet00},
from the measurement of the U-band excess luminosity ~\citep{gullbring} to the luminosity
of optical and infrared emission lines~\citep[as, e.\,g., H$\alpha$, \ion{Ca}{ii}, Pa$\beta$, Br$\gamma$; e.\,g.,][]{muzerolle98,calvet04,natta04,natta06},
which are thought to be mainly produced in the magnetospheric accretion flow. The inferred relationships have been
successfully tested for a wide range of YSO masses, from substellar-mass to intermediate-mass YSOs~\citep{calvet04,natta04,garcia06},
but have been rarely tested simultaneously on the same YSO sample~\citep{muzerolle98,rigliaco}, because of the wide wavelength spread between
the considered lines and the intrinsic YSO variability, which require simultaneous coverage of the optical and infrared spectral 
wavelength range. 

In \textit{paper-1} we analysed various empirical line-$L_{acc}$ relationships from five different tracers, namely [\ion{O}{i}] at 6300\AA, 
H$\alpha$, \ion{Ca}{ii} at 8542\AA, Pa$\beta$, and Br$\gamma$, critically discussing the various determinations in the light of the source properties.
As a result, we showed that the Br$\gamma$ and Pa$\beta$ lines give the smallest dispersion of $L_{acc}$ over the entire range of $L_*$, whereas the other tracers, especially
the H$\alpha$ and [\ion{O}{i}] lines, provide much more scattered $L_{acc}$ results, not expected for the homogeneous sample of targets observed.

Here, we use our 0.6-2.5\,$\mu$m spectra to derive $L_{acc}$ simultaneously from the \ion{Ca}{ii}, Pa$\beta$, Br$\gamma$ 
line luminosities.

As in \textit{paper-1}, $L_{acc}$ from these lines is derived from the following empirical relationships~\citep{hercz,calvet00,calvet04}:

\begin{equation}
\label{Caii:eq}
Log(L_{acc}/L_{\sun}) = 1.02 \times Log(L_{CaII 8542}/L_{\sun}) + 2.5
\end{equation}

\begin{equation}
\label{PaB:eq}
Log(L_{acc}/L_{\sun}) = 1.03 \times Log(L_{Pa\beta}/L_{\sun}) + 2.80
\end{equation}

\begin{equation}
\label{BrG:eq}
Log(L_{acc}/L_{\sun}) = 0.90 \times Log(L_{Br\gamma}/L_{\sun}) + 2.90
\end{equation}

Line luminosities were obtained after de-reddening the observed fluxes by the adopted $A_\mathrm{V}$ (Table~\ref{av:tab}),
using the standard ~\citet{R&L} extinction law, and assuming a distance of 450\,pc.
Results are reported in Table~\ref{lacc:tab}, where source ID, $L_{acc}$(Br$\gamma$),
$L_{acc}$(Pa$\beta$), and $L_{acc}$(\ion{Ca}{ii}$_{0.854}$) are listed.
The derived values range from $\sim$0.1 to $\sim$60\,L$_{\sun}$. We assigned an upper limit to source \#\,23,
computed from the upper limit for Br$\gamma$ emission.

\begin{figure}
 \centering
\includegraphics [width=8.5cm] {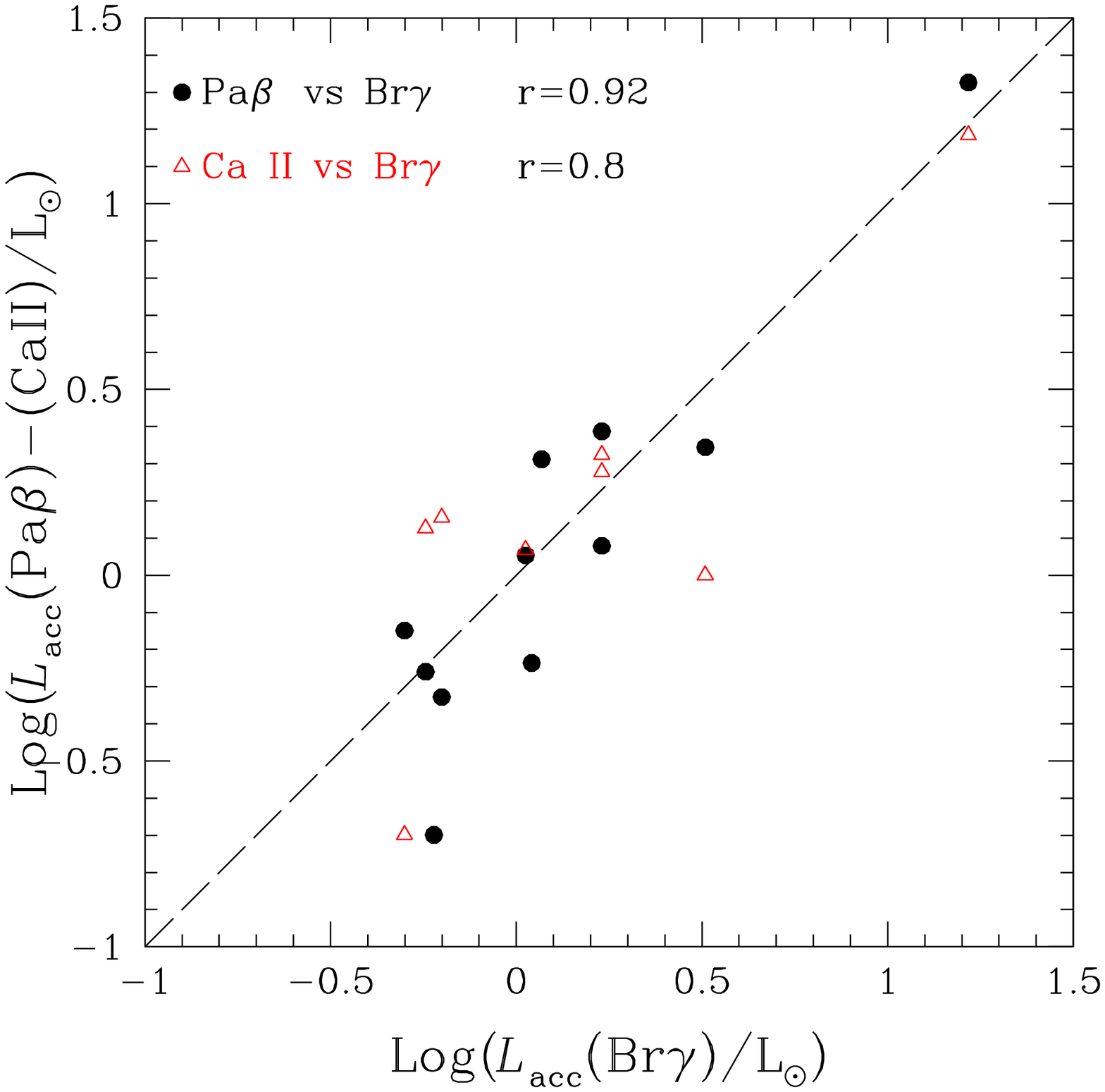}\\
  \caption{Comparison between the $L_{acc}$ values derived from different lines:
Pa$\beta$ vs Br$\gamma$ (black circles), and \ion{Ca}{ii} vs Br$\gamma$ (red open triangles).
Correlation coefficients are also reported.
\label{lacc:fig}}
\end{figure}

In Figure~\ref{lacc:fig} we compare the obtained results, plotting 
$Log (L_{acc}(Pa\beta))$ vs $Log (L_{acc}(Br\gamma))$ (black circles), and
$Log (L_{acc}(\ion{Ca}{ii}))$ vs $Log (L_{acc}(Br\gamma))$ (red triangles).
The dashed line indicates the equivalent locus.
In the first case, Pa$\beta$ vs Br$\gamma$, there is a very good agreement with a correlation coefficient
$r$ equal to 0.92, whereas the correlation of the second dataset (\ion{Ca}{ii} vs Br$\gamma$) gives a worst match with $r=0.8$,
as can also be noted from the larger scatter in the plot. Such a difference cannot be attributed to a wrong extinction
estimate, with the possible exception of source \#\,11 (see also Tab.~\ref{lacc:tab}), otherwise, due to differential extinction, 
we would observe the data points below the dotted line with the circles positioned above the triangles (overestimated extinction),
or the opposite (underestimated extinction).
It is worth noting that our low resolution spectroscopy does not allow us to de-blend the \ion{Ca}{ii} (0.854\,$\mu$m) and the
Pa\,15 (0.855\,$\mu$m) lines, thus the observed \ion{Ca}{ii} 
flux at 0.854\,$\mu$m might be overestimated up to $\sim$20\%.
As a consequence, we prefer to discard the values from the \ion{Ca}{ii} lines.
For each source we adopt an average $L_{acc}$ value, obtained by averaging the accretion luminosities inferred from the Br$\gamma$
and Pa$\beta$ lines. Results are reported in column 7 of Table~\ref{parameters:tab}.


\begin{table}
\begin{minipage}[t]{\columnwidth}
\caption{ Accretion luminosity estimates from Br$\gamma$, Pa$\beta$, and {\ion{Ca}{ii}}\,0.854\,$\mu$m lines.}
\label{lacc:tab}
\centering
\renewcommand{\footnoterule}{}
\begin{tabular}{cccc}
\hline \hline
Source  &  $L_{acc}$(Br$\gamma$)& $L_{acc}$(Pa$\beta$) & $L_{acc}$({\ion{Ca}{ii}}\,$_{0.854}$) \\
ID      &  (L$_{\sun}$)      &  (L$_{\sun}$)      & (L$_{\sun}$)    \\
\hline\\[-5pt]
1       &  0.6   	   &  0.2	      & $\cdots$      \\
2       &  1.7  	   &  1.2	      & 1.9	      \\
3       &  16.5 	   &  21.2 	      & 15.3	      \\
4       &  0.5   	   &  0.71	      & 0.2	      \\
5       &  1.06 	   &  1.13	      & 1.17	      \\
6       &  0.5    	   &  $\cdots$        & $\cdots$      \\
7       &  1.2  	   &  $\cdots$        & $\cdots$      \\
8       &  61.3 	   &  $\cdots$        & $\cdots$      \\
9       &  1.17 	   &  2.05	      & $\cdots$      \\
10      &  0.57 	   &  0.55	      & 1.34	      \\
11      &  3.23 	   &  2.21	      & 1.00	      \\
12      &  0.15 	   &  $\cdots$        & $\cdots$      \\
13      &  0.3  	   &  $\cdots$        & $\cdots$      \\
14      &  3.1  	   &  $\cdots$        & $\cdots$      \\
15      &  0.63 	   &  0.47	      & 1.43	      \\
16      &  1.0  	   &  $\cdots$        & $\cdots$      \\
17      &  1.78 	   &  $\cdots$        & $\cdots$      \\
18      &  1.70 	   &  2.44	      & 2.11	      \\
19      &  0.31 	   &  $\cdots$        & $\cdots$      \\
20      &  2.5  	   &  $\cdots$        & $\cdots$      \\
21      &  0.8  	   &  $\cdots$        & $\cdots$      \\
22      &  $<$0.06 	   &  $\cdots$        & $\cdots$      \\
23      &  2.8  	   &  $\cdots$        & $\cdots$      \\
24      &  2.1   	   &  $\cdots$        & $\cdots$      \\
25      &  0.31 	   &  $\cdots$        & $\cdots$      \\
26      &  1.1   	   &  0.58            & $\cdots$      \\
27    	&  0.18 	   &  $\cdots$        & $\cdots$      \\
\hline
\end{tabular}
\end{minipage}
\end{table}

\subsubsection{Stellar spectral types}
\label{SpT:sec}

We obtain the spectral types (SpTs) of our targets by combining various methods.
The derived SpTs along with the method used and the effective temperatures ($T_{\rm eff}$) are reported in Table~\ref{parameters:tab} (Columns 6 and 7, respectively). 
The adopted $T_{\rm eff}$ is converted from the spectral type adopting the relations given by \citet{Kenyon95} (for SpTs earlier than M0), and by \citet{Luhman03} 
(for M0 or later SpTs). 

For the 14 sources with optical spectra (see Sect.~\ref{sofoscspec:sec} and Tab.~\ref{obs_spec:tab}) we employ the spectral classification code \emph{SPTclass} developed 
by \citet{hernandez}\footnote{http://www.astro.lsa.umich.edu/~hernandj/SPTclass/sptclass.html}. 
This code uses empirical relationships between the equivalent widths (EWs) of many atomic/molecular absorption/emission lines and $T_{\rm eff}$. 
As a result, this automated 
code can classify optical spectra to a precision of about one subtype. Additionally,
spectral classifications for many of these optical sources (9 out of 14) already exist in the literature~\citep{strom,fang,connelley}. There 
is a good match ($\sim$1 subtype) with our results, except for the possible FU-Ori object \object{[CTF93]50} (our source \#2, see also Sect.~\ref{OptIRspec:sec}), 
which shows a later spectral type (K6$\pm$1) with respect to previous measurements from \cite{strom} (G5-G7, source KMS\,31) and \cite{fang} (K3-4.5, source \#105 in their work). 
Indeed these differences are not surprising, because FU-Ori sources may show large spectral type variabilities~\citep[see e.\,g.][]{hartmann96}.

On the other hand, the thirteen remaining NIR spectra (i.\,e. sources \#\,6, 12, 13, 14, 17, 19, 20, 21, 22, 23, 24, 25, 27)
show a steeply rising flat continuum that is almost featureless or with absorption water bands (in the $J$, $H$ and $K$ bands, 
between 1.3-1.5\,$\mu$m, and 1.7-2.1\,$\mu$m). 
These bands arise at $T_{eff}\le$3800\,K (M0 or later spectral types), and thus are typical of M type objects.
Thus, these thirteen spectral types have been identified following the prescription given by \citet{gorlova}, i.\,e. dividing
the spectra in two groups, which either show or do not the absorption water bands in the NIR. 

H$_2$O broad band features are clearly detected in eight out of thirteen spectra (namely \#\,12, 13, 19, 21, 22, 23, 24, 25).
The depth of these bands increases with decreasing temperatures, thus the values of their EWs, once corrected for the proper $A_\mathrm{V}$, 
can be related to the spectral subtype \citep[see, e.\,g.,][]{Kleinmann,greene,Luhman03,gorlova}. Usually the 1.3-1.5\,$\mu$m absorption feature is detected for 
$A_V<$20\,mag, and the EW of the 1.7-2.1\,$\mu$m feature decreases by about a factor of 2, as the visual extinction increases from 0 to 30\,mag
($A_V$=25\,mag is the maximum extinction value in our sample, see also Tab.~\ref{parameters:tab}, column 5).
Along with the absorption water bands, source \#12 (\object{Meag31}) also shows VO bands in absorption at $\sim$1.15\,$\mu$m~\citep[typical of
M6 or later SpTs; see, e.\,g.,][]{cushing}. On the other hand the lack of strong narrow band features, mostly due to the veiling and the low resolution of our spectra, 
prevents us from an accurate spectral classification. For these objects, the uncertainties on the SpT range from one to two subtypes, 
i.\,e. up to $\sim$200-300\,K.

The five remaining spectra (namely sources \#\,6, 14, 17, 20, 27) are featureless 
with steeply rising SEDs, and without any sign of overturn in the $H$ or $K$ band, 
which means either SpT later than M0, or SpT earlier or equal to M0 and $A_\mathrm{V}$$>$30-40\,mag \citep[see, e.\,g.,][]{greene,gorlova}. 
Our $A_\mathrm{V}$ estimates exclude the latter hypothesis, thus we safely classify them as K spectral types. 
One of them (\object{[CTF93]245B-1}, i.e. source \#27) was already classified as K6 by \citet{strom}, thus we keep this classification.
The lack of features in the remaining four sources precludes any detailed spectral analysis.
Therefore, we use both our spectral and photometric data (see also Fig.~\ref{SEDs:fig}) 
to model the four YSOs with the SED fitting tool from \citet{robitailleFT} and provide some tighter constraints on the underlying stellar objects.
Indeed, the SED fitting tool alone cannot confidently predict the stellar temperature in embedded sources. 
Even knowing the $L_*$ value, there is still degeneracy between $T_*$ and $R_*$ values. This degeneracy can be partially removed, given $L_*$, $L_{bol}$, extinction, and distance.
In fact a substantial increase (decrease) of $T_*$ (500\,K or more) produces a decrease (increase) of $R_*$, generating an older (younger) YSO and substantially modifying the SED shape. 

\begin{figure}
 \centering
\includegraphics [width=8.5cm] {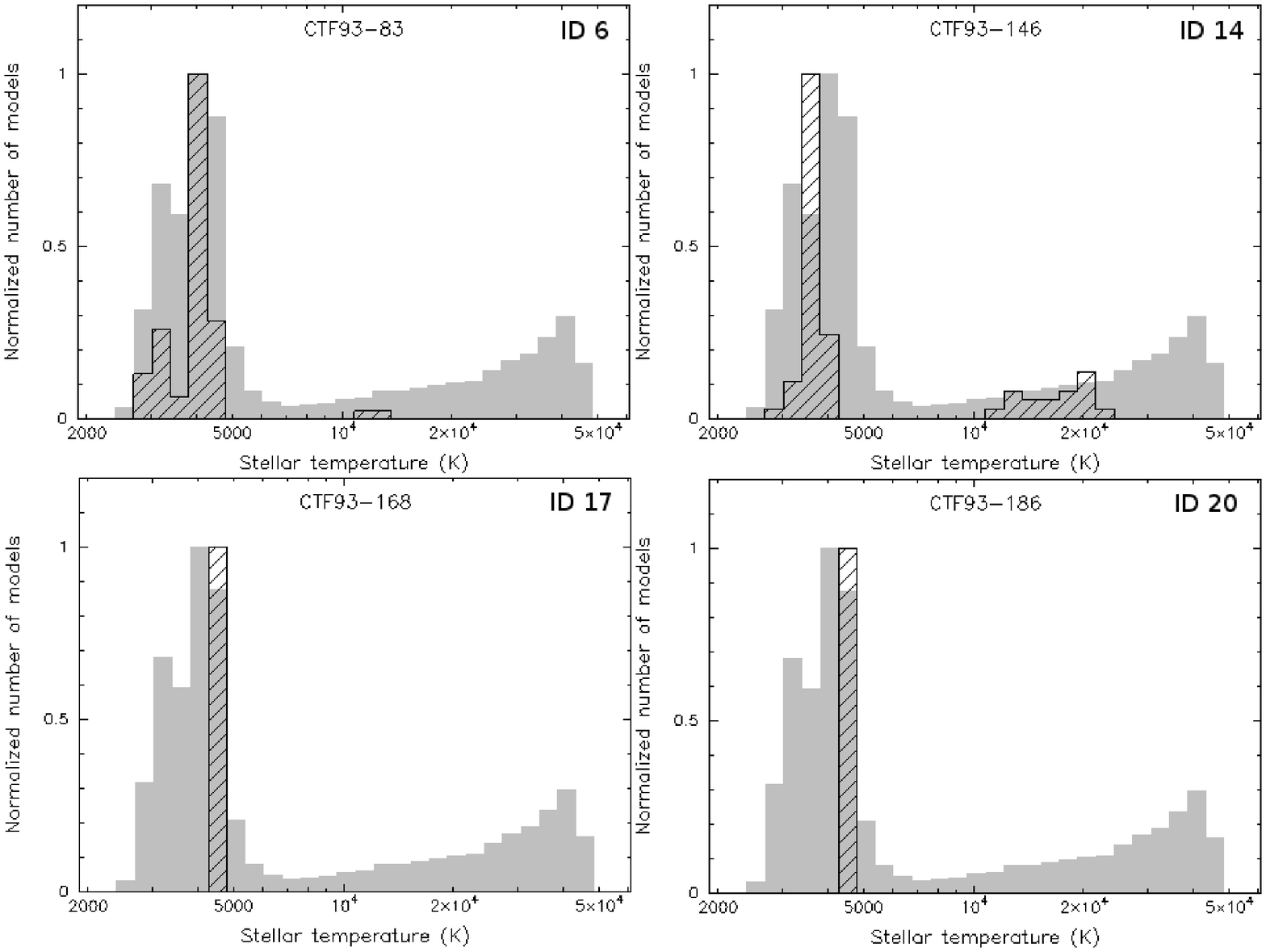}
  \caption{Temperatures inferred from SED fitting. For each YSO the stellar temperature distribution (hashed histogram), normalised to its maximum,
is plotted over the entire grid of models (in grey).
\label{temp:fig}}
\end{figure}

Thus, as further constraints for the fitting tool, distance, extinction, and bolometric luminosity of each source were used (as listed in Tab.~\ref{parameters:tab}). An additional 
constraint is given by the YSO spectral type (K), and thus by the temperature range. 
Then, for each object, we obtained a grid of possible models, listed as a function of their $\chi^2$ values. We then select the best 50 models for each YSO, plot the output 
stellar temperature distribution, and infer the most likely temperature. The distribution spread gives us a rough estimate of the error.
Our results are reported in Fig.~\ref{temp:fig}, where, for each object, the stellar-temperature distribution (hashed histogram), normalised to its maximum,
is plotted over the entire grid of models (in grey).

The derived spectral types for the entire sample are reported in Table~\ref{parameters:tab} (column 6), and range from B7 to M7.5.
Most of the objects ($\sim$74\%) are low-mass YSOs with SpT between K5 and M7.5, four ($\sim$15\%) are intermediate mass YSOs (F7 to K4),
and three ($\sim$11\%) are Herbig Ae/Be stars (A5 to B7).

\subsubsection{H-R diagram and stellar parameters}
\label{Sparam:sec}

\begin{figure}
 \centering
\includegraphics [width=8.5cm] {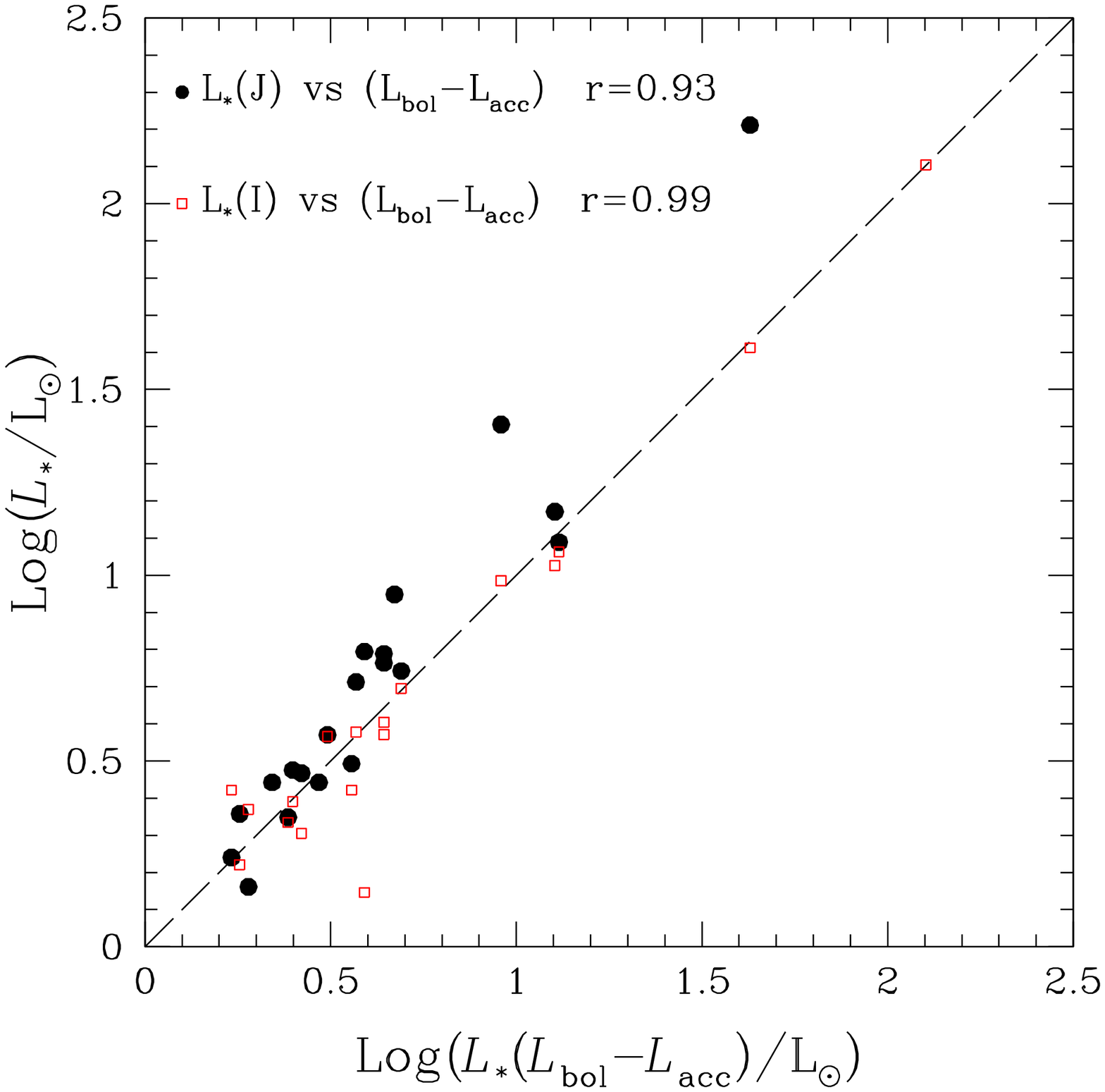}
  \caption{$L_*$ values derived from the $I_J$ and $J$ band photometry (red triangles and black dots, respectively) versus
$L_*$ values obtained from $L_* = L_{bol}-L_{acc}$. The dashed line marks the equivalue locus.
\label{Lstar:fig}}
\end{figure}

\begin{figure}
 \centering
\includegraphics [width=8.5cm] {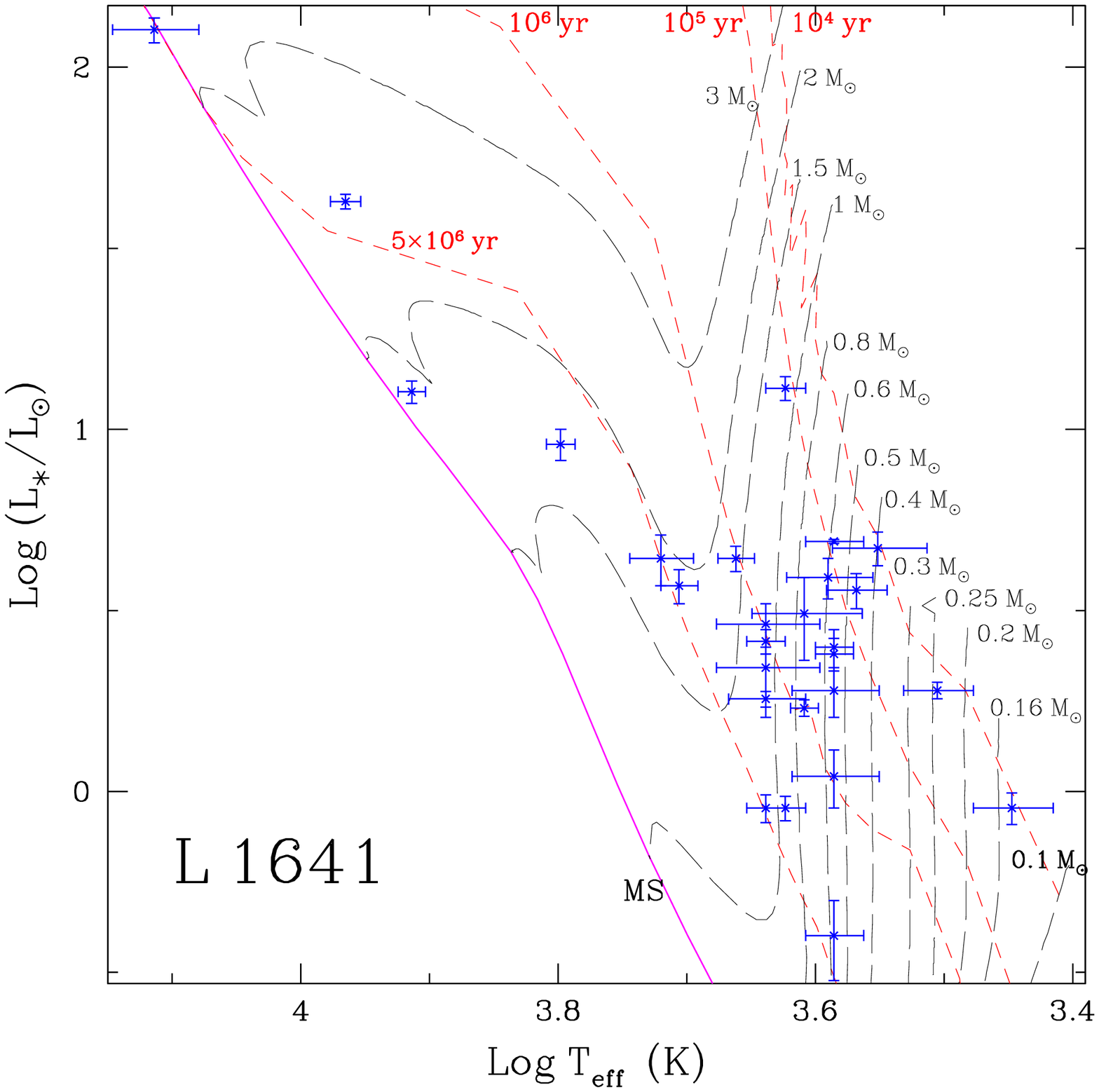}\\
  \caption{HR diagram for the whole sample. The evolutionary models of \citet{siess} are
shown with isochrones at 10$^4$, 10$^5$, 10$^6$, and 5$\times$10$^6$\,yr (red dashed lines),
and mass tracks from 0.1 to 3\,M$_{\sun}$ (black dashed lines).
\label{HR:fig}}
\end{figure}

\begin{figure}
 \centering
\includegraphics [width=8.5cm] {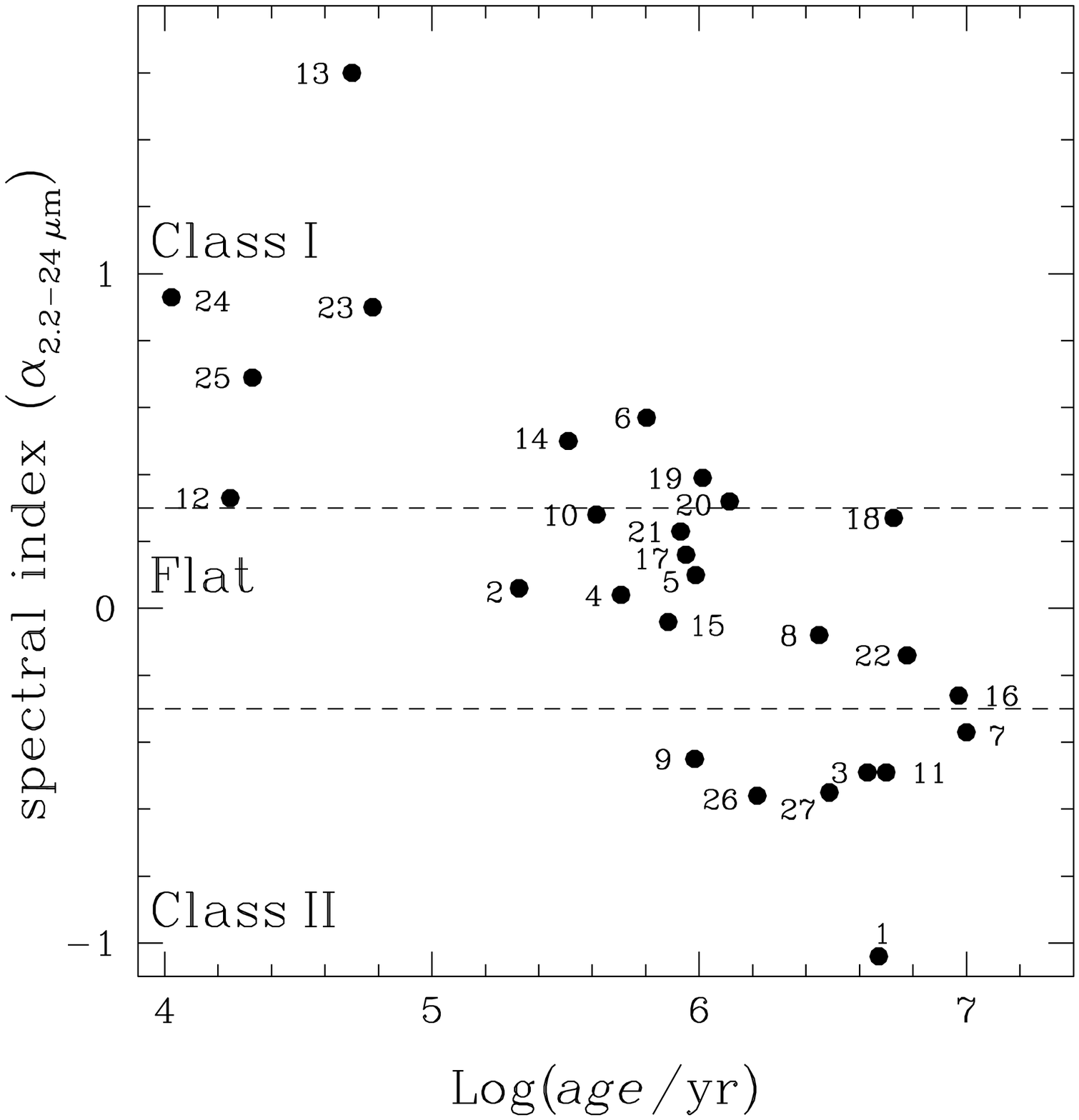}\\
  \caption{Spectral index ($\alpha$) versus age. 
The two dashed lines delineate the Class\,I-Flat ($\alpha$ = 0.3) and Flat-Class\,II ($\alpha$ = -0.3) boundaries.
The object IDs are also reported.
\label{alfa_age:fig}}
\end{figure}

To characterise our YSOs it is necessary to derive the stellar parameters ($L_*$, $M_*$, $R_*$).
$L_*$ of each source is inferred from $L_* = L_{bol}-L_{acc}$, i.\,e. assuming
that the observed $L_{bol}$ is simply the sum of accretion and stellar luminosities.
The resulting $L_*$ values are reported in Table~\ref{parameters:tab} (column 8).
Assuming that the YSO SED is well modelled by our data, i.\,e. that the derived $L_{bol}$ has a relatively
small uncertainty, the error on $L_*$ is due to the uncertainty of the $L_{acc}$ estimates.

To test the consistency of these results we derive $L_*$ from the stellar bolometric magnitude~\citep[see, e.\,g.,][]{gorlova}, obtained
from the dereddened $I_J$ and $J$ magnitudes (Table~\ref{photometry:tab}, Columns 5 and 6, respectively) 
for 22 out of 27 and 26 out of 27 objects of the sample:

\begin{equation}
\label{LI:eq}
Log(L_{*}/L_{\sun}) = 1.86 - 0.4(I - 0.482 A_\mathrm{V} + BC_I - 5 log(450) + 5)
\end{equation}

\begin{equation}
\label{LJ:eq}
Log(L_{*}/L_{\sun}) = 1.86 - 0.4(J - 0.265 A_\mathrm{V} + BC_J - 5 log(450) + 5)
\end{equation}

where the bolometric correction in $I_J$ and $J$ ($BC_I$ and $BC_J$, respectively),
for each SpT of Table~\ref{parameters:tab}, are inferred from the dwarf colours in \citet{Kenyon95}.

Figure~\ref{Lstar:fig} compares in logarithmic scale $L_*$ values obtained from $I_J$ and $J$ magnitudes (red triangles and black dots, respectively) against $L_* = L_{bol}-L_{acc}$ 
values of our Table~\ref{parameters:tab}. The dashed line marks the equivalue locus.
Both samples show some scattering around the  equivalue locus, but, as expected, values derived from the $J$ band 
have a larger scatter and tend to be located above the dashed line. This is because eq.~\ref{LJ:eq} does not take into account 
the infrared excess in the $J$ band, which is relevant in young stellar objects~\citep[see e.\,g.][]{cieza}, thus $L_*$ is overestimated.
On the other hand, the $I_J$ band should be much less affected by the infrared excess, and thus the derived $L_*$ value should be more
reliable, if the extinction has been correctly estimated.

Our $L_*$ and $T_{eff}$ estimates can be plotted on a HR diagram to infer $M_*$, $R_*$, and age of each
sampled source. We adopt the evolutionary tracks from \citet{siess}, 
with a metallicity of $Z=0.02$, and we use the on-line tool\footnote{http://www-astro.ulb.ac.be/~siess/database.html} 
to infer these results and produce the HR diagram. The HR diagram for the sample is shown in Figure~\ref{HR:fig}, and the 
inferred $M_*$, $R_*$, and age values are reported in Table~\ref{parameters:tab} (Columns 9, 10 , and 11, respectively). 
For self-consistency, isochrones and ages down to $\sim$10$^4$\,yr are reported in both Figure~\ref{HR:fig} and 
Column\,11 of Table~\ref{parameters:tab}, however, we stress that age estimates below $\sim$10$^5$\,yr are \textit{not reliable}.
Thus, the reader should consider these values more like a qualitative indication of stellar youth, rather than a real estimate.
Moreover, it is worth noting that adopting different sets of evolutionary tracks would provide different values for the derived stellar 
masses and ages, up to a factor of 2-3~\citep[see, e.\,g.,][]{hillen,spezzi,fang}, where the largest discrepancies are in the age estimates of 
objects with age $\le$1\,Myr. For example, for low-mass objects, the morphology of \citet{siess} tracks is quite similar to those of \citet{baraffe}, 
thus there is good agreement between ages, masses and radii given by the two models ($\sim$20-40\,\%). On the other hand, for example, tracks from \citet{dant} give, on 
average for our sample, lower mass and younger age estimates than \citet{siess}. As a consequence, these quantities and those later inferred, like, e.\,g., the mass accretion rate in 
Sect.~\ref{macc:sec}, can differ up to a factor of 2-3, depending on the model. 

According to the \citet{siess} model, the sampled stellar masses range from 0.4 to 3.4\,M$_{\sun}$, with an average and median 
values of $\sim$1.1 and 0.9\,M$_{\sun}$, respectively. 
The average and median age of the sample is about 2 and 1\,Myr, respectively, with values ranging from $\le$10$^5$ to 10$^7$\,yr.
In Figure~\ref{alfa_age:fig} the measured $\alpha$ values, derived in Sect.~\ref{phot:sec}, are plotted against stellar ages, indicating that, on average,
$\alpha$ decreases with stellar age. However, there is no straightforward correlation between spectral index and age.
Six out of nine Class\,I YSOs have an age of $\sim$10$^5$, while the remaining three (namely \#\,6, 19, and 20) appear to be older (5$\times$10$^5$ to 10$^6$\,yr).
The estimated age for the flat spectrum sources ranges from  $\sim$10$^5$ to $\sim$10$^7$\,yr, with an average value of 2\,Myr,
whereas the age of Class\,II sources spans from $\sim$10$^6$ to $\sim$10$^7$\,yr, with an average value of 4\,Myr. 
This indicates that the standard SED classification ($\alpha$ computed between 2.2 and 24\,$\mu$m) might not properly reflect YSO age, 
i.\,e. that the SED slope might not be a good indicator of the stellar age.
Indeed, geometrical effects may play a role in modelling the YSO SED~\citep[see, e.\,g.,][]{whitney03-1,robitailleFT}, altering the slope and generating misclassifications. 
This is particularly clear when the SEDs show a double peak (at optical/NIR and MIR wavelengths), often indicative of edge-on discs or transitional 
discs~\citep[see, e.\,g.,][see also Fig.~\ref{SEDs:fig}]{cieza07,merin,williams}.

Finally, we also note that two massive sources, namely \#7 and \#16, are $\sim$10$^7$\,yr old. These are the only objects showing strong \ion{H}{i}
absorption lines in the optical/NIR sampled spectra (see Sect.~\ref{OptIRspec:sec} and Fig.~\ref{spec1:fig} and \ref{spec2:fig}).
Since they seem to be older than L\,1641~\citep[$\sim$5$\times$10$^6$\,yr; e\,g.,][]{allendavis}, this might indicate that they are not part 
of this star forming region.

\subsubsection{Mass accretion rates}
\label{macc:sec}

Mass accretion rates ($\dot{M}_{acc}$) for the whole sample can be derived once the accretion luminosities and the stellar parameters have
been inferred. 
Since the accretion luminosity corresponds to the energy released by the accreting matter onto the YSO, assuming that the 
free fall starts at the co-rotational radius ($R_{in}$), i.\,e. at $\sim$5\,$R_*$~\citep{gullbring}, $L_{acc}$ is then: 

\begin{equation}
\label{Macc1:eq}
L_{acc} \sim G M_* \dot{M}_{acc} (1 - R_*/R_{in}) / R_* 
\end{equation}

and thus $\dot{M}_{acc}$ is given by:

\begin{equation}
\label{Macc2:eq}
\dot{M}_{acc} = L_{acc} * 1.25 R_* / G M_*
\end{equation}

The derived $\dot{M}_{acc}$ values are reported in Table~\ref{parameters:tab} (Column\,14). 
These values span four orders of magnitude, ranging from 3.6$\times$10$^{-9}$ to 1.2$\times$10$^{-5}$\,M$_{\sun}$\,yr$^{-1}$,
with the highest $\dot{M}_{acc}$ given by the low-mass outbursting source \#\,25~\citep[\object{[CTF93]216-2};][]{caratti11},
and the lowest value (source \#\,23) being an upper limit.
Error estimates are particularly difficult, because they depend on both observational and theoretical uncertainties, namely on $L_{acc}$,
$R_*$, and $M_*$. Therefore an error bar up to one order of magnitude can be expected.

Mass accretion estimates for five targets of the sample have been previously reported in literature (see also notes
in Table~\ref{parameters:tab}). \citet{fang} characterised four sources, namely \#\,1, \#\,2, \#\,4, and \#\,18, by means of optical spectroscopy.
Estimates for two sources (\#\,1 and \#\,18) are in good agreement with our results, whereas the others (sources \#\,2, and \#\,4), differ of about one 
order of magnitude, possibly because of the different $A_\mathrm{V}$ and $L_{bol}$ estimates. On the other hand, source \#\,2 has also been analysed by \citet{fischer}, 
who modelled photometric and spectroscopic data (\textit{Spitzer} and \textit{Herschel}), infer an $\dot{M}_{acc}$ value very similar to ours.
Finally, \citet{hillenbrand} derive source \#\,3 (e.\,g. \object{V* V380 Ori}) stellar properties from optical and IRAS photometry, inferring a mass accretion rate of $\sim$ 
one order of magnitude higher than ours, likely because they obtain larger stellar mass and luminosity.

\subsubsection{Jet detection and mass ejection rates}
\label{mout:sec}

Accretion and ejection activity are complementary and interconnected processes of the stellar birth. In particular,
part of the accreting material is ejected by the YSO in the form of a bipolar jet, which, in turn, sweeps up the circumstellar and interstellar medium,
producing a molecular outflow~\citep[see, e.\,g.,][]{reipurth01}. Strong accretion is usually accompanied by strong ejection processes,
thus the detection and study of jets/outflows give us insights into YSO activity.
The YSO outflow-activity is evinced from typical jet-tracing lines in the spectra (see also Sect.~\ref{OptIRspec:sec} and Tab.~\ref{lines2:tab}),
narrow-band imaging centred on these lines, and/or CO maps tracing wider outflows.

With these ideas in mind we inspected our \textit{Spitzer} images in search of possible signatures of jets from the sampled sources. 
Indeed, the \textit{IRAC} bands contain both molecular hydrogen and ionic lines, which may be shock-excited in
protostellar outflows. In particular, band 2 (4.5\,$\mu$m) contains bright molecular hydrogen lines and can be used to detect 
shock spots (knots)~\citep[see, e.\,g.,][]{peterson}. 
To better identify these features in the \textit{IRAC/Spitzer} mosaics, we thus constructed \textit{IRAC} three-colour
images, using 3.6\,$\mu$m, 4.5\,$\mu$m, and 8\,$\mu$m (i.\,e. bands 1, 2, and 4, in blue green, and red,
respectively), and identified the knots by means of their colours and morphologies. 
We detect jets from eleven sources (namely sources \#\,2, 3, 5, 6, 8, 10, 11, 15, 18, 24, 25), nine of them were already observed by \citet{davis09} 
in their H$_2$ survey of L\,1641, whereas sources \#\,24, and \#\,25 were not covered by that survey. A faint H$_2$ emission was discovered by \citet{connelley07} around \#\,24.
The two jets from \#\,24, and \#\,25 detected in the \textit{IRAC/Spitzer} mosaics are described and shown in Appendix~\ref{appendixB:sec}.
\textit{Spitzer} images usually reveal bright jets emitting in the mid-IR, but, can fail to detect faint jets, only visible in narrow-band images.
Therefore, we also searched the literature for indications of further jets/outflows from our YSOs.
In Column\,16 of Table~\ref{parameters:tab} we indicate the presence of jets and outflows driven by our sources, 
as reported by \citet[]{allendavis,davis09,connelley07} (jets), by \citet[]{morgan,dent} (CO outflows), and by this paper, whereas in Column\,17 we 
indicate the presence of jet-tracing lines in our optical/NIR spectra.

Additionally, to compare accretion and ejection activity in a statistical way, 
we also infer a crude estimate of the mass ejection rates from the different lines tracing jets 
(namely [\ion{O}{i}], [\ion{S}{ii}], [\ion{Fe}{ii}], and  H$_2$ lines reported in Tab.~\ref{lines2:tab}), 
and observed in our optical/NIR spectra (see Sect.~\ref{OptIRspec:sec}). 
As these lines are optically thin, their luminosity gives us an estimate of the total mass ($M$). Thus the
mass ejection rate ($\dot{M}_{out}$) can be inferred if the tangential velocity ($v_{\rm t}$) and jet extension ($l_{\rm t}$) are known,
i.\,e. $\dot{M}_{out}= M v_{\rm t} / l_{\rm t}$. 
Indeed our low resolution spectra do not allow us to measure the radial velocity, neither do
we know the inclination angle of the jets, thus we assume an average $v_{\rm t}$~\citep[150\,km\,s$^{-1}$ and 50\,km\,s$^{-1}$, for the
atomic and H$_2$ lines, respectively; see, e.\,g.,][]{hartigan95,wh04,caratti09}. 
Moreover, the jet size is assumed equal to the measured seeing, projected to the L\,1641 distance. 
This assumption rests on the fact that the aperture-extraction width in our spectra is defined by the seeing 
limited width of the stellar continuum, thus the jet is not spatially resolved, and on the fact that the spectroscopic
absolute flux calibrations were done using the photometry.

For the atomic species we use the following relationship, $\dot{M}_{out} = \mu m_{\rm H} (N_{\rm H} V) v_{\rm t} / l_{\rm t}$~\citep[e.\,g.][]{nisini2005,caratti09}, with
$N_{\rm H} V = L_{\rm X} (h \nu A_{\rm i} f_{\rm i} \frac{\rm X^{i}}{\rm X} \frac{\rm [X]}{\rm [H]})^{-1}$, where $L_{\rm X}$ is the luminosity of the element X, 
for the selected transition, $A_{\rm i}$ and $f_{\rm i}$ are the radiative rate and the fractional population of the upper level of the transition, 
$\frac{\rm X^{i}}{X}$ is the ionisation fraction of the considered species with a total abundance of $\rm\frac{[X]}{[H]}$ with respect to
hydrogen.
We assume that the element X is completely ionised, and element abundances for Orion from \citet{esteban1} and \citet{esteban2}.
For the  [\ion{O}{i}] (6300\AA) and [\ion{S}{ii}] (6731\AA) lines, we follow prescriptions given by \citet{hartigan95},
using their equations A8 and A10, respectively. 
To derive the [\ion{Fe}{ii}] line (1.64\,$\mu$m) intensities, for all the sources we assume an electron density of $n_{\rm e}$=10$^5$\,cm$^{-3}$, i.\,e. close to the
[\ion{Fe}{ii}] critical density. This particular value is typical of the jet base~\citep{takami,garcia08,garcia10}, and it has been obtained from source \#\,24, using
the different [\ion{Fe}{ii}] line ratios observed (namely 1.64/1.53, 1.64/1.60, 1.64/1.66, and 1.64/1.68\,$\mu$m), and
adopting the technique used by \citet{nisini02} and \citet{takami}.
Fluxes of optical and NIR lines were dereddened using the adopted $A_{\rm V}$ values reported in Tab.~\ref{parameters:tab} and 
the dereddening law form \citet{R&L}.

In a similar way, the value of $\dot{M}_{\rm out}$(H$_2$) can be written as $\dot{M}_{out}(H_2) = 2 \mu m_{\rm H} N_{\rm H_2} A v_{\rm t} / l_{\rm t}$~\citep[e.\,g.][]{nisini2005,caratti09}, 
where $\mu $ is the average atomic weight, $m_{\rm H}$ the proton mass, $N_{\rm H_2}$ the H$_2$ column density, A 
the area of the H$_2$ knot (i.\.e. encompassed by the slit), $v_{\rm t}$ the tangential velocity, and $l_{\rm t}$ the projected length of the knot
(in this case the slit width). 
The $N_{\rm H_2}$ value has been obtained from the dereddened intensity of the 1-0\,S(1) line (2.12\,$\mu$m), assuming a typical shock 
temperature of 2\,000\,K.

We infer an average $\dot{M}_{out}$ value for those sources with more than one mass-loss rate estimate. 
The derived $\dot{M}_{out}$ values are reported in Column\,15 of Table~\ref{parameters:tab}. 
Because of the several assumptions made, these are very crude estimates with at least one dex uncertainties, 
and they should not be considered accurate for single sources.
The $\dot{M}_{out}$ estimated values range from 4$\times$10$^{-10}$ to 6$\times$10$^{-9}$\,M$_{\sun}$\,yr$^{-1}$, i.\,e. from one to two orders of magnitude lower than 
those of well known powerfull jets and HH objects~\citep[e.\,g.,][]{podio,antoniucci,caratti09}. 
The mass ejection rates estimated here are thus more consistent with less powerful outflows,
driven by less powerful accretors, as also observed by \citet{hartigan95} in Taurus or by \citet{wh04} in the Taurus and Auriga regions.

\subsubsection{Statistics}

Once we have derived the main physical parameters of the YSOs (Tab.~\ref{parameters:tab}), we can then analyse
our sample in a statistical way, and investigate whether, and how, these
quantities vary depending on different types of subsamples. 

To this aim, we have selected six subsamples: \textit{a)} three subsamples were selected according to the class of the sources (Class\,I, Flat, and Class\,II);
\textit{b)} a subsample called `Outflow YSOs', that includes the sampled YSOs with an outflow/jet signature in the images (Col.\,16 of Tab.~\ref{parameters:tab}) and/or
in the optical/NIR spectra (Col.\,17 of Tab.~\ref{parameters:tab});
\textit{c)} a subsample called `Jetless YSOs', that includes the sampled YSOs with no outflow/jet signature;
\textit{d)} a subsample called `Very young YSOs', that includes the sampled YSOs with age $\leq$5$\times$10$^5$\,yr (Col.\,11 of Tab.~\ref{parameters:tab}).
The full sample as well as the subsamples \textit{do not include} the outburst source \#\,25, because of its peculiar characteristics,
which would affect the statistics.
As main observables we take into account $\dot{M}_{acc}$, $\dot{M}_{out}$, $\dot{M}_{acc}/\dot{M}_{out}$ ratio, stellar age, $L_{acc}$/$L_{bol}$, $A_\mathrm{V}$,
$\alpha$, and $M_*$, for which an average value is derived. The number of each subsample elements ($n$) is in round brackets, and it differs
for $\dot{M}_{out}$, where values are reported according to Column 15 of Tab.~\ref{parameters:tab}
Although some of these parameters have large uncertainties (e.\,g., $\dot{M}_{out}$) when 
considering single objects, they are still significant for statistical purposes.
Results are reported in Table~\ref{stat:tab}. 
$\dot{M}_{acc}$, as well as $A_\mathrm{V}$,
$\alpha$, and $M_*$ decrease with YSO class, whereas the average age increases, as expected.
The mass ejection rate decreases from Class\,I to Flat, while it increases in the Class\,II sample.
However, this value is computed from only two Class\,II sources (which show jet lines), thus it does not represent a proper statistical sample.
Sources with outflows have mass accretion rates $\sim$1 dex higher than `jetless' YSOs, which are older, indicating
that accretion/ejection activity decreases with time.
The $\dot{M}_{acc}/\dot{M}_{out}$ ratio seems to be constant ($\sim$0.01), although, due to the uncertainties, this value is not very indicative.
On the other hand, the $L_{acc}$/$L_{bol}$ ratio is slightly higher in younger and active sources.

\begin{landscape}
\begin{table}
\caption{Stellar Parameters}
\label{parameters:tab}
\centering
\begin{tabular}{ccccccccccccccccc}
\hline \hline\\[-8pt]
ID & $\alpha$ & Class & $L_{bol}$  & $A_\mathrm{V}$ & SpT & $T_{\rm eff}$  & $L_*$  & $M_*$ & $R_*$ & Age & $L_{acc}$ & $L_{acc}$/$L_{bol}$ & $\dot{M}_{acc}$ &  $\dot{M}_{out}$ & Jet/Outflow\tablefootmark{a} & Jet lines\tablefootmark{b} \\
 & & &(L$_{\sun}$) & (mag) & & (K) & (L$_{\sun}$) &  (M$_{\sun}$) &  (R$_{\sun}$) & (yr) & (L$_{\sun}$) & &(M$_{\sun}$\,yr$^{-1}$) & (M$_{\sun}$\,yr$^{-1}$) & & \\
\hline\\[-5pt]
1\tablefootmark{aa}   & -1.04  &  II  & 1.3 &  2  & K5\tablefootmark{d,e}$\pm$1 & 4350$\pm$150  & 0.9 & 1.17  & 1.95 & 4.71E+6 & 0.4 &  0.31 &  2.1E-8 & $\cdots$   & n  & n \\
2\tablefootmark{bb}   &  0.06  &  F  & 14.5  &  8.5  & K6\tablefootmark{e}$\pm$1   & 4200$\pm$150  & 13.0 & 1.13  & 6.45 & 2.12E+5 & 1.5 &  0.10 &  2.7E-7 & 2E-09      & y  & y \\
3\tablefootmark{cc}   & -0.49  &  II & 61.4  &  2  & A1\tablefootmark{c,e}$\pm$1 & 9230$\pm$250  & 42.6 & 2.68  & 2.68 & 4.26E+6 & 18.8 &  0.31 &  6.0E-7 & $\cdots$   & y  & n \\
4\tablefootmark{dd}   &  0.04  &  F  & 3.1  &  3    & M0\tablefootmark{c,d,e}$\pm$1 & 3850$\pm$130 & 2.5 & 0.55  & 3.42 & 5.19E+5 & 0.6 &  0.19 & 1.2E-7 & 5E-09      & n  & y \\
5   &  0.10  &  F  & 3.7   &  10   & K5\tablefootmark{e}$\pm$1  & 4350$\pm$150  & 2.6 & 1.1   & 2.91 & 9.71E+5 & 1.1 &  0.3 &   9.3E-8 & 2E-09      & y  & y \\
6   &  0.57  &  I  & 3.6  &  24   & K7\tablefootmark{g}$\pm$3  & 4060$\pm$400  & 3.1 & 0.69  & 3.56 & 6.36E+5 & 0.5 &  0.14 &   8.2E-8 & 4E-10      & y  & y \\
7   & -0.37  &  II  & 13.1 &  0    & A5\tablefootmark{c,e}$\pm$1 & 8200$\pm$200  & 11.9 & 1.88  & 1.84 & 1.00E+7 & 1.2 &  0.09 &  3.7E-8 & $\cdots$   & n  & n \\
8   & -0.08  &  F  & 188.3 &  19   & B7\tablefootmark{c,e}$\pm$1 & 13000$\pm$1000 & 127 & 3.43  & 2.79 & 2.81E+6 & 61.3 &  0.33 & 1.6E-6 & $\cdots$   & y  & n \\
9   & -0.45  &  II & 6  &  2  & K4\tablefootmark{c,e}$\pm$1 & 4590$\pm$150  & 4.4 & 1.51  & 3.23 & 9.62E+5 & 1.6 &  0.27 &  1.1E-7 & 2E-09      & y  & y \\
10  &  0.28  &  F  & 3.0  &  8.2  & M0\tablefootmark{e,f}$\pm$1 & 3850$\pm$130  & 2.4 & 0.55  & 3.69 & 4.13E+5 & 0.6 &  0.20 &  1.3E-7 & 8E-10      & y  & y \\
11  & -0.49  &  II & 7.1  &  2  & K0\tablefootmark{e}$\pm$2   & 5250$\pm$300  & 4.4 & 1.89  & 2.39 & 5.01E+6 & 2.7 &  0.38 &  1.1E-7 & 2E-09      & y  & y \\
12  &  0.33  &  I  & 1.1  &  8    & M7.5\tablefootmark{f}$\pm$1.5 & 2800$\pm$200 & 0.9 & 0.15  & 3.71 & 1.76E+4 & 0.2  &  0.18 & 1.6E-7 & 8E-10      & n  & y \\
13  &  1.60  &  I  & 3.9  &  25   & M1\tablefootmark{f}$\pm$2   & 3705$\pm$300  & 3.6 & 0.47  & 4.40 & 5.01E+4 & 0.3 &  0.08 &  9.0E-8 & $\cdots$   & n  & n \\
14  &  0.50  &  I  & 7.0  &  24 & K7\tablefootmark{g}$\pm$3   & 3890$\pm$400  & 3.9 & 0.60  & 4.10 & 3.24E+5 & 3.1  &  0.44 &  6.8E-7 & 5E-09      & y  & y \\
15  & -0.04  &  F  & 2.3 &  8.5    & K7\tablefootmark{d}$\pm$1  & 4060$\pm$100  & 1.7  & 0.73  & 2.87 & 7.66E+5 & 0.6 &  0.26 &   7.5E-8 & 7E-10      & y  & y \\
16\tablefootmark{ee}   & -0.26  &  F  & 10.2 &  4    & F7\tablefootmark{c,d,e}$\pm$2  & 6280$\pm$160  & 9.1 & 1.70  & 2.54 & 9.35E+6 & 1.1 &  0.11 & 5.2E-8 & $\cdots$   & n  & n \\
17  &  0.16  &  F  & 4.7  &  23   & K5\tablefootmark{g}$\pm$2  & 4350$\pm$300  & 2.9 & 1.12 & 2.94 & 8.93E+5 & 1.8 &  0.38 &    1.5E-7 &  $\cdots$    & n  & n \\
18\tablefootmark{ff}   &  0.27  &  F  & 5.8  &  2    & K1\tablefootmark{c,d,e}$\pm$1 & 5080$\pm$170  & 3.7 & 1.74  & 2.14 & 5.34E+6 & 2.1  &  0.36 & 8.3E-8 & 6E-09   & y  & y \\
19  &  0.39  &  I  & 1.4  &  20   & M0\tablefootmark{f}$\pm$2 & 3850$\pm$300  & 1.1 & 0.56  & 2.25 & 1.00E+6 & 0.3  &  0.21 &    3.8E-8 & $\cdots$   & n  & n \\
20  &  0.32  &  I  & 4.7   &  23   & K5\tablefootmark{g}$\pm$2 & 4350$\pm$300  & 2.2 & 1.12  & 2.31 & 1.30E+6 & 2.5 &  0.53 &    2.1E-7 & $\cdots$   & n  & n \\
21  &  0.23  &  F  & 2.0  &  22.5 & M0\tablefootmark{f}$\pm$2 & 3850$\pm$300  & 1.25 & 0.56  & 2.39 & 8.52E+5 & 0.75 &  0.37 &    1.0E-7 & $\cdots$   & n  & n \\
22  & -0.14  &  F  & 0.32  &  12   & M0\tablefootmark{f}$\pm$2   & 3850$\pm$300  & 0.26 & 0.59  & 1.09 & 6.00E+6 & $<$0.06 &  0.19 &  $<$3.6E-9 & $\cdots$   & n  & n \\
23  &  0.90  &  I  & 7.7   &  21  & M0\tablefootmark{f}$\pm$2   & 3850$\pm$300  & 4.9 & 0.60  & 5.26 & 6.01E+4 & 2.8 &  0.36 &  7.9E-7 & 4E-09      & n  & y \\
24  &  0.93  &  I  & 6.8  &  25 & M2\tablefootmark{f}$\pm$2   & 3560$\pm$300  & 4.7 & 0.4   & 5.73 & 1.06E+4 & 2.1 &  0.31 &  9.6E-7 & 6E-09       & y  & y \\
25  &  0.69/0.25\tablefootmark{f} &  I/F  & 2/22\tablefootmark{h} &  18 & M5\tablefootmark{f}$\pm$1.5 & 3200$\pm$200  & 1.9 & 0.24  & 4.40 & 2.13E+4 & 20 &  0.91 &  1.2E-5 &  $\cdots$ & y  & n \\
26  & -0.56  &  II & 2.6  &  10    & K5\tablefootmark{e,g}$\pm$2 & 4350$\pm$300  & 1.8 & 1.11 & 2.25 & 1.63E+6 & 0.8 &  0.31 &   6.3E-8 & $\cdots$      & y  & n \\
27  & -0.55  &  II & 1.1  &  8    & K6\tablefootmark{c,g}$\pm$1  & 4200$\pm$150  & 0.9 & 0.92  & 1.67 & 3.07E+6 & 0.2 &  0.18 & 1.1E-8 & $\cdots$      & n  & n \\
\hline
\end{tabular}
\tablefoot{
\tablefoottext{a} {Indicates whether the YSO drives a known jet or CO outflow~\citep[][and this paper]{morgan,dent,allendavis,davis09}.}
\tablefoottext{b} {Indicates whether jet lines have been detected in our spectra.}
Used reference/method:
\tablefoottext{c} {from \citet{strom};}
\tablefoottext{d} {from \citet{fang};}
\tablefoottext{e} {obtained with SpTclass;}
\tablefoottext{f} {obtained with the H$_2$O broad band features - water breaks;}
\tablefoottext{g} {SED modelling.}
\tablefoottext{h} {Pre- and outburst values.}
\tablefoottext{aa} {\citet{fang}: $L_{bol}$ = 1.9\,L$_{\sun}$, SpT = K5, $M_*$ = 0.9\,M$_{\sun}$, Age = 7E+5\,yr, $\dot{M}_{acc}$ = 1.5\,E-8\,M$_{\sun}$\,yr$^{-1}$.}
\tablefoottext{bb} {\citet{fang}: $L_{bol}$ = 3.6\,L$_{\sun}$, SpT = K4.5, $M_*$ = 0.9\,M$_{\sun}$, Age = 3E+5\,yr, $\dot{M}_{acc}$=1.9\,E-8\,M$_{\sun}$\,yr$^{-1}$. \citet{fischer}:
$L_{bol}$ = 23\,L$_{\sun}$, $M_*$ = 2.2\,M$_{\sun}$, Age = 3E+5\,yr, $\dot{M}_{acc}$ = 4\,E-7\,M$_{\sun}$\,yr$^{-1}$. \citet{strom}: SpT=G5--G7.}
\tablefoottext{cc} {\citet{alecian}: $L_{*}$ = 97.7\,L$_{\sun}$, SpT = B9, $M_*$ = 1.9\,M$_{\sun}$, $R_*$ = 3\,R$_{\sun}$, Age = 2E+6\,yr. \citet{manoj}: $L_{*}$ = 93\,L$_{\sun}$, 
SpT = A1, $M_* >$ 4.9\,M$_{\sun}$. \citet{hillenbrand}: SpT = B9, $M_*$ = 3.6\,M$_{\sun}$, $R_*$ = 2.8\,R$_{\sun}$, $\dot{M}_{acc}$ = 5\,E-6\,M$_{\sun}$\,yr$^{-1}$.}
\tablefoottext{dd} {\citet{fang}: $L_{bol}$ = 4.2\,L$_{\sun}$, SpT = M0, $M_*$ = 0.4\,M$_{\sun}$, Age = 3E+4\,yr, $\dot{M}_{acc}$ = 1.5\,E-6\,M$_{\sun}$\,yr$^{-1}$.}
\tablefoottext{ee} {\citet{fang}: $L_{bol}$ = 60.7\,L$_{\sun}$, SpT = F7, $M_*$ = 3\,M$_{\sun}$, Age = 1.9E+6\,yr.}
\tablefoottext{ff} {\citet{fang}: $L_{bol}$ = 4.9\,L$_{\sun}$, SpT = K1, $M_*$ = 1.9\,M$_{\sun}$, Age = 1.4E+6\,yr, $\dot{M}_{acc}$ = 4\,E-7\,M$_{\sun}$\,yr$^{-1}$.}}
\end{table}
\end{landscape}

\begin{table*}[!h]
\caption{Statistics for the all sample and selected subsamples. The outbursting source \#\,25 has been excluded from any statistic.}
\label{stat:tab}
\centering
\begin{scriptsize}
\begin{tabular}{ccccccccc}
\hline \hline
ID   &  Avg.\,($\dot{M}_{acc}$) & Avg.\,($\dot{M}_{out}$) & Avg.\,($\dot{M}_{out} / \dot{M}_{acc}$) & Avg.\,(Age)  &  Avg.\,($L_{acc}$/$L_{bol}$) & Avg.\,($A_\mathrm{V}$) & Avg.\,($\alpha$)& Avg.\,($M_*$)\\
 (n)\tablefootmark{a}  & 10$^{-7}$ (M$_{\sun}$\,yr$^{-1}$) & 10$^{-8}$ (M$_{\sun}$\,yr$^{-1}$) (n)\tablefootmark{a} &  &  (Myr) &                  & (mag)    & & M$_{\sun}$        \\
\hline\\[-8pt]
All sample (26)       &  2.5   &  0.29 (13) & 0.01 & 2.3 & 0.26  &  12   &  0.1 & 1.1   \\
Class\,I (8)        &  3.8   &  0.34 (5) & 0.01 & 0.4 & 0.3   &  21   &  0.7  & 0.6  \\
Flat (11)             &  2.6   &  0.27 (6) & 0.01 & 2.5 & 0.25  &  11   &  0.1 & 1.2   \\
Class\,II (7)         &  1.3   &  0.20 (2) & 0.03 & 4.2 & 0.25  &  4   &  -0.6 & 1.6   \\
Outflow YSOs\tablefootmark{b} (18)     &  3.4   &  0.29 (13) & 0.01 & 1.4 & 0.28  &  13   &  0.1 & 1.1                     \\
Jetless YSOs\tablefootmark{c} (8)     &  0.4   &  N/A  &  N/A & 4.4 & 0.19  &  12   &  0  & 1.0                        \\
Very young YSOs\tablefootmark{d} (7)   &  4.5   &  0.32 (6) & 0.01 & 0.2 & 0.28  &  17   &  0.7  & 0.5                    \\
\hline
\end{tabular}
\tablefoot{
\tablefoottext{a} {Number of elements per subsample}
\tablefoottext{b} {This subsample includes the YSOs with an outflow/jet signature in the images (Col.\,16 of Tab.~\ref{parameters:tab}) and/or
in the optical/NIR spectra (Col.\,17 of Tab.~\ref{parameters:tab}); source \#\,25 was not included. }
\tablefoottext{c} {This subsample includes the YSOs with no outflow/jet signature.}
\tablefoottext{d} {This subsample includes the YSOs with age $\leq$5$\times$10$^5$\,yr (Col.\,11 of Tab.~\ref{parameters:tab}); source \#\,25 was not included.}
}
\end{scriptsize}
\end{table*}


\section{Discussion}
\label{discussion:sec}

On the basis of the results from our spectral and photometric survey, we can characterise both
stellar and cirumstellar properties of our flux-limited sample, addressing several issues related to the evolution
of very young and embedded stellar objects, in particular their accretion and ejection activity.  

\subsection{Accretion properties of the sample}

\subsubsection{Accretion luminosities}
\begin{figure}
 \centering
\includegraphics [width=9cm] {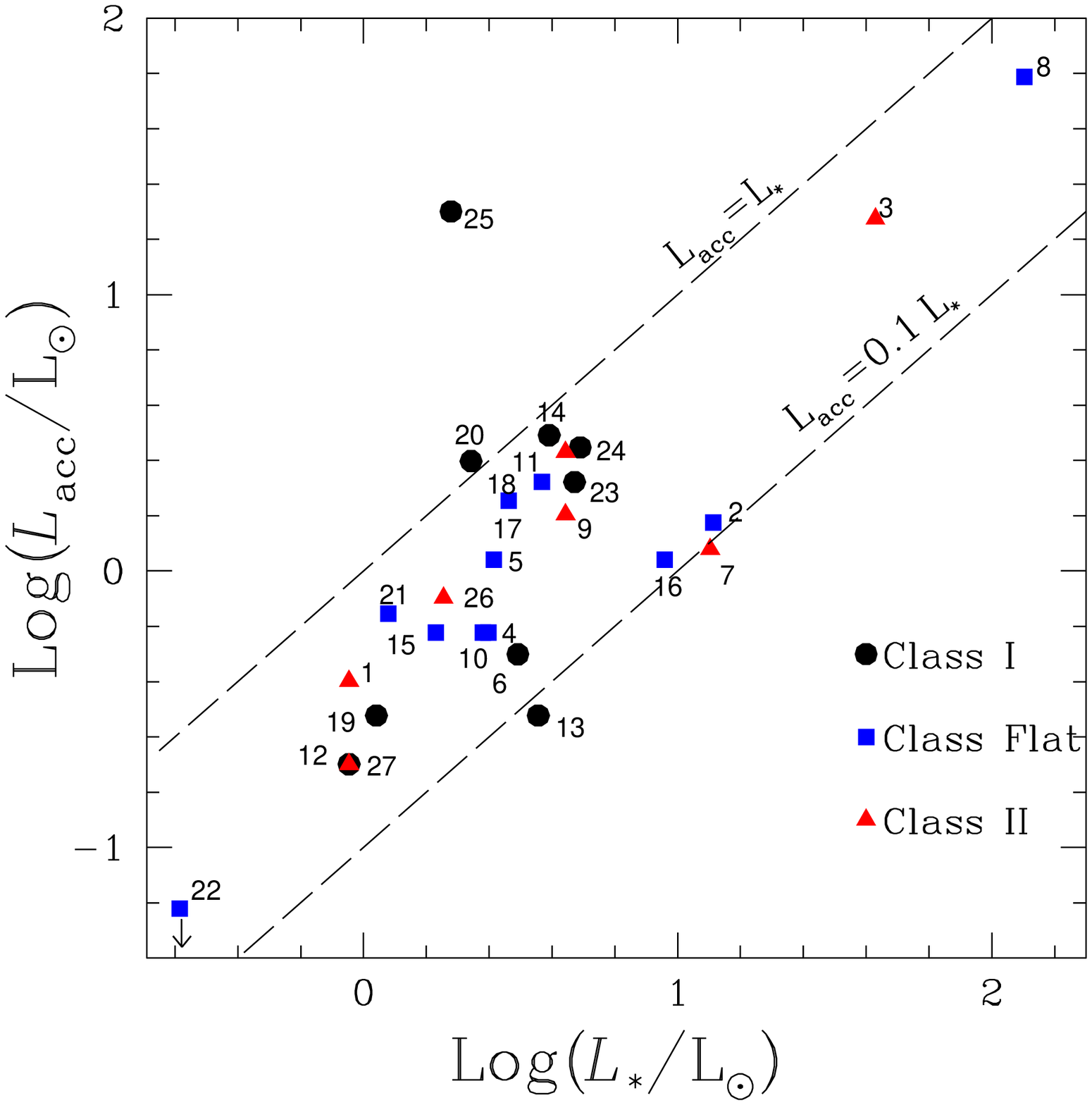}
  \caption{$L_{acc}$ values plotted as a function of $L_*$ for the all sample.
The two dashed lines show the loci of $L_{acc} = L_*$ and  $L_{acc} = 0.1 L_*$.
The different symbols indicate the class of the objects: Class\,I, Flat, or Class\,II
(dots, squares, and triangles, respectively). The object IDs are also reported.
\label{Lacc_vs_Lstar:fig}}
\end{figure}

Figure~\ref{Lacc_vs_Lstar:fig} shows the $L_{acc}$ values plotted as a function of $L_*$ for the whole sample.
The various symbols indicate the different class of object, as derived from the standard SED classification 
(see Sect.~\ref{phot:sec} and Tab.~\ref{parameters:tab}), with Class\,I, Flat, and Class\,II represented 
as dots, squares, and triangles, respectively. 
As already observed by several authors \citep[e.\,g.,][]{muzerolle98,natta06,antoniucci11}, $L_{acc}$ increases with $L_*$. 
In our sample there is a large scatter of $L_{acc}/L_*$ values for objects with similar $L_*$ values,  
i.\,e. about one order of magnitude, excluding the outbursting source \#\,25. 
This spread is similar to what we observe in \textit{paper-1} for the Cha I and II sample, 
which have the same selection and observational constraints like ours. As a comparison, the \citet{natta06} sample of Class\,II and III YSOs in Ophiuchus 
shows a larger scatter, ranging over more than two orders of magnitude.

Excluding the outbursting source \#\,25, which stands out from the others with $L_{acc}/L_* \sim 10$, the $L_{acc}/L_*$ ratio for our sample ranges
from $\sim$0.1 to $\sim$1, indicating that none of the observed objects, except source \#\,25 and, marginally, \#\,20 ($L_{acc}/L_{bol}\sim$0.53), 
is accretion dominated (i.\,e. has $L_{acc}>L_*$).
This result confirms previous findings in other star forming regions~\citep[see e.\,g.][]{muzerolle98,wh04,nisini05,antoniucci}, which indicate 
that several Class\,I sources are not in the main accretion phase. Moreover, there is no clear trend between the $L_{acc}/L_*$ ratio 
and the YSO class, with the various classes randomly distributed between 0.1 and 1, 
indicating that high and low accretors are homogeneously present in Class\,I, Flat, and Class\,II YSOs.
It is worth noting that three out of the four sources with $L_{acc}/L_* \lesssim 0.1$  (all except source \#\,2) do not show ejection activity in
our spectra or images (Columns 16 and 17, Tab.~\ref{parameters:tab}), indicating that accretion and ejection activity are directly related,
and confirming the expected trend that the higher the accretion rate the higher is the ejection rate.

\subsubsection{Mass accretion rates and accretion evolution}

\begin{figure}[h!]
 \centering
\includegraphics [width=9cm] {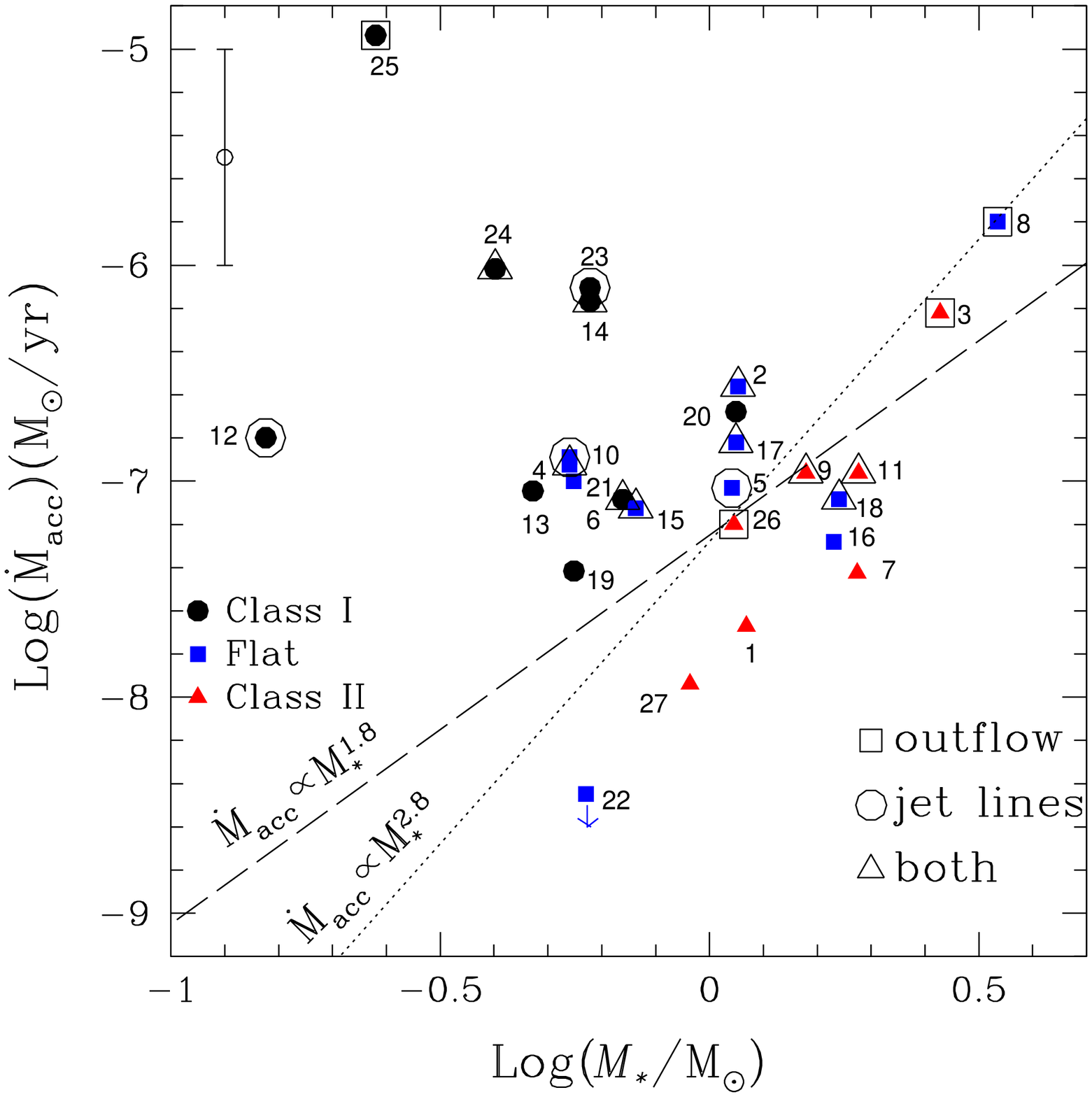}\\
\includegraphics [width=9cm] {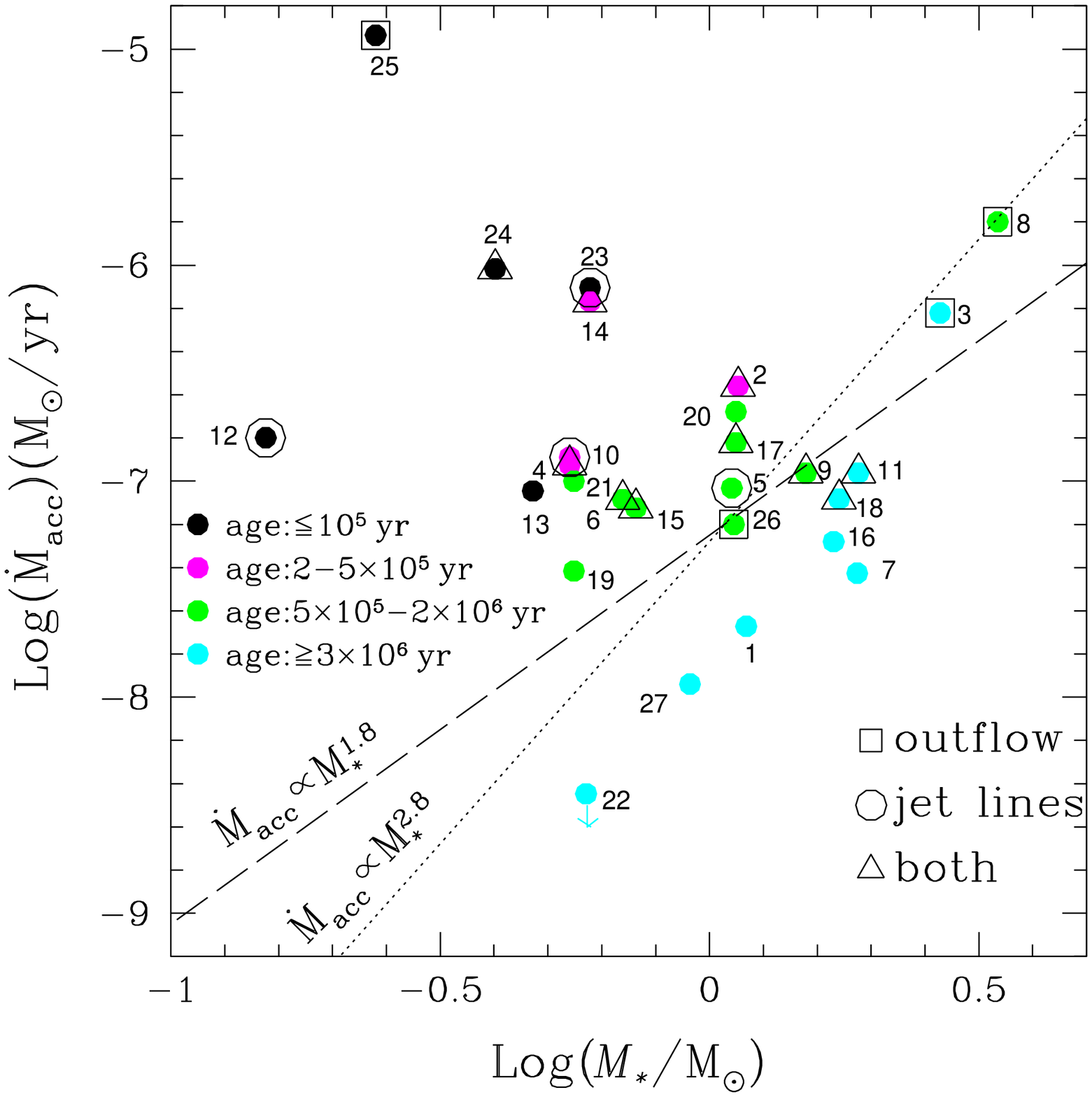}\\
 \caption{\textbf{Top panel:} Mass accretion rate versus stellar mass for the all sample.
The different symbols indicate the class of the objects: Class\,I, Flat, or Class\,II
(dots, squares, and triangles, respectively). The object ID is also reported.
The presence of jets/outflows in the images, lines tracing jets in the spectra, or both are indicated
by open squares, circles, or triangles, respectively.
The dashed line shows the $\dot{M}_{acc} \propto M_{*}^{1.8}$ relationship obtained by
\citet{natta06} for the Ophiuchus sample, whereas the dotted line displays the $\dot{M}_{acc} \propto M_{*}^{2.8}$ 
relationship obtained by \citet{fang} for the L\,1634 and the L\,1641 samples.
The lines are merely indicative and do not fit the data. 
\textbf{Bottom panel:} Same as top panel, with colours indicating the different age of the objects, ranging from
$\le$10$^5$ to 10$^7$\,yr.
\label{Macc_vs_Mstar:fig}}
\end{figure}

The top panel of Figure~\ref{Macc_vs_Mstar:fig} shows the mass accretion rates of the sampled objects 
versus their stellar masses. Again, the various symbols indicate the different classification of the objects (Class\,I, Flat, and Class\,II are represented 
as dots, squares, and triangles, respectively), as also reported in Fig.~\ref{Lacc_vs_Lstar:fig}.
The presence of jets/outflows in the images, lines tracing jets in the spectra, or both is indicated
by open squares, circles, or triangles, respectively, over-plotted on each object.

Several authors have pointed out the existence of an empirical relationship between the mass accretion rate and the stellar mass, i.\,e. 
$\dot{M}_{acc} \propto M_*^{k}$, with $k$ ranging from $\sim$2~\citep[][]{calvet04,muzerolle03,natta04,natta06} to $\sim$3~\citep[][]{fang}.
Our data marginally indicate such a trend, because of the large spread in $\dot{M}_{acc}$ (more than 3 orders of magnitude) 
and YSO age, the small $M_*$ range, and the small number of objects, thus our data do not allow us to obtain a better estimate for $k$.
In the figure we also show the $\dot{M}_{acc} \propto M_{*}^{1.8}$ relationship derived by \citet{natta06}
(dashed line) for their Ophiuchus sample, and the $\dot{M}_{acc} \propto M_{*}^{2.8}$ relationship from \citet{fang} (dotted line)
for their L\,1634 and L\,1641 samples. The lines serve as a guide to the eye and are not intended to fit the data.

As mentioned in Sect.~\ref{macc:sec}, the outbursting source \#\,25 shows the highest mass accretion rate, and, in general, 
when considering a power-law relationship, Class\,I objects have mass accretion rates higher than the other YSOs,
being positioned mostly in the top-left area of the panel. The remaining Class\,I YSOs are located in the central part of the
plot, together with the majority of Flat YSOs. Finally, moving towards the bottom-right part of the panel, there 
are Class\,II and a few Flat YSOs. The bottom end is occupied by YSOs, that do not show any evidence of outflow activity.
In conclusion, there is a marginal trend between the mass accretion rate evolution and the
YSO class. Indeed, this is related to the fact that YSO class is not strictly related to YSO age, 
as already shown in Sect.~\ref{Sparam:sec} and Fig.~\ref{alfa_age:fig} 

On the other hand, there is a more clear correlation between the mass accretion evolution and the age of the YSOs,
as shown in the bottom panel of Figure~\ref{Macc_vs_Mstar:fig}, which displays the same $\dot{M}_{acc}$ vs $M_{*}$ plot, 
with different colours indicating different age ranges. In this case a clear age stratification is
visible in the figure, going from the top-left (youngest sources) to the bottom-right (oldest sources) of the panel.
Moreover, if we consider objects in the same age range (i.\,e. YSOs with the same colour in the plot), 
we note some spread, with high accretors (i.\,e. with large $L_{acc}/L_{bol}$ ratio) 
usually positioned at the top-left of their group, and low accretors at the bottom-right. 
For example the black points (age$\lesssim$10$^5$\,yr) have source \#\,25 with $L_{acc}/L_{bol}\sim0.9$ at one end, and source \#\,13
with $L_{acc}/L_{bol}\sim0.1$ at the other. Notably, this low-accreting Class\,I source does not show any relevant ejection activity
in our images and spectra. The violet subsample (age $\sim$2--5$\times$10$^5$\,yr) has source \#\,14 with $L_{acc}/L_{bol}\sim0.4$ and source \#\,2 with $L_{acc}/L_{bol}\sim0.1$, and so on.
This spread is usually less than one order of magnitude for the various considered age-bins, except for the black points
with the outbursting source \#\,25 ($\sim$ 2 orders of magnitude). 
Although the largest spread is due to the outburst,
the small spreads ($\leq 1$ dex) might be attributed to several causes:
\textit{i)} uncertainties on both $\dot{M}_{acc}$ and $M_*$ values; \textit{ii)} uncertainties on the stellar age;
\textit{iii)} short-time scale variability; 
\textit{iv)} differences in the initial and environmental conditions~\citep[see,][for a detailed discussion]{natta06}.
As stated in Sect.~\ref{macc:sec}, $\dot{M}_{acc}$ errors can be up to one dex and might cause the spread.
There are large uncertainties on the determination of the stellar age, due to our uncertainties on $L_*$ and $T_{eff}$.
On the other hand, Tab.~\ref{mags:tab} indicates that, with the exception of the outbursting source \#\,25, short-time scale variability
cannot be responsible for such a large spread. Indeed, the maximum $\sim$10\,yr
variability in the $K_s$ band is $\lesssim$0.5\,mag, and, on average, $<$0.2\,mag. This type of variability is 
extremely common in YSOs~\citep[see, e.\,g.,][]{beck}. Even assuming that extinction has not changed and the observed variability in the $K_s$ band
is entirely due to Br$\gamma$ variation (i.\,e. to accretion), this translates into a Br$\gamma$ flux change of $\sim$1.6, i.\,e. less than a factor of 4 in $\dot{M}_{acc}$, that is
well below the desired one dex.

\begin{figure}[h!]
 \centering
\includegraphics [width=9cm] {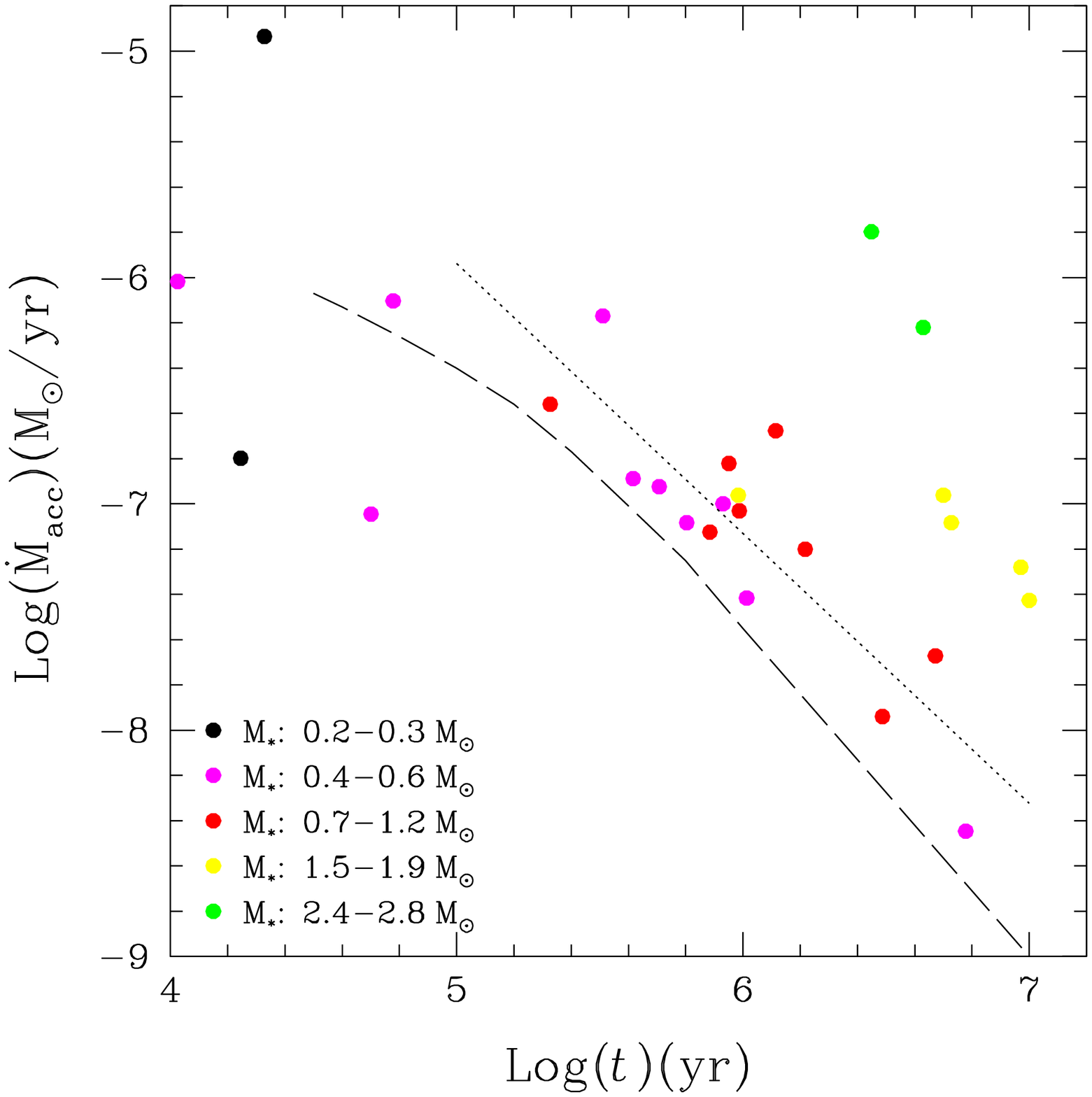}\\
\includegraphics [width=9cm] {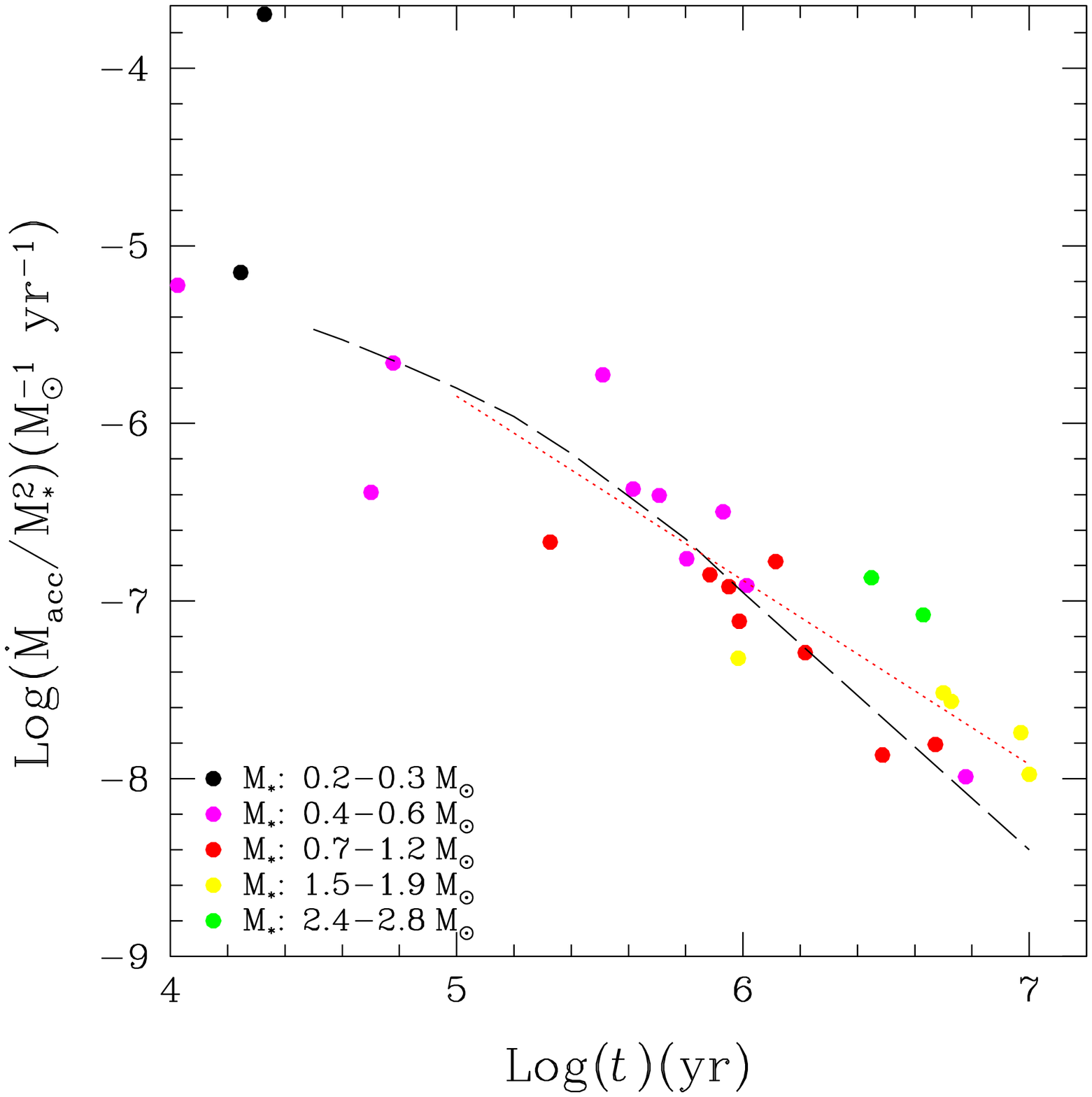}\\
 \caption{\textbf{Top panel:} Mass accretion rate as a function of stellar age for the observed sample. 
The various colours indicate different mass values. 
The dashed line displays one of the viscous models of \citet{hartmann98}, for a 
$M_*=0.5$\,$M_{\sun}$, an initial disc mass $M_d =0.1$\,$M_{\sun}$, viscosity parameter $\alpha$ = 10$^{-2}$,
and viscosity exponent $\gamma=1$, i.\,e. $\eta=1.5$ in the $\dot{M}_{acc} \propto t^{-\eta}$ relationship.
The red dotted line shows the best linear fit for data points with $0.4\leq M_*\leq1.2$\,$M_{\sun}$ and
$t > 10^5$\,yr, i.\,e. $\dot{M}_{acc} \propto t^{-1.2}$.
\textbf{Bottom panel:} Same as top panel, but with the $\dot{M}_{acc}$ values normalised to the stellar $M_*^2$, to take into account $\dot{M}_{acc}$ dependence
from $M_*$.
\label{Macc_vs_time:fig}}
\end{figure}

Despite the large uncertainties on $\dot{M}_{acc}$, $M_{*}$, and age, Figure~\ref{Macc_vs_Mstar:fig} (bottom panel) indicates that the large
scatter can be, at least partially, explained in terms of YSO age and long-time scale variability. 
Mass accretion evolution has been studied for T Tauri discs~\citep[][]{hartmann98}, and mass accretion rates are
thought to decrease with time as $\dot{M}_{acc} \propto t^{-\eta}$, consistently with the evolution of a 
viscous disc~\citep[][]{hartmann98,muzerolle00,sicilia06,sicilia10}. 
According to these viscous disc models, the main infall process from envelope to disc is over, thus the envelope is not providing a significant amount 
of mass and angular momentum to the disc.
Additionally, from this stage on, the quantity of angular momentum removed by the jet/wind is considered irrelevant. Therefore, the disc evolution is mostly driven 
by viscous processes, which control the mass accretion rate. Once the initial disc mass ($M_d$) and viscosity parameter are set,
these models predict that the mass accretion decreases exponentially with time. 
Top panel of Figure~\ref{Macc_vs_time:fig} shows the accretion rates of the sampled objects as a function of the stellar age, with different colours indicating different
mass ranges. Again, the data show a clear trend of the mass accretion decaying with time. This trend matches well the results of other 
works~\citep[][]{hartmann98,sicilia06,sicilia10}, with our data having a smaller $\dot{M}_{acc}$ spread for a given mass interval. 
For comparison, the black dashed line shows the fiducial model of \citet{hartmann98} viscous disc. The model
assumes a stellar mass $M_*$ of 0.5\,$M_{\sun}$, a disc temperature at 100\,AU of 10\,K, an initial disc mass $M_d$ of 0.1\,$M_{\sun}$, a
viscosity parameter $\alpha$ = 10$^{-2}$, and viscosity exponent $\gamma=1$, i.\,e. $\eta=1.5$ in the $\dot{M}_{acc} \propto t^{-\eta}$ relationship. 
The curve reproduces well the distribution of our YSOs with $0.4\leq M_*\leq1.2$\,$M_{\sun}$ (i.\,e. violet and red dots). 
Additionally, more massive and less massive YSOs are positioned above and below the curve, respectively. This is expected, since we observe a correlation 
between $\dot{M}_{acc}$ and $M_*$. 
The \citet{hartmann98} model would probably better match our data using a slightly smaller initial disc mass, which would shift the curve downwards, 
and a slightly smaller $\eta$ value, which would decrease the curve steepness (-$\eta$ is the curve slope). 
This last parameter is particularly important, because it regulates the accretion rate, giving us indication of the disc lifetime.
To get a better estimate of the $\eta$ parameter, we then fit those data points with age $t > 10^5$\,yr, i.\,e.
in a temporal range where the viscous processes likely control the mass accretion rate.
Additionally, we constrain our sample to stellar masses $0.4\leq M_*\leq1.2$\,$M_{\sun}$, to be consistent with 
models and data from the literature. The red dotted line shows the best linear fit 
to the considered data points (namely violet and red dots with $t > 10^5$\,yr), giving $\eta$ = 1.2$\pm$0.4. 
An identical result ($\eta \sim 1.2$) was obtained by \citet{sicilia10} from their Cep\,OB2 sample. The authors
remark that the mass accretion decay appears to be slower than previously assumed in the models, where $\eta$ values
range from 1.5 to 2.8~\citep{hartmann98}. 
To take into account the $\dot{M}_{acc}$ dependence from $M_*$ and reduce the scatter in the plot, 
we normalise $\dot{M}_{acc}$ dividing by $M_*^2$. The results are shown in the bottom panel of Figure~\ref{Macc_vs_time:fig}.
The data scatter is smaller, although the most massive stars are still positioned well above the average. 
Considering the data points with $t > 10^5$\,yr, we obtain $\eta$=1.0$\pm$0.3, or, discarding the most massive objects ($M_* \ge$1.5\,$M_{\sun}$, i.\,e.
yellow and green data points), $\eta$=1.3$\pm$0.2.

\begin{figure}
 \centering
\includegraphics [width=8.5cm] {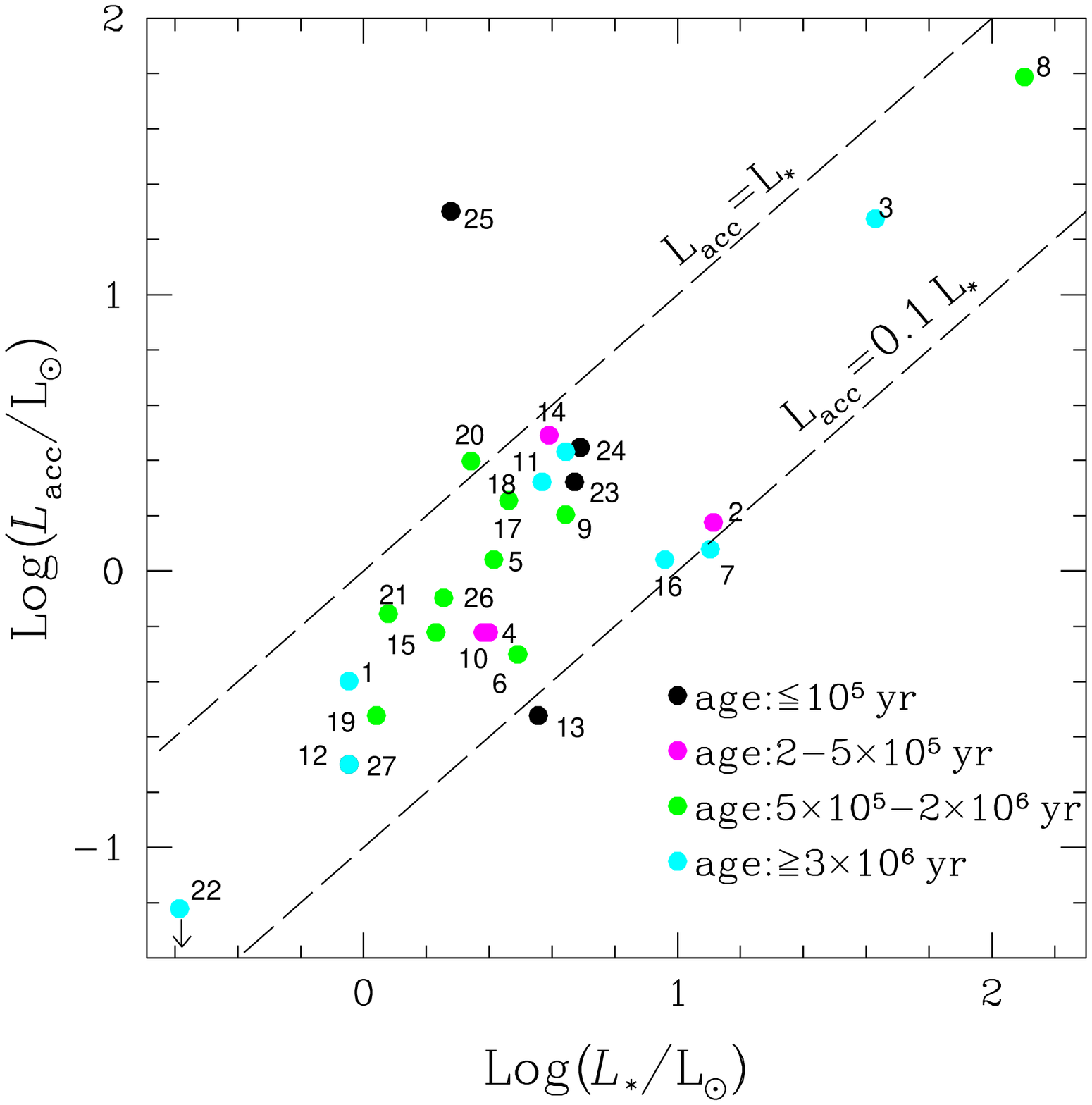}
\includegraphics [width=8.5cm] {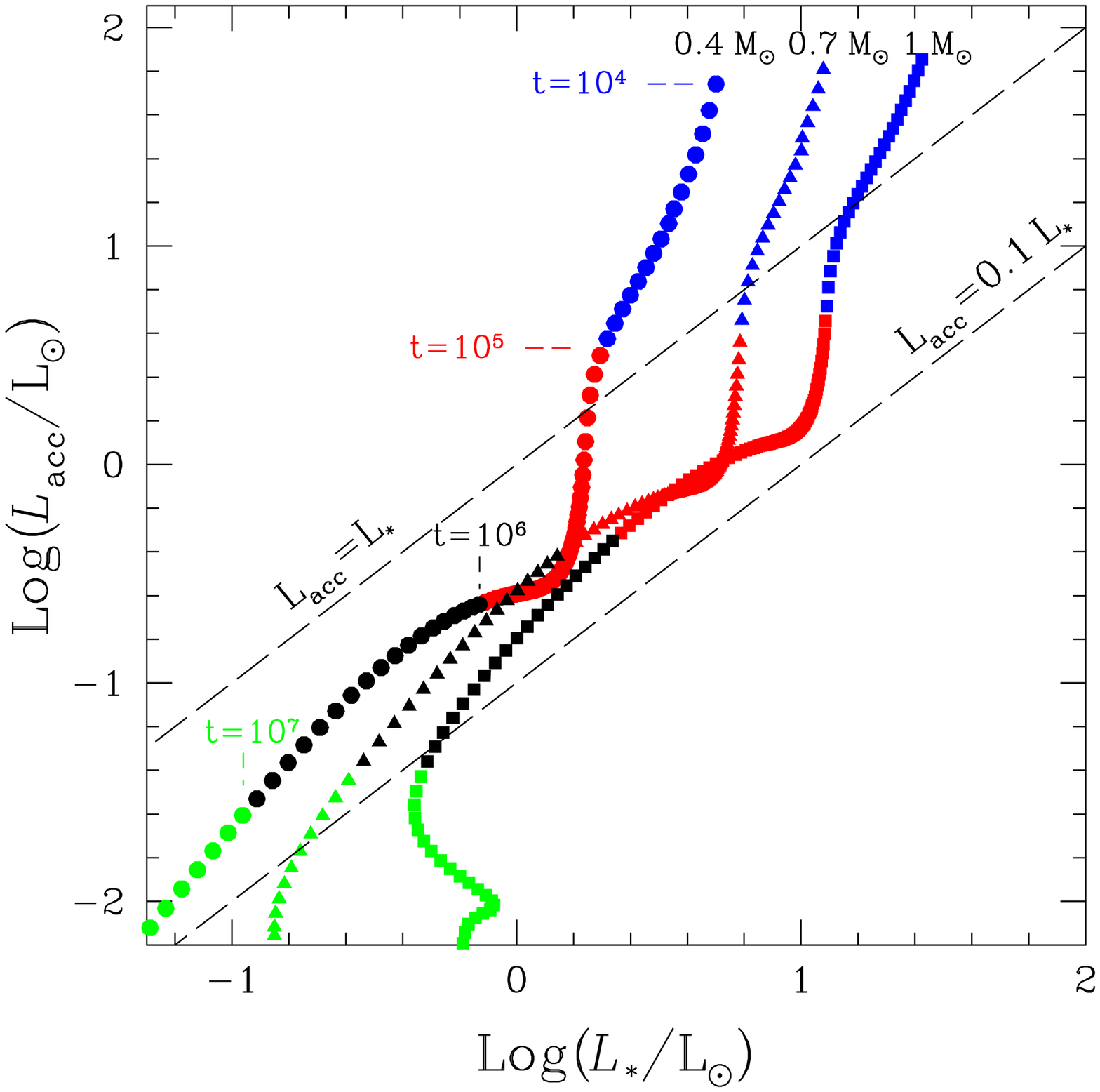}
  \caption{\textbf{Top panel:} $L_{acc}$ vs $L_*$ as in Fig.~\ref{Lacc_vs_Lstar:fig}, with colours indicating the different age of the objects
as bottom panel of Fig.~\ref{Macc_vs_Mstar:fig}.
\textbf{Bottom panel:} Simulation of the $L_{acc}$ vs $L_*$ temporal evolution for YSOs of different masses (0.4, 0.7, and 1\,$M_{\sun}$),
derived considering the evolutionary tracks from \citet{siess} and a viscous disc evolution with
$\dot{M}_{acc} \propto t^{-\eta}$, with $\eta$=1.2. Different colours indicate different age, ranging from 10$^4$
to 10$^7$\,yr.
\label{laccsimulato:fig}}
\end{figure}

Finally, on the basis of our previous results, it is reasonable to question whether the scatter observed in the $L_{acc}$ vs $L_*$ plot of
Figure~\ref{Lacc_vs_Lstar:fig} might be related to the inferred mass accretion evolution. Naively, we might expect an age stratification
in the plot similar to what we observe in the $\dot{M}_{acc}$ vs $M_*$, with younger objects showing larger $L_{acc}/L_*$ ratios 
and the older YSOs with smaller ratios. However, if we plot $L_{acc}$ vs $L_*$ as a function of the age as in Fig.~\ref{Macc_vs_Mstar:fig}, bottom
panel, there is not any stratification (see Fig.~\ref{laccsimulato:fig}, top panel). 
Besides the uncertainties on $L_*$ and $L_{acc}$, which indeed affect our results, 
this is because $L_{acc}$ does not depend only on $\dot{M}_{acc}$, but also on the $M_*/R_*$ ratio, 
which is not linear.

To clarify this point, we simulate the temporal evolution of $L_{acc}$ vs $L_*$, for YSOs with 0.4, 0.7, and 1\,$M_{\sun}$ stellar masses, 
using \citet{siess} evolutionary tracks, and assuming an $\dot{M}_{acc}$ evolution with $\dot{M}_{acc} \propto t^{-1.2}$.
The bottom panel of Figure~\ref{laccsimulato:fig} shows the results of our simulation, with different colours and labels indicating
various age bins, from 10$^4$ to 10$^7$\,yr. According to this plot, objects with the same age but with different stellar mass usually have
different $L_{acc}/L_*$ values, which increase with $M_*$. More notably, for a given $M_*$, the $L_{acc}/L_*$ ratio does not decrease monotonically with time,
but, between 10$^5$ and 10$^6$\,yr, the function shows a local minimum (coloured red in the plot), which appears at earlier times with increasing 
stellar mass (see Figure~\ref{laccsimulato:fig}, bottom panel). 
Indeed, our oversimplified simulation does not take into account other important factors like mass accretion variability, and different 
environmental and initial conditions, which may strongly affect the $L_{acc}$ value, producing a more confused outcome.
Nevertheless, it might explain, at least partially, the $L_{acc}/L_*$ spreads so far observed in the literature and here.

\subsubsection{Mitigating the luminosity problem?}

Several observational studies have already shown that a significant discrepancy
exists between the accretion rates observed in embedded YSOs and those predicted by theoretical models.
This inconsistency was first noted by \citet[][]{kenyon90}, and it is also known as the \textit{luminosity problem}~\citep{Kenyon95},
because the early-YSO luminosities, which should be accretion dominated, appear to be fainter than expected.
The latest \textit{Spitzer Space Telescope} surveys~\citep[see, e.\,g.,][]{evans09,enoch} support these findings,   
revealing that more than 50\% of the embedded YSOs have bolometric luminosities, and thus inferred mass accretion rates, 
considerably lower than those predicted by the free-fall theoretical  
models~\citep[i.\,e. $\dot{M}_{acc}<<$10$^{-6}$\,$M_{\sun}$\,yr$^{-1}$ vs $\sim$2$\times$10$^{-6}$\,$M_{\sun}$\,yr$^{-1}$, for a Class\,I YSO with M$\sim$1\,$M_{\sun}$;][]{shu,terebey}. 
More evidence comes from recent NIR spectral studies of Class\,I YSOs~\citep[][]{wh04,nisini05,doppmann,antoniucci,beck,prato}, 
aimed at measuring both stellar and accretion luminosities, and at estimating the stellar age. 
These studies have shown that, on average, a large fraction of Class\,I YSOs has mass accretion and ejection rates lower 
than theoretically expected (from one to two orders of magnitude), i.\,e. more similar to the accretion rates of the 
classical T Tauri stars ($\sim$10$^6$\,yr).
To reconcile theory with observations, these studies provide various motivations.

For example, misclassification might be one of the causes, and some Class\,I objects are likely to be misclassified Class\,II YSOs, 
seen through optically thick discs. Indeed, geometrical effects can easily modify the SED shape and thus the observed spectral index~\citep[see, e.\,g.,][]{whitney03-1}.
Nevertheless, radiative transfer models of \citet{whitney03-1} clearly indicate that misclassified objects are about five times fainter than real
Class\,I YSOs, and we actually observe these difference in our sample (see, e.\,g., source \#\,19 vs \#\,23 in Table~\ref{parameters:tab}). 
\citet{wh07} suggest that at least one-third of Class\,I YSOs in Taurus have been misclassified, and we find the same result in our small 
sample, i.\,e. three (namely \#\,6, \#\,19, and \#\,20) out of nine Class\,I YSOs have ages $\sim$1\,Myr, i.\,e. they are likely edge-on CTT stars.

However, results from this and other works indicate that several Class\,I YSOs are genuine young stellar objects ($\le$5$\times$10$^5$\,yr)
with accretion rates ranging  from $\sim$$\dot{M}_{acc}\sim$10$^{-6}$ to 10$^{-8}$\,$M_{\sun}$\,yr$^{-1}$. Therefore, to reach their final
masses, these YSOs cannot accrete mass in a steady way, but they need to acquire most of their mass through short episodic outbursts, 
during which $\dot{M}_{acc}$ increases by some order of magnitudes up to $\dot{M}_{acc}$$\sim$10$^{-5}$\,$M_{\sun}$\,yr$^{-1}$ for YSOs
with $\sim$1\,$M_{\sun}$. 

The results of our survey, although biased by the relatively small number of targets and by the sample selection criteria,
are consistent with this picture. A small number of very young objects (`true' Class\,I YSOs) shows high mass accretion rates. 
Among them we detected an outbursting YSO (source \#\,25), which has recently increased its mass accretion rate by $\sim$2 orders of magnitude~\citep{caratti11}, 
whereas the remaining sources have, on average, $\dot{M}_{acc}\sim$5$\times$10$^{-7}$\,$M_{\sun}$\,yr$^{-1}$ (six objects with $M_* \lesssim$1\,$M_{\sun}$) 
and $\dot{M}_{acc}\sim$10$^{-7}$\,$M_{\sun}$\,yr$^{-1}$ (one object with $M_* \sim$0.1\,$M_{\sun}$). 
Considering different ranges of masses, these $\dot{M}_{acc}$ values are almost one order of magnitude higher than those reported by \citet{wh07},
implying that these young objects should spend $\sim$5\% of their 
time in an outburst state to reach typical CTT masses, i.\,.e. this percentage is slightly smaller than \citet{wh07} estimates.
Therefore, statistically, we should observe that among `true' Class\,I sources about $\sim$5\% of these objects are in a outburst phase, while the remaining are quiescent, 
and thus `under-luminous'.
So far our statistics is pretty limited and a larger number of deep-NIR imaging surveys is needed.
Nevertheless, the results of the latest NIR surveys, which discovered several embedded outbursting  YSOs~\citep[see, e.\,g.,][]{covey2010,kospal,caratti11},
seem to point in this direction.




\section{Conclusions}

\label{conclusion:sec}
As part of the POISSON project (Protostellar Optical-Infrared Spectral Survey On NTT), 
we present the results of a multi-wavelength spectroscopic and photometric survey of 27
embedded YSOs in the L\,1641 star forming region, aimed at deriving the stellar parameters and evolutionary stage, 
as well as inferring their accretion and ejection properties.
Our multi-wavelength database includes low-resolution optical-IR spectra from \textit{NTT} and \textit{Spitzer} (0.6-40\,$\mu$m), as well as 
photometric data covering a spectral range from 0.4 to 1100\,$\mu$m, which allow us to construct the YSOs spectral energy distributions (SEDs) 
and to characterise the object parameters (visual extinctions, spectral types, accretion and bolometric luminosities, mass accretion and ejection rates).

The main observational results of this work are the following:

\begin{enumerate}
\item[-] Photometric variability of $\Delta m_K\geq$0.3\,mag is detected in $\sim$22\% of the YSOs ($\sim$10\,yr baseline).

\item[-] The optical-NIR spectra show several emission lines, mostly associated with YSO accretion and ejection activity (e.\,g., 
\ion{H}{i}, \ion{Ca}{ii}, \ion{He}{i}, [\ion{S}{ii}], [\ion{Fe}{ii}], H$_2$), as well as other emission 
lines (e.\,g. \ion{O}{i}, \ion{Fe}{i} \ion{Na}{i}, \ion{Mg}{ii}, \ion{C}{i}) characteristic of active young stars, broad-band molecular absorption features, 
typical of cool objects (VO, TiO and H$_2$O bands) and CO overtone bands both in absorption and emission.  
Also \textit{Spitzer} spectra show emission lines, mostly H$_2$ pure rotational lines, [\ion{Fe}{ii}], [\ion{Si}{ii}] and [\ion{Si}{iii}], and [\ion{S}{iii}], largely 
originating from shocks along the YSOs jets, or tracing the disc. Ice absorption features at 5 to 8\,$\mu$m (H$_2$O) and 15.2\,$\mu$m (CO$_2$), as well as the amorphous silicate absorption feature at 9.7\,$\mu$m have 
been also detected in some of the sources. 


\item[-] SEDs were constructed using both photometric and spectroscopic data, allowing to classify the 27 targets as nine Class\,I, eleven Flat, and seven Class\,II YSOs.
The inferred $L_{bol}$ values range from $\sim$0.3 to 180\,$L_{\sun}$.

\item[-] The spectral types for the sample range from B7 to M7.5, with 74\% of the objects being low-mass YSOs (SpT$\geq$K5).
The average and median age of the sample are about 2 and 1\,Myr, respectively, with values ranging from $\le$10$^5$ to 10$^7$\,yr.
The average ages of Class\,I YSOs, Flat, and Class\,II YSOs are 0.4, 2, and 4\,Myr. Six out of nine Class\,I YSOs have an age of $\sim$10$^5$,
while the remaining three appear to be older (5$\times$10$^5$ to 10$^6$\,yr).
Also among the Flat sources there are three objects (out of eleven) with age $\geq$5$\times$10$^6$\,yr, i.\,e.
consistent with more evolved sources. 

\item[-] Accretion luminosities, obtained by averaging luminosities from Pa$\beta$, and Br$\gamma$ lines, range from $\sim$0.1 to $\sim$60\,$L_{\sun}$,
whereas mass accretion rates range from 3.6$\times$10$^{-9}$ to 1.2$\times$10$^{-5}$\,M$_{\sun}$\,yr$^{-1}$.
\end{enumerate}

From these results we draw the following conclusions:

\begin{enumerate}

\item[-] Geometrical effects can significantly modify the SED shapes, and the standard SED classification ($\alpha$ computed between 2.2 and 24\,$\mu$m)
may not properly reflect YSO age.

\item[-] Mass accretion and ejection rates, extinction, and spectral indices decrease with YSO age. The youngest YSOs have the highest mass accretion rates, 
whereas the oldest YSOs do not show detectable jet activity in both images and spectra. Nevertheless, apart from the outbursting source \#\,25 and, marginally, 
source \#\,20, none of the remaining YSOs is accretion dominated (i.\,e. having $L_{acc} > L_*$).

\item[-] There is a marginal trend between the mass accretion rate evolution and the YSO class, whereas 
there is a more clear correlation between the mass accretion evolution and the age of the YSOs. Despite the large uncertainties on $\dot{M}_{acc}$, $M_{*}$, and age, 
our analysis indicates that the large spread of mass accretion rates can be mostly explained in terms of YSO age and long-time scale (non-periodic) variability. 

\item[-] We obtain a clear correlation between $\dot{M}_{acc}$ and age ($\dot{M}_{acc} \propto t^{-\eta}$, with $\eta$=1.2), 
for YSOs with $t > 10^5$\,yr, consistent with the mass accretion evolution in viscous disc models. These results indicate 
that the mass accretion decay appears to be slower than previously assumed in the classical models~\citep{hartmann98} (where the $\eta$ value
ranges from 1.5 to 2.8) and coincides with the latest results.

\item[-] The majority of very young objects in our sample (`true' Class\,I YSOs, $t\le$5$\times$10$^5$\,yr) shows, on average, high mass accretion rates. 
Among them, we detected an outbursting YSO (source \#\,25), which has recently increased its mass accretion rate by $\sim$2 orders of magnitude,
whereas the remaining sources have, on average, $\dot{M}_{acc}\sim$5$\times$10$^{-7}$\,$M_{\sun}$\,yr$^{-1}$ (six objects with $M_* \lesssim$1\,$M_{\sun}$) 
or $\dot{M}_{acc}\sim$10$^{-7}$\,$M_{\sun}$\,yr$^{-1}$ when considering less massive objects ($M_* \sim$0.1\,$M_{\sun}$). 
Considering different ranges of masses, these $\dot{M}_{acc}$ values are almost one order of magnitude higher than those previously studied, e.\,g. 
in Taurus and Auriga star forming regions, implying that these young objects should spend 1--5\% of their lifetime in an outburst state, to reach typical CTTs masses.

\end{enumerate}


\begin{acknowledgements}
We are grateful to Tom Megeath and Min Fang for providing us with optical SDSS, IRAC, and MIPS 24\,$\mu$m photometry;
to Jesus Hernandez for providing us with his SPTclass code; to Antonella Natta, Malcolm Walmsley, and Lori Allen for fruitful discussions and insightful comments.
We wish to thank an anonymous referee for useful insights and comments.
ACG acknowledges support from the Science Foundation of Ireland, grant 07/RFP/PHYF790, and the European Commision, grant FP7/MC/ERG249157.
This publication has made use of data from the ``From Molecular Cores to Planet-forming Disks'' (c2d) Legacy project.
This research has also made use of NASA's Astrophysics Data System Bibliographic Services and the SIMBAD database, operated
at the CDS, Strasbourg, France, the 2MASS data, obtained as part of the Two Micron All Sky Survey,
a joint project of the University of Massachusetts and the Infrared Processing and Analysis Center/California Institute of Technology,
funded by the National Aeronautics and Space Administration and the National Science Foundation, and the UKIDSS data (UKIRT
Infrared Deep Sky Survey).

\end{acknowledgements}

\bibliographystyle{aa}
\bibliography{ref}

\Online

\begin{appendix}
\section{Photometry \& Spectroscopy}
\label{appendixA:sec}

\begin{landscape}
\begin{table}
\begin{scriptsize}
\caption{Available photometry from 0.44\,$\mu$m to 8\,$\mu$m (Jy)}
\label{photometry:tab}
\centering
\begin{tabular}{ccccccccccccc}
\hline\hline             
Source & B  &  V  & R & I$_B$ & I$_J$ & J & H & Ks & IRAC\,1 & IRAC\,2 &  IRAC\,3 &  IRAC\,4 \\
ID     & 0.44\,$\mu$m & 0.55\,$\mu$m & 0.64\,$\mu$m & 0.79\,$\mu$m &  0.9\,$\mu$m & 1.25\,$\mu$m &  1.65\,$\mu$m &  2.2\,$\mu$m & 3.6\,$\mu$m & 4.5\,$\mu$m & 5.8\,$\mu$m & 8.0\,$\mu$m\\
\hline\\
1      & $\cdots$          & $\cdots$          & 0.003$\pm$0.001      & 0.004$\pm$0.001     & 0.021$\pm$0.005 & 0.0224$\pm$0.0009 & 0.066$\pm$0.002 & 0.089$\pm$0.002 &  0.122$\pm$0.007 & 0.115$\pm$0.006 & 0.103$\pm$0.006 & 0.097$\pm$0.005 \\
2      & 6e-5$\pm$2e-5     & $\cdots$          & 0.001$\pm$0.0003     & 0.005$\pm$0.001     &  0.0055$\pm$0.0005 & 0.048$\pm$0.001 & 0.166$\pm$0.005 & 0.347$\pm$0.007 &  0.651$\pm$0.006 &  0.83$\pm$0.001 & 0.97$\pm$0.01 & 1.17$\pm$0.01 \\
3      & 0.12$\pm$0.03     & 0.23$\pm$0.05     & $\cdots$             & 0.43$\pm$0.06       &0.25$\pm$0.03  & 0.91$\pm$0.02 & 1.68$\pm$0.04  &  2.79$\pm$0.06 & 2.09$\pm$0.09 & $\cdots$ & 4.1$\pm$0.1 & 5.0$\pm$0.1\\
4      & 0.0016$\pm$0.0003 & 0.0043$\pm$0.0008 & 0.007$\pm$0.0019     & 0.018$\pm$0.003     & 0.018$\pm$0.003 & 0.059$\pm$0.002 & 0.089$\pm$0.003 & 0.104$\pm$0.003 & 0.116$\pm$0.002 & 0.137$\pm$0.004 & 0.156$\pm$0.004 & 0.221$\pm$0.006 \\
5      & $\cdots$          & $\cdots$          & 4e-5$\pm$1e-5        &  23e-5$\pm$6e-5     & 0.0004$\pm$0.0002 & 0.0092$\pm$0.0003 & 0.03911$\pm$0.0009 & 0.076$\pm$0.002 & 0.147$\pm$0.003 & 0.203$\pm$0.004 & 0.245$\pm$0.007 & 0.382$\pm$0.007 \\
\hline 
\end{tabular}
\caption{Available photometry from 24\,$\mu$m to 1100\,$\mu$m (Jy)}
\label{photometry2:tab}
\begin{tabular}{ccccccccc}
\hline
Source &  MIPS\,1 &  MIPS\,2 &  MIPS\,3  & UKT14  &  SCUBA  & UKT14  & SCUBA & UKT14 \\
ID     &  24\,$\mu$m & 70\,$\mu$m & 160\,$\mu$m & 350\,$\mu$m & 450\,$\mu$m & 800\,$\mu$m & 850\,$\mu$m & 1100\,$\mu$m\\
\hline\\
1      &  0.172$\pm$0.005 & $\cdots$ & $\cdots$ & $\cdots$ & $\cdots$  & $\cdots$ & $\cdots$ &  $\cdots$ \\
2      &  4.5$\pm$0.3     & 11.2$\pm$0.5 &  $\cdots$ & 6.52$\pm$2.1 & 3.75$\pm$0.27 & 0.635$\pm$0.032 & $\cdots$ & 0.239$\pm$0.039  \\
3      &  6.6$\pm$0.8     & 6.5$\pm$2.2 & $\cdots$ & $\cdots$ & 2.540$\pm$0.06 & 0.9$\pm$0.06 &  $\cdots$ \\
4      &  0.88$\pm$0.08   & 1.5$\pm$0.2 & $\cdots$ & $\cdots$ & $\cdots$  & $\cdots$ & $\cdots$ &  $\cdots$  \\
5      &  0.97$\pm$0.04   & 1.4$\pm$0.1 & 2.5$\pm$1 & $\cdots$ & $\cdots$  & $\cdots$ & $\cdots$ &  $\cdots$  \\
\hline
\end{tabular}
\end{scriptsize}
\end{table}
\end{landscape}



\longtab{3}{
\begin{longtable}{cccccc}
\caption{Observed emission lines in \object{[CSH2001]13811} - ID\,1.}
\label{tab:ctf29}\\
\hline \hline 
Line id. & $\lambda$ &  F & $\Delta$F & FWHM & W$_\lambda$ \\
 ~       &  (\um)    & \multicolumn{2}{c}{(erg cm$^{-2}$ s$^{-1}$)} & \AA & \AA \\
\hline\\[-3pt] 
\endfirsthead
\caption{continued.}\\
\hline\hline
Line id. & $\lambda$ &  F & $\Delta$F & FWHM & W$_\lambda$ \\
 ~       &  (\um)    & \multicolumn{2}{c}{(erg cm$^{-2}$ s$^{-1}$)} & \AA & \AA \\
\hline
\endhead
\hline\\[-3pt]
\endfoot
\multicolumn{6}{c}{EFOSC \& SofI - NTT}
\\
H$\alpha$                    	& 0.656 &    9.53e-14 & 1.00e-14 &  11.0&  -11.01 \\
Pa\,$\delta$            	& 1.005 &    8e-15 & 3e-15 &  16.8&  -2.0 \\
\cip        			& 1.069 &    7.42e-15 & 2.00e-15 &  18.0&  -18.0 \\
Pa\,$\gamma$			& 1.094 &    1.09e-14 & 0.30e-14 &  19.1&  -2.8 \\
Pa\,$\beta$			& 1.282 &    1.8e-14 & 0.30e-14 &  21.6&   -4.0  \\ 
\brg				& 2.166 &    2.96e-14 & 0.50e-14 &  55.0&   -5.1 \\
\hline\\[2pt]
\hline \hline\\[-4pt]
Line id. & $\lambda$ &  F & $\Delta$F & FWHM & W$_\lambda$ \\
 ~       &  (\um)    & \multicolumn{2}{c}{(Jy\,\um)} & (\um)&  (\um)\\
\hline\\[-5pt]
\multicolumn{6}{c}{IRS - Spitzer}\\
H$_2$               &   17.0 & 0.010 & 0.002 & 0.25 &  -0.07\\
\fe                 &   22.9 & 0.014 & 0.003 & 0.24 &  -0.08 \\ 
\fe                 &   24.5 & 0.007 & 0.003 & 0.30 &  -0.04 \\ 
\SiIII              &   25.6 & 0.028 & 0.005 & 0.40 &  -0.16 \\
H$_2$               &   28.2 & 0.031 & 0.005 & 0.40 &  -0.16 \\
\siii               &   33.5 & 0.053 & 0.008 & 0.6  &  -0.27 \\
\fe                 &   35.8 & 0.023 & 0.007 & 0.24 &  -0.09 \\
\fii                &   36.5 & 0.068 & 0.008 & 0.36 &  -0.27 \\
\hline 
\end{longtable}
}

\longtab{4}{
\begin{longtable}{cccccc}
\caption{\label{tab:ctf50} Observed emission lines in \object{[CTF93]\,50} - ID\,2.  }\\
\hline \hline 
Line id. & $\lambda$ &  F & $\Delta$F & FWHM & W$_\lambda$ \\
 ~       &  (\um)    & \multicolumn{2}{c}{(erg cm$^{-2}$ s$^{-1}$)} & \AA & \AA \\
\hline\\[-3pt] 
\endfirsthead
\caption{continued.}\\
\hline\hline
Line id. & $\lambda$ &  F & $\Delta$F & FWHM & W$_\lambda$ \\
 ~       &  (\um)    & \multicolumn{2}{c}{(Jy\,\um)} & (\um)&  (\um)\\
\hline\\[-3pt]
\endhead
\hline
\endfoot
\multicolumn{6}{c}{EFOSC \& SofI - NTT}
\\
\oi\,$^1$D$_2 - ^3$P$_2$ 		& 0.630 &  7.66e-15  &  0.16e-15  &  8.322 &  -7.065 \\
\feii                                   & 0.635 &  0.96e-15  &  0.30e-15  &  11.8  &  -0.87  \\
\oi\,$^1$D$_2 - ^3$P$_1$ 		& 0.636 &  2.66e-15  &  0.30e-15  &  9.8   &  -2.343 \\
\feii                                  & 0.646 &  0.78e-15  &  0.30e-15  &  8.90  &  -0.564 \\
\fei                                   & 0.649 &   1.06e-15  &  0.40e-15  &  12.80 &  -0.857 \\
\feii                                  & 0.651 &   1.10e-15  &  0.40e-15  &  11.00 &  -0.6   \\
\Nii\,$^1$D$_2 - ^3$P$_0$               & 0.653 &  4.43e-14 &  0.50e-14	&  11.8  &-1.246     \\
\hline 
\end{longtable}
}

\longtab{5}{
\begin{longtable}{cccccc}
\caption{\label{tab:ctf47} Observed emission lines in \object{[CTF93]\,47} - ID\,3.  }\\
\hline \hline 
Line id. & $\lambda$ &  F & $\Delta$F & FWHM & W$_\lambda$ \\
 ~       &  (\um)    & \multicolumn{2}{c}{(erg cm$^{-2}$ s$^{-1}$)} & \AA & \AA \\
\hline\\[-3pt] 
\endfirsthead
\caption{continued.}\\
\hline\hline
Line id. & $\lambda$ &  F & $\Delta$F & FWHM & W$_\lambda$ \\
 ~       &  (\um)    & \multicolumn{2}{c}{(erg cm$^{-2}$ s$^{-1}$)} & \AA & \AA \\
\hline\\[-3pt] 
\endhead
\hline
\endfoot
\multicolumn{6}{c}{EFOSC \& SofI - NTT}
\\
\feii                                  & 0.615 &  3.17e-13  & 0.10e-13   &  9.77   & -1.755   \\
\feii                                  & 0.625 &  6.69e-13  & 0.10e-13   &  14.4   & -3.68    \\ 
\feii                        		& 0.635 & 3.13e-13  & 0.10e-13   &  13.10  & -1.781   \\
\feii                       		& 0.637 & 3.17e-13  & 0.10e-13   &  13.00  & -1.754   \\
\feii                       		& 0.642 & 2.32e-13  & 0.10e-13   &  11.10  & -1.27    \\
\fei                                   & 0.643 &  3.04e-13  & 0.10e-13   &  11.80  & -1.67    \\
\feii                                  & 0.646 & 6.78e-13  & 0.10e-13   &  10.00  & -3.725   \\
\hline
\end{longtable}
}

\longtab{6}{
\begin{longtable}{cccccc}
\caption{\label{tab:ctf32} Observed emission lines in \object{[CTF93]\,32} - ID\,4.}\\
\hline \hline 
Line id. & $\lambda$ &  F & $\Delta$F & FWHM & W$_\lambda$ \\
 ~       &  (\um)    & \multicolumn{2}{c}{(erg cm$^{-2}$ s$^{-1}$)} & \AA & \AA \\
\hline\\[-3pt]  
\endfirsthead
\caption{continued.}\\
\hline\hline
Line id. & $\lambda$ &  F & $\Delta$F & FWHM & W$_\lambda$ \\
 ~       &  (\um)    & \multicolumn{2}{c}{(erg cm$^{-2}$ s$^{-1}$)} & \AA & \AA \\
\hline\\[-3pt] 
\endhead
\hline
\endfoot
\multicolumn{6}{c}{EFOSC \& SofI - NTT}
\\
\oi\,$^1$D$_2 - ^3$P$_2$                & 0.630 &     1.30e-14  &  0.15e-14  &    9.28&  -2.728\\
\oi\,$^1$D$_2 - ^3$P$_1$                & 0.636 &    5.85e-15  &  1.50e-15  &   13.90&  -1.164\\
H$\alpha$                    		& 0.656 &     8.23e-13  &  0.20e-14  &   10.4 &  -143.1\\
\hei\,$^3$Po$_2-0 - ^3$S$_1$               & 0.667 &     7.15e-15  &  0.20e-14  &   11.50&  -1.35\\
Pa\,18  +                              & 0.844 & ~ & ~ & \\
\oip\,$^3$P$_0 - ^3$S$_1^*$		& 0.845 &     1.28e-14  &  0.40e-14  &   10.4 &  -2.274\\
\oip\,$^3$P$_2 - ^3$S$_1^*$		& ~	&   ~	   &   ~ &	\\
\hline
\end{longtable}
}

\longtab{7}{
\begin{longtable}{cccccc}
\caption{\label{tab:ctf72} Observed emission lines in \object{[CTF93]\,72} - ID\,5.  }\\
\hline \hline 
Line id. & $\lambda$ &  F & $\Delta$F & FWHM & W$_\lambda$ \\
 ~       &  (\um)    & \multicolumn{2}{c}{(erg cm$^{-2}$ s$^{-1}$)} & \AA & \AA \\
\hline\\[-3pt] 
\endfirsthead
\caption{continued.}\\
\hline\hline
Line id. & $\lambda$ &  F & $\Delta$F & FWHM & W$_\lambda$ \\
 ~       &  (\um)    & \multicolumn{2}{c}{(erg cm$^{-2}$ s$^{-1}$)} & \AA & \AA \\
\hline\\[-3pt]
\endhead
\hline
\endfoot
\multicolumn{6}{c}{EFOSC \& SofI - NTT}
\\
H$\alpha$                    		& 0.656 &  1.33e-15   &  0.20e-15  &    7.75 &     \\
Pa\,20                                 & 0.839 &  1.58e-15   &  0.30e-15  &   11.9  & -24.5      \\
Pa\,18                                 & 0.844 &   1.46e-15   &  0.30e-15  &   15.14 & -26.3  \\
Pa\,16                                & 0.850 & 4.53e-16   &  0.30e-15  &   10.7  & -5.7 \\
\caii\,$^2$P$_{3/2}-^2$D$_{3/2}$	& 0.850 & 4.15e-15   &  0.30e-15  &   10.6  & -19.0     \\
\caii\,$^2$P$_{3/2}-^2$D$_{5/2}$ + 	& 0.854 & 4.41e-15 &  0.30e-15  &  11.1 & -54.5\\
Pa\,15                                  & 0.855 &  3.77e-16   &  0.30e-16  &   12.1  & -4.5 \\
\hline
\end{longtable}
}

\longtab{8}{
\begin{longtable}{cccccc}
\caption{\label{tab:ctf83} Observed emission lines in \object{[CTF93]\,83} - ID\,6.}\\
\hline \hline 
Line id. & $\lambda$ &  F & $\Delta$F & FWHM & W$_\lambda$ \\
 ~       &  (\um)    & \multicolumn{2}{c}{(erg cm$^{-2}$ s$^{-1}$)} & \AA & \AA \\
\hline\\[-3pt] 
\endfirsthead
\caption{continued.}\\
\hline\hline
Line id. & $\lambda$ &  F & $\Delta$F & FWHM & W$_\lambda$ \\
 ~       &  (\um)    & \multicolumn{2}{c}{(erg cm$^{-2}$ s$^{-1}$)} & \AA & \AA \\
\hline\\[-3pt]
\endhead
\hline
\endfoot
\multicolumn{6}{c}{EFOSC \& SofI - NTT}
\\
H$_2$~1--0\,S(1)$^a$             & 2.122 &    2e-15 &   1e-15	& 36.00 &  -5.4 \\
\brg				& 2.166  &    4e-15  &  1e-15 & 82.35 & -6.5 \\
\hline\\[2pt]
\hline \hline\\[-4pt]
Line id. & $\lambda$ &  F & $\Delta$F & FWHM & W$_\lambda$ \\
 ~       &  (\um)    & \multicolumn{2}{c}{(Jy\,\um)} & (\um)&  (\um)\\
\hline\\[-5pt]
\multicolumn{6}{c}{IRS - Spitzer}\\
\SiIII              &   25.6 & 0.067 & 0.02 & 0.53  &  -0.04 \\
H$_2$               &   28.2 & 0.041 & 0.01 & 0.36 &  -0.02 \\
\siii + ?           &   33.5 & 0.12  & 0.02 & 0.56  &  -0.06 \\
\SiII               &   35.3 & 0.14 & 0.03 & 0.47  &  -0.06 \\
\hline
\end{longtable}
}
\longtab{9}{
\begin{longtable}{cccccc}
\caption{\label{tab:ctf62} Observed emission lines in \object{[CTF93]\,62} - ID\,7.}\\
\hline \hline 
Line id. & $\lambda$ &  F & $\Delta$F & FWHM & W$_\lambda$ \\
 ~       &  (\um)    & \multicolumn{2}{c}{(erg cm$^{-2}$ s$^{-1}$)} & \AA & \AA \\
\hline\\[-3pt]  
\endfirsthead
\caption{continued.}\\
\hline\hline
Line id. & $\lambda$ &  F & $\Delta$F & FWHM & W$_\lambda$ \\
 ~       &  (\um)    & \multicolumn{2}{c}{(erg cm$^{-2}$ s$^{-1}$)} & \AA & \AA \\
\hline\\[-3pt] 
\endhead
\hline
\endfoot
\multicolumn{6}{c}{EFOSC \& SofI - NTT}
\\
H$\alpha$                    	& 0.656 &    1.18e-13 &  1.00e-14 &  9.2&  -5.7 \\
\brg				& 2.166 &    8.47e-14 &  1.70e-14 &  53.4&  -3.1 \\
\hline\\[2pt]
\hline \hline\\[-4pt]
Line id. & $\lambda$ &  F & $\Delta$F & FWHM & W$_\lambda$ \\
 ~       &  (\um)    & \multicolumn{2}{c}{(Jy\,\um)} & (\um)&  (\um)\\
\hline\\[-5pt]
\multicolumn{6}{c}{IRS - Spitzer}\\
\SiIII              &   25.6 & 0.057 & 0.009 & 0.36  &  -0.02 \\
H$_2$               &   28.2 & 0.05  & 0.01 & 0.37   &  -0.02 \\
\siii               &   33.5 & 0.05  & 0.02 & 0.32  &  -0.06 \\
\SiII               &   34.8 & 0.06  & 0.02 & 0.4  &  -0.03 \\
\fii               &   36.5 & 0.11 & 0.03 & 0.5  &  -0.04 \\
\hline
\end{longtable}
}

\longtab{10}{
\begin{longtable}{cccccc}
\caption{\label{tab:ctf79} Observed emission lines in \object{[CTF93]\,79} - ID\,8.}\\
\hline \hline 
Line id. & $\lambda$ &  F & $\Delta$F & FWHM & W$_\lambda$ \\
 ~       &  (\um)    & \multicolumn{2}{c}{(erg cm$^{-2}$ s$^{-1}$)} & \AA & \AA \\
\hline \\[-3pt]
\endfirsthead
\caption{continued.}\\
\hline\hline
Line id. & $\lambda$ &  F & $\Delta$F & FWHM & W$_\lambda$ \\
 ~       &  (\um)    & \multicolumn{2}{c}{(erg cm$^{-2}$ s$^{-1}$)} & \AA & \AA \\
\hline
\endhead
\hline
\endfoot
\multicolumn{6}{c}{EFOSC \& SofI - NTT}
\\
\hei                 	       & 2.058 &     1.04e-12  &  0.30e-13 & 34.60 & -3.346 \\
\brg				& 2.166 &   1.38e-12  &  1.00e-13 & 45.42 & -3.639 \\
\hline
\end{longtable}
}

\longtab{11}{
\begin{longtable}{cccccc}
\caption{\label{tab:ctf99} Observed emission lines in \object{[CTF93]\,99} - ID\,9.  }\\
\hline \hline 
Line id. & $\lambda$ &  F & $\Delta$F & FWHM & W$_\lambda$ \\
 ~       &  (\um)    & \multicolumn{2}{c}{(erg cm$^{-2}$ s$^{-1}$)} & \AA & \AA \\
\hline \\[-3pt] 
\endfirsthead
\caption{continued.}\\
\hline\hline
Line id. & $\lambda$ &  F & $\Delta$F & FWHM & W$_\lambda$ \\
 ~       &  (\um)    & \multicolumn{2}{c}{(erg cm$^{-2}$ s$^{-1}$)} & \AA & \AA \\
\hline
\endhead
\hline
\endfoot
\multicolumn{6}{c}{EFOSC \& SofI - NTT}
\\
H$\alpha$                    		& 0.656 &   1.34e-13   &  0.02e-13   &  9.02  & -16.99   \\
\cip\,(1)			& 1.069 &   3.92e-14   &  1.00e-14   &  33.2  & -2.097   \\
\hei 		                   & 1.083 &     3.81e-14   &  0.80e-14   &  14.95 & -1.93   \\
\hei  PCygni	                & 1.082 &   -2.30e-14 &   &       	 10.2 &  1.2 \\
\cip                    ?        & 1.167 &  3.11e-14   &  1.00e-14   &  19.05 & -1.359   \\
\cip\,(24)			& 1.175 &   3.08e-14   &  1.00e-14   &  16.0  & -1.349   \\
?                               & 1.263 &   5.06e-14   &  1.20e-14   &  36.2  &  -1.99   \\
\hei ?                           & 1.276  &      3.14e-14   &  1.00e-14   &  17.5  &  -1.22   \\
Pa\,$\beta$			& 1.282     &      4.20e-13   &  1.20e-14   &  28.03 &  -1.64   \\ 
\hei                             & 2.058 &  5.42e-14   &  1.00e-14   &  28.86 &  -2.02   \\
H$_2$~1--0\,S(1)               & 2.122  &   3.05e-14   &  1.00e-14   &  28.86 &  -1.3   \\
\brg				& 2.166 &   9.80e-14   &  1.20e-14   &  64.21 &  -3.79   \\
\hline\\[2pt]
\hline \hline\\[-4pt]
Line id. & $\lambda$ &  F & $\Delta$F & FWHM & W$_\lambda$ \\
 ~       &  (\um)    & \multicolumn{2}{c}{(Jy\,\um)} & (\um)&  (\um)\\
\hline\\[-5pt]
\multicolumn{6}{c}{IRS - Spitzer}\\
\fe                 &   22.9 & 0.05 & 0.01 & 0.37 &  -0.04 \\
\hline
\end{longtable}
}

\longtab{12}{
\begin{longtable}{cccccc}
\caption{\label{tab:ctf87} Observed emission lines in \object{[CTF93]\,87} - ID\,10.}\\
\hline \hline 
Line id. & $\lambda$ &  F & $\Delta$F & FWHM & W$_\lambda$ \\
 ~       &  (\um)    & \multicolumn{2}{c}{(erg cm$^{-2}$ s$^{-1}$)} & \AA & \AA \\
\hline \\[-3pt] 
\endfirsthead
\caption{continued.}\\
\hline\hline
Line id. & $\lambda$ &  F & $\Delta$F & FWHM & W$_\lambda$ \\
 ~       &  (\um)    & \multicolumn{2}{c}{(erg cm$^{-2}$ s$^{-1}$)} & \AA & \AA \\
\hline \\[-3pt]
\endhead
\hline
\endfoot
\multicolumn{6}{c}{EFOSC \& SofI - NTT}
\\
\oi\,$^1$D$_2 - ^3$P$_2$         	& 0.630 &    2.64e-15   &   0.20e-15   &    4.2  & -17.25      \\
PCygni                                  & 0.629 &    -1.50e-15  &            &    4.9 &  7.75 \\
\oi\,$^1$D$_2 - ^3$P$_1$         	& 0.636 &  7.61e-16   &   2.00e-16   &   4.40  & -4.14       \\
PCygni                                  & 0.635 &      -3.40e-16 &           &     3.6 &  1.75  \\
H$\alpha$                    		& 0.656 &     3.05e-14   &   0.03e-14   &   10.12 & -96.09      \\
\hei\,$^3$Po$_2-0 - ^3$S$_1$               & 0.667 &      6.00e-16   &   3.00e-16   &   9.6   & -2.18       \\
\sii\,$^2$D$_{5/2} - ^4$S$_{3/2}$ +	& 0.672 &      0.72e-15   &   0.30e-15   &  10.00  & -2.561      \\
\hline
\end{longtable}
}

\longtab{13}{
\begin{longtable}{cccccc}
\caption{\label{tab:ctf104} Observed emission lines in \object{[CTF93]\,104} - ID\,11.}\\
\hline \hline 
Line id. & $\lambda$ &  F & $\Delta$F & FWHM & W$_\lambda$ \\
 ~       &  (\um)    & \multicolumn{2}{c}{(erg cm$^{-2}$ s$^{-1}$)} & \AA & \AA \\
\hline \\[-3pt] 
\endfirsthead
\caption{continued.}\\
\hline\hline
Line id. & $\lambda$ &  F & $\Delta$F & FWHM & W$_\lambda$ \\
 ~       &  (\um)    & \multicolumn{2}{c}{(erg cm$^{-2}$ s$^{-1}$)} & \AA & \AA \\
\hline \\[-3pt]
\endhead
\hline
\endfoot
\multicolumn{6}{c}{EFOSC \& SofI - NTT}
\\
\feii                                  & 0.615 &  2.70e-15  &   1.00e-15   &   11.3  & -0.462   \\
\oi\,$^1$D$_2 - ^3$P$_2$ 		& 0.630 &  2.00e-14  &   0.10e-14   &    9.61 & -3.142   \\
\oi\,$^1$D$_2 - ^3$P$_1$ 		& 0.636 &  6.47e-15  &   1.20e-15   &   12.00 & -0.937   \\
\feii                                  & 0.646 &   3.77e-15  &   1.20e-15   &   10.00 & -0.522   \\
\feii                                  & 0.651 &  2.90e-15  &   1.00e-15   &   12.00 & -0.376   \\
H$\alpha$                    		& 0.656 & 6.99e-13  &   0.01e-13   &   11.08 & -88.21   \\
\hei\,$^3$Po$_2-0 - ^3$S$_1$               & 0.667 &  1.69e-14  &   0.11e-14   &   10.95 & -2.037   \\
\hline
\end{longtable}
}

\longtab{14}{
\begin{longtable}{cccccc}
\caption{\label{tab:meag31} Observed emission lines in \object{Meag\,31} - ID\,12.}\\
\hline \hline 
Line id. & $\lambda$ &  F & $\Delta$F & FWHM & W$_\lambda$ \\
 ~       &  (\um)    & \multicolumn{2}{c}{(erg cm$^{-2}$ s$^{-1}$)} & \AA & \AA \\
\hline 
\endfirsthead
\caption{continued.}\\
\hline\hline
Line id. & $\lambda$ &  F & $\Delta$F & FWHM & W$_\lambda$ \\
 ~       &  (\um)    & \multicolumn{2}{c}{(erg cm$^{-2}$ s$^{-1}$)} & \AA & \AA \\
\hline \\[-3pt]
\endhead
\hline
\endfoot
\multicolumn{6}{c}{EFOSC \& SofI - NTT}
\\
H$_2$\,1--0 S(1)                & 2.122 &   7.2e-15   & 1.5e-15 &	 33.0 & -4.8 \\
\brg    			& 2.166  &  6.4e-15 &   1.5e-15 &	 30.2 & -4.3 \\
\hline
\end{longtable}
}

\longtab{15}{
\begin{longtable}{cccccc}
\caption{\label{tab:ctf146-2} Observed emission lines in \object{[CTF93]\,146-2} - ID\,13.}\\
\hline \hline 
Line id. & $\lambda$ &  F & $\Delta$F & FWHM & W$_\lambda$ \\
 ~       &  (\um)    & \multicolumn{2}{c}{(erg cm$^{-2}$ s$^{-1}$)} & \AA & \AA \\
\hline \\[-3pt] 
\endfirsthead
\caption{continued.}\\
\hline\hline
Line id. & $\lambda$ &  F & $\Delta$F & FWHM & W$_\lambda$ \\
 ~       &  (\um)    & \multicolumn{2}{c}{(erg cm$^{-2}$ s$^{-1}$)} & \AA & \AA \\
\hline \\[-3pt]
\endhead
\hline
\endfoot
\endfoot
\multicolumn{6}{c}{EFOSC \& SofI - NTT}
\\
\brg$^a$				& 2.166  &  2.00e-15  &    1.00e-15  &    ?  &  ?    \\
\hline\\[2pt]
\hline \hline\\[-4pt]
Line id. & $\lambda$ &  F & $\Delta$F & FWHM & W$_\lambda$ \\
 ~       &  (\um)    & \multicolumn{2}{c}{(Jy\,\um)} & (\um)&  (\um)\\
\hline\\[-5pt]
\multicolumn{6}{c}{IRS - Spitzer}\\
\fe                 &   23.3 & 0.03 & 0.01 & 0.25 &  -0.02 \\
H$_2$               &   28.2 & 0.05 & 0.01 & 0.4 &  -0.02 \\
\siii               &   33.5 & 0.15 & 0.02 & 0.4  &  -0.04 \\
\SiII               &   35.3 & 0.15 & 0.02 & 0.5  &  -0.05 \\
\fe                 &   35.8 & 0.16 & 0.01 & 0.4  &  -0.04 \\
\fii                &   36.5 & 0.14 & 0.07 & 0.3 &  -0.03 \\
\hline
$^a$ S/N ratio below 3$\sigma$\\
\end{longtable}
}

\longtab{16}{
\begin{longtable}{cccccc}
\caption{\label{tab:ctf146-1} Observed emission lines in \object{[CTF93]\,146-1} - ID\,14.}\\
\hline \hline 
Line id. & $\lambda$ &  F & $\Delta$F & FWHM & W$_\lambda$ \\
 ~       &  (\um)    & \multicolumn{2}{c}{(erg cm$^{-2}$ s$^{-1}$)} & \AA & \AA \\
\hline \\[-3pt]  
\endfirsthead
\caption{continued.}\\
\hline\hline
Line id. & $\lambda$ &  F & $\Delta$F & FWHM & W$_\lambda$ \\
 ~       &  (\um)    & \multicolumn{2}{c}{(erg cm$^{-2}$ s$^{-1}$)} & \AA & \AA \\
\hline \\[-3pt] 
\endhead
\hline
\endfoot
\endfoot
\multicolumn{6}{c}{EFOSC \& SofI - NTT}
\\
\fe\,$a^4\!D_{7/2}-a^4\!F_{9/2}$ & 1.644 &   5.69e-15  &  0.70e-15 & 36.66 & -6.626  \\
\fe\,$a^4\!D_{1/2}-a^4\!F_{5/2}$$^a$ & 1.664 &    1.91e-15  &    0.70e-15  &     33.23 &   -2.074   \\
\fe\,$a^4\!D_{5/2}-a^4\!F_{7/2}$ & 1.677 &   3.6e-15 &   0.50e-15 &	 39.0 & -3.8 \\
\hei                            & 2.058 &  1.71e-14  &    0.50e-14  &    46.45  &  -6.946    \\
\brg				& 2.166  &  3.05e-14  &    0.30e-14  &    45.68  &  -11.06    \\
\tii   ?                         &  2.185 &  6.65e-15  &    2.50e-15  &    32.10  &  -2.375    \\
\nai\,$^2$D$_{3/2-5/2}-^2$P$^0$(m)  & 2.207 &  1.60e-14  &    2.50e-15  &    43.30  &  -5.63     \\
\end{longtable}
}

\longtab{17}{
\begin{longtable}{cccccc}
\caption{\label{tab:ctf211} Observed emission lines in \object{[CTF93]\,211} - ID\,15.}\\
\hline \hline 
Line id. & $\lambda$ &  F & $\Delta$F & FWHM & W$_\lambda$ \\
 ~       &  (\um)    & \multicolumn{2}{c}{(erg cm$^{-2}$ s$^{-1}$)} & \AA & \AA \\
\hline\\[-3pt]   
\endfirsthead
\caption{continued.}\\
\hline\hline
Line id. & $\lambda$ &  F & $\Delta$F & FWHM & W$_\lambda$ \\
 ~       &  (\um)    & \multicolumn{2}{c}{(erg cm$^{-2}$ s$^{-1}$)} & \AA & \AA \\
\hline\\[-3pt]  
\endhead
\hline
\endfoot
\multicolumn{6}{c}{EFOSC \& SofI - NTT}
\\
\oi\,$^1$D$_2 - ^3$P$_2$                & 0.630 &   5.44e-15   &   0.50e-15    &     7.81  &  -12.59   \\
\oi\,$^1$D$_2 - ^3$P$_1$                & 0.636 &   2.66e-15   &   0.50e-15    &    10.44  &  -5.7	   \\
H$\alpha$                    		& 0.656 &    5.18e-14   &   0.05e-14    &    11.11  &  -95.61   \\
\Nii\,$^1$D$_2 - ^3$P$_2$               & 0.659 &  2.95e-15   &   0.50e-15    &    15.1   &  -4.81    \\
\sii\,$^2$D$_{5/2} - ^4$S$_{3/2}$ +	& 0.672 &   2.63e-15   &   0.40e-15    &     9.8   &  -4.49    \\
\sii\,$^2$D$_{3/2} - ^4$S$_{3/2}$	& 0.673 &    3.22e-15   &   0.40e-15    &    12.1   &  -5.51    \\
Pa\,18  +                              & 0.844 & ~ & ~ & \\
\hline
\end{longtable}
}
\longtab{18}{
\begin{longtable}{cccccc}
\caption{\label{tab:ctf187} Observed emission lines in \object{[CTF93]\,187} - ID\,16.}\\
\hline \hline 
Line id. & $\lambda$ &  F & $\Delta$F & FWHM & W$_\lambda$ \\
 ~       &  (\um)    & \multicolumn{2}{c}{(erg cm$^{-2}$ s$^{-1}$)} & \AA & \AA \\
\hline\\[-3pt]  
\endfirsthead
\caption{continued.}\\
\hline\hline
Line id. & $\lambda$ &  F & $\Delta$F & FWHM & W$_\lambda$ \\
 ~       &  (\um)    & \multicolumn{2}{c}{(erg cm$^{-2}$ s$^{-1}$)} & \AA & \AA \\
\hline
\endhead
\hline
\endfoot
\multicolumn{6}{c}{EFOSC \& SofI - NTT}
\\
\brg				& 2.166  &  7.00e-14 &   1.00e-14 &	 43.2 & -2.5 \\
\hline\\[2pt]
\hline \hline\\[-4pt]
Line id. & $\lambda$ &  F & $\Delta$F & FWHM & W$_\lambda$ \\
 ~       &  (\um)    & \multicolumn{2}{c}{(Jy\,\um)} & (\um)&  (\um)\\
\hline\\[-5pt] 
\multicolumn{6}{c}{IRS - Spitzer}\\
\sii ?              &   25.2 & 0.2 & 0.06  & 0.8  & -0.07 \\
\SiIII              &   25.6 & 0.2 & 0.06 & 0.6  &  -0.05 \\
?                   &   26.9 & 0.3 & 0.06 & 0.9 & -0.08\\
\SiII               &   34.8 & 0.13 & 0.04 & 0.4  &  -0.04 \\
\SiII               &   35.3 & 0.13 & 0.04 & 0.5  &  -0.04 \\
\hline
\end{longtable}
}

\longtab{19}{
\begin{longtable}{cccccc}
\caption{\label{tab:ctf168} Observed emission lines in \object{[CTF93]\,168} - ID\,17.}\\
\hline \hline 
Line id. & $\lambda$ &  F & $\Delta$F & FWHM & W$_\lambda$ \\
 ~       &  (\um)    & \multicolumn{2}{c}{(erg cm$^{-2}$ s$^{-1}$)} & \AA & \AA \\
\hline\\[-3pt]  
\endfirsthead
\caption{continued.}\\
\hline\hline
Line id. & $\lambda$ &  F & $\Delta$F & FWHM & W$_\lambda$ \\
 ~       &  (\um)    & \multicolumn{2}{c}{(erg cm$^{-2}$ s$^{-1}$)} & \AA & \AA \\
\hline\\[-3pt] 
\endhead
\hline
\endfoot
\multicolumn{6}{c}{EFOSC \& SofI - NTT}
\\
\brg				& 2.166  &  1.80e-14 &   0.30e-14 &	 45.6 & -6.9 \\
\hline\\[2pt]
\hline \hline\\[-4pt]
Line id. & $\lambda$ &  F & $\Delta$F & FWHM & W$_\lambda$ \\
 ~       &  (\um)    & \multicolumn{2}{c}{(Jy\,\um)} & (\um)&  (\um)\\
\hline\\[-5pt] 
\multicolumn{6}{c}{IRS - Spitzer}\\
\sii ?              &   25.2 & 0.05 & 0.01  & 0.3  & -0.05 \\
?                   &   29.7 & 0.06 & 0.02  & 0.4 & -0.01\\
\siii               &   33.5 & 0.09 & 0.02 &  0.5  &  -0.08 \\
\SiII               &   34.8 & 0.13 & 0.04 & 0.4  &  -0.04 \\
\hline
\end{longtable}
}
\longtab{20}{
\begin{longtable}{cccccc}
\caption{\label{tab:ctf191} Observed emission lines in \object{[CTF93]\,191} - ID\,18.  }\\
\hline \hline 
Line id. & $\lambda$ &  F & $\Delta$F & FWHM & W$_\lambda$ \\
 ~       &  (\um)    & \multicolumn{2}{c}{(erg cm$^{-2}$ s$^{-1}$)} & \AA & \AA \\
\hline\\[-3pt] 
\endfirsthead
\caption{continued.}\\
\hline\hline
Line id. & $\lambda$ &  F & $\Delta$F & FWHM & W$_\lambda$ \\
 ~       &  (\um)    & \multicolumn{2}{c}{(erg cm$^{-2}$ s$^{-1}$)} & \AA & \AA \\
\hline\\[-3pt]
\endhead
\hline
\endfoot
\multicolumn{6}{c}{EFOSC \& SofI - NTT}
\\
\feii                                  & 0.615 &  1.16e-13 &  5.00e-15	&  21.3  &-3.432     \\
\feii                                  & 0.617 &  3.55e-14 &  5.00e-15	&  11.5  &-1.053     \\
\fei                                   & 0.619 &  1.92e-13 &  5.00e-15	&  11.5  &-1.053     \\   
\feii                                  & 0.625 &  8.67e-14 &  5.00e-15	&  21.0  &-4.95      \\
\oi\,$^1$D$_2 - ^3$P$_2$ 		& 0.630 &  9.12e-14 &  0.40e-14	&  11.42 &-2.649     \\
\oi\,$^1$D$_2 - ^3$P$_1$ 		& 0.636 &  2.98e-14 &  0.40e-14	&  11.30 &-0.871     \\
\fei                                   & 0.643 &  6.61e-14 &  0.40e-14	&  10.8  &-1.926     \\
\hline
\end{longtable}
}
\longtab{21}{
\begin{longtable}{cccccc}
\caption{\label{tab:ctf246A} Observed emission lines in \object{[CTF93]\,246\,A} - ID\,19.}\\
\hline \hline 
Line id. & $\lambda$ &  F & $\Delta$F & FWHM & W$_\lambda$ \\
 ~       &  (\um)    & \multicolumn{2}{c}{(erg cm$^{-2}$ s$^{-1}$)} & \AA & \AA \\
\hline\\[-3pt] 
\endfirsthead
\caption{continued.}\\
\hline\hline
Line id. & $\lambda$ &  F & $\Delta$F & FWHM & W$_\lambda$ \\
 ~       &  (\um)    & \multicolumn{2}{c}{(erg cm$^{-2}$ s$^{-1}$)} & \AA & \AA \\
\hline\\[-3pt]
\endhead
\hline
\endfoot
\multicolumn{6}{c}{EFOSC \& SofI - NTT}
\\
\brg 				& 2.166  &   3.0e-15  &  1.00e-15   &   46.3 &  -9.8   \\
\hline\\[2pt]
\hline \hline\\[-4pt]
Line id. & $\lambda$ &  F & $\Delta$F & FWHM & W$_\lambda$ \\
 ~       &  (\um)    & \multicolumn{2}{c}{(Jy\,\um)} & (\um)&  (\um)\\
\hline\\[-5pt] 
\multicolumn{6}{c}{IRS - Spitzer}\\
\SiIII              &   25.6 & 0.05 & 0.01 & 0.4  &  -0.12 \\
\siii               &   33.5 & 0.04 & 0.01 &  0.5  &  -0.08 \\
\fii                &   36.5 & 0.06 & 0.02 & 0.4 &  -0.13 \\
\hline
\end{longtable}
}

\longtab{22}{
\begin{longtable}{cccccc}
\caption{\label{tab:ctf186} Observed emission lines in \object{[CTF93]\,186} - ID\,20.}\\
\hline \hline 
Line id. & $\lambda$ &  F & $\Delta$F & FWHM & W$_\lambda$ \\
 ~       &  (\um)    & \multicolumn{2}{c}{(erg cm$^{-2}$ s$^{-1}$)} & \AA & \AA \\
\hline\\[-3pt] 
\endfirsthead
\caption{continued.}\\
\hline\hline
Line id. & $\lambda$ &  F & $\Delta$F & FWHM & W$_\lambda$ \\
 ~       &  (\um)    & \multicolumn{2}{c}{(erg cm$^{-2}$ s$^{-1}$)} & \AA & \AA \\
\hline\\[-3pt]
\endhead
\hline
\endfoot
\endfoot
\multicolumn{6}{c}{EFOSC \& SofI - NTT}
\\
\brg				& 2.166  &  1.90e-14 &   0.30e-14 &	 54.2 & -7.1 \\
\hline\\[2pt]
\hline \hline\\[-4pt]
Line id. & $\lambda$ &  F & $\Delta$F & FWHM & W$_\lambda$ \\
 ~       &  (\um)    & \multicolumn{2}{c}{(Jy\,\um)} & (\um)&  (\um)\\
\hline\\[-5pt] 
\multicolumn{6}{c}{IRS - Spitzer}\\
\siii               &   33.5 & 0.15 & 0.02 &  0.5  &  -0.05 \\
\hline
\end{longtable}
}
\longtab{23}{
\begin{longtable}{cccccc}
\caption{\label{tab:ctf246B} Observed emission lines in \object{[CTF93]\,246\,B} - ID\,21.}\\
\hline \hline 
Line id. & $\lambda$ &  F & $\Delta$F & FWHM & W$_\lambda$ \\
 ~       &  (\um)    & \multicolumn{2}{c}{(erg cm$^{-2}$ s$^{-1}$)} & \AA & \AA \\
\hline\\[-3pt] 
\endfirsthead
\caption{continued.}\\
\hline\hline
Line id. & $\lambda$ &  F & $\Delta$F & FWHM & W$_\lambda$ \\
 ~       &  (\um)    & \multicolumn{2}{c}{(erg cm$^{-2}$ s$^{-1}$)} & \AA & \AA \\
\hline\\[-3pt]
\endhead
\hline
\endfoot
\multicolumn{6}{c}{EFOSC \& SofI - NTT}
\\
\brg 				& 2.166  & 7.5e-15  &  1.00e-15   &   51.7 &  -7.3   \\
\hline\\[2pt]
\hline \hline\\[-4pt]
Line id. & $\lambda$ &  F & $\Delta$F & FWHM & W$_\lambda$ \\
 ~       &  (\um)    & \multicolumn{2}{c}{(Jy\,\um)} & (\um)&  (\um)\\
\hline\\[-5pt] 
\multicolumn{6}{c}{IRS - Spitzer}\\
\SiIII              &   25.6 & 0.015 & 0.02 & 0.3  &  -0.03 \\
\SiII               &   34.8 & 0.09 & 0.02 & 0.3  &  -0.2 \\
\hline
\end{longtable}
}

\longtab{24}{
\begin{longtable}{cccccc}
\caption{\label{tab:ctf237-2} Observed emission lines in \object{[CTF93]\,237-2} - ID\,22.}\\
\hline \hline 
Line id. & $\lambda$ &  F & $\Delta$F & FWHM & W$_\lambda$ \\
 ~       &  (\um)    & \multicolumn{2}{c}{(erg cm$^{-2}$ s$^{-1}$)} & \AA & \AA \\
\hline\\[-3pt] 
\endfirsthead
\caption{continued.}\\
\hline\hline
Line id. & $\lambda$ &  F & $\Delta$F & FWHM & W$_\lambda$ \\
 ~       &  (\um)    & \multicolumn{2}{c}{(erg cm$^{-2}$ s$^{-1}$)} & \AA & \AA \\
\hline
\endhead
\hline
\endfoot
\endfoot
\multicolumn{6}{c}{EFOSC \& SofI - NTT}
\\
H$_2$\,1--0 S(1)                & 2.122 &   2.0e-15   & 1.0e-15 &	 38.0 & -2.8 \\
\brg$^a$			& 2.166  &  $<$1.0e-15 & $\cdots$ &	$\cdots$ & $\cdots$ \\
\hline\\
$^a$ S/N ratio below 3$\sigma$\\

\end{longtable}
}

\longtab{25}{
\begin{longtable}{cccccc}
\caption{\label{tab:ctf237-1} Observed emission lines in \object{[CTF93]\,237-1} - ID\,23.}\\
\hline \hline 
Line id. & $\lambda$ &  F & $\Delta$F & FWHM & W$_\lambda$ \\
 ~       &  (\um)    & \multicolumn{2}{c}{(erg cm$^{-2}$ s$^{-1}$)} & \AA & \AA \\
\hline\\[-3pt] 
\endfirsthead
\caption{continued.}\\
\hline\hline
Line id. & $\lambda$ &  F & $\Delta$F & FWHM & W$_\lambda$ \\
 ~       &  (\um)    & \multicolumn{2}{c}{(erg cm$^{-2}$ s$^{-1}$)} & \AA & \AA \\
\hline
\endhead
\hline
\endfoot
\multicolumn{6}{c}{EFOSC \& SofI - NTT}
\\
H$_2$\,1--0 S(1)                & 2.122 &   8.1e-15   & 2.5e-15 &	 38.0 & -3.4 \\
\brg    			& 2.166  &  1.20e-14 &   0.30e-14 &	 29.0 & -5.0 \\
\hline
\end{longtable}
}

\longtab{26}{
\begin{longtable}{cccccc}
\caption{\label{tab:ctf216-1} Observed emission lines in \object{[CTF93]\,216-1} - ID\,24.}\\
\hline \hline 
Line id. & $\lambda$ &  F & $\Delta$F & FWHM & W$_\lambda$ \\
 ~       &  (\um)    & \multicolumn{2}{c}{(erg cm$^{-2}$ s$^{-1}$)} & \AA & \AA \\
\hline\\[-3pt] 
\endfirsthead
\caption{continued.}\\
\hline\hline
Line id. & $\lambda$ &  F & $\Delta$F & FWHM & W$_\lambda$ \\
 ~       &  (\um)    & \multicolumn{2}{c}{(erg cm$^{-2}$ s$^{-1}$)} & \AA & \AA \\
\hline\\[-3pt]
\endhead
\hline
\endfoot
\endfoot
\multicolumn{6}{c}{EFOSC \& SofI - NTT}
\\
\fe\,$a^4\!D_{5/2}-a^4\!F_{9/2}$ & 1.534 &  7.09e-15 &   0.60e-15 & 33.11 & -15.3  \\
\fe\,$a^4\!D_{3/2}-a^4\!F_{7/2}$ & 1.600 &  8.00e-15  &  0.80e-15 &	 33.11 & -15.3 \\
\fe\,$a^4\!D_{7/2}-a^4\!F_{9/2}$ & 1.644 &   2.48e-14 &   0.50e-15 &	 29.43 & -45.8 \\
\fe\,$a^4\!D_{1/2}-a^4\!F_{5/2}$ & 1.664 &   3.34e-15 &   0.40e-15 &	 27.12 & -6.03 \\
\fe\,$a^4\!D_{5/2}-a^4\!F_{7/2}$ & 1.677 &   7.46e-15 &   0.50e-15 &	 33.01 & -13.2 \\
\fe\,$a^4\!D_{3/2}-a^4\!F_{5/2}$ & 1.712 &  3.01e-15 &   0.50e-15 &	 35.21 &  -4.89 \\
H$_2$\,1--0 S(8)$^a$  & 1.714 &    1.00e-15 &    0.40e-15 	 & 17.00 &  -4.89  \\
\end{longtable}
}

\longtab{27}{
\begin{longtable}{cccccc}
\caption{\label{tab:ctf216-2} Observed emission lines in \object{[CTF93]\,216-2} - ID\,25.}\\
\hline \hline 
Line id. & $\lambda$ &  F & $\Delta$F & FWHM & W$_\lambda$ \\
 ~       &  (\um)    & \multicolumn{2}{c}{(erg cm$^{-2}$ s$^{-1}$)} & \AA & \AA \\
\hline\\[-3pt] 
\endfirsthead
\caption{continued.}\\
\hline\hline
Line id. & $\lambda$ &  F & $\Delta$F & FWHM & W$_\lambda$ \\
 ~       &  (\um)    & \multicolumn{2}{c}{(erg cm$^{-2}$ s$^{-1}$)} & \AA & \AA \\
\hline\\[-3pt]
\endhead
\hline
\endfoot
\multicolumn{6}{c}{EFOSC \& SofI - NTT}
\\
\brg				& 2.166  &  9.6e-14 &   1.0e-14 &  59.05 & -3.6 \\
\hline
\end{longtable}
}

\longtab{28}{
\begin{longtable}{cccccc}
\caption{\label{tab:ctf245B2} Observed emission lines in \object{[CTF93]\,245\,B-2} - ID\,26.}\\
\hline \hline 
Line id. & $\lambda$ &  F & $\Delta$F & FWHM & W$_\lambda$ \\
 ~       &  (\um)    & \multicolumn{2}{c}{(erg cm$^{-2}$ s$^{-1}$)} & \AA & \AA \\
\hline\\[-3pt]
\endfirsthead
\caption{continued.}\\
\hline\hline
Line id. & $\lambda$ &  F & $\Delta$F & FWHM & W$_\lambda$ \\
 ~       &  (\um)    & \multicolumn{2}{c}{(erg cm$^{-2}$ s$^{-1}$)} & \AA & \AA \\
\hline\\[-3pt]
\endhead
\hline
\endfoot
\endfoot
\multicolumn{6}{c}{EFOSC \& SofI - NTT}
\\
\fe\,$a^4\!D_{7/2}-a^4\!F_{9/2}$ & 1.644 &  2.32e-14  &  0.50e-15   &   28.00 &  -41.21   \\
\fe\,$a^4\!D_{1/2}-a^4\!F_{5/2}$ & 1.664 &  3.58e-14  &  0.50e-15   &   28.93 &  -6.50    \\
\fe\,$a^4\!D_{5/2}-a^4\!F_{7/2}$ & 1.677 &   6.67e-15  &  0.50e-15   &   29.87 &  -11.59   \\
\fe\,$a^4\!D_{3/2}-a^4\!F_{5/2}$ & 1.712 &    2.94e-15  &  0.50e-15   &   36.17 &  -4.685   \\
H$_2$\,1--0 S(8)$^a$  & 1.714 &     1.00e-15  &  0.40e-15   &   17.00 &  -4.685   \\
\fe\,$a^4\!D_{1/2}-a^4\!F_{3/2}$ + & 1.745 & & & & \\
H$_2$\,1--0 S(7)            & 1.749 &  1.68e-15  &  0.50e-15   &   30.90 &  -2.166   \\
H$_2$\,1--0 S(1)                & 2.122 &    6.35e-15  &  0.60e-15   &   26.60 &  -4.51    \\
\fe\,$a^2\!P_{3/2}-a^4\!P_{3/2}$ & 2.133 &    2.12e-15  &  0.60e-15   &   36.72 &  -1.498   \\  
\brg$^a$				& 2.166  &   1.56e-14  &  0.80e-14   &   44.72 &  -11.06   \\
\hline\\
$^a$ S/N ratio below 3$\sigma$\\
\end{longtable}
}

\longtab{29}{
\begin{longtable}{cccccc}
\caption{\label{tab:ctf245B1} Observed emission lines in \object{[CTF93]\,245\,B-1} - ID\,27.}\\
\hline \hline 
Line id. & $\lambda$ &  F & $\Delta$F & FWHM & W$_\lambda$ \\
 ~       &  (\um)    & \multicolumn{2}{c}{(erg cm$^{-2}$ s$^{-1}$)} & \AA & \AA \\
\hline\\[-3pt] 
\endfirsthead
\caption{continued.}\\
\hline\hline
Line id. & $\lambda$ &  F & $\Delta$F & FWHM & W$_\lambda$ \\
 ~       &  (\um)    & \multicolumn{2}{c}{(erg cm$^{-2}$ s$^{-1}$)} & \AA & \AA \\
\hline\\[-3pt]
\endhead
\hline
\endfoot
\endfoot
\multicolumn{6}{c}{EFOSC \& SofI - NTT}
\\
Pa\,$\delta$            	& 1.005 &   2.31e-15  &    0.80e-15  &    24.25 &  -3.78	\\
\hei    			& 1.083 &  3.96e-15   &   0.90e-15   &    20.5  &  -3.94	\\
Pa\,$\gamma$			& 1.094 &   3.13e-15   &   1.00e-15   &    24.1  &  -2.96	\\
?                               & 1.135 &  4.56e-15   &   1.00e-15   &    23.17 &  -3.37	\\
\cip ?                               & 1.166 & 4.25e-15   &   1.00e-15   &    28.19 &  -2.76	\\
\cip                            & 1.175 &  2.97e-15   &   1.00e-15   &    20.0  &  -1.88	\\
?                               & 1.278 &  8.24e-15   &   2.00e-15   &    29.5  &   -3.3	\\
Pa\,$\beta$			& 1.282     &  1.43e-14   &   2.00e-15   &    28.33 &   -5.5	\\
\brg				& 2.166 &   4.37e-14   &   0.60e-14   &    58.97 &   -5.2	\\
\hline
\end{longtable}
}

\end{appendix}

\begin{appendix}
\section{Spitzer-IRAC flows in \object{[CTF93]\,216} }
\label{appendixB:sec}

\begin{figure}
 \centering
\includegraphics [width=8cm] {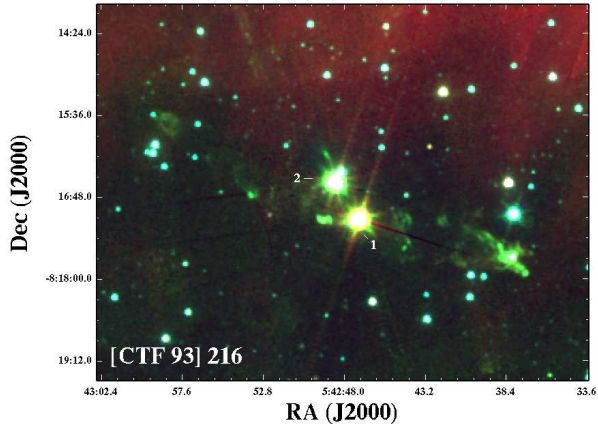}\\
  \caption{\textit{Spitzer/IRAC} three-colour image (3.6, 4.5, and 8.0\,$\mu$m - blue, red, and green, respectively) of the \object{[CTF93]\,216} system,
sources \#\,24 (-1), and \#\,25 (-2), along with their jets (in green). These images were taken in 2006, before the outburst of source \#\,25.
 \label{CTF216_img:fig}}
 \end{figure}

Figure~\ref{CTF216_img:fig} shows the \textit{Spitzer/IRAC} three-colour image (3.6, 4.5, and 8.0\,$\mu$m - blue, red, and green, respectively)
of the \object{[CTF93]\,216} binary flow system. The two precessing jets from source \#\,24 ([CTF93]\,216-1), and \#\,25 ([CTF93]\,216-2) are visible in green,
and they are roughly E-W, and NE-SW oriented, respectively.

\end{appendix}

\end{document}